\documentclass{svjour3}

\smartqed

\usepackage{alltt                                    % I like these
          , multirow
          , booktabs
          , listings
          , graphicx
          ,float
          ,verbatim
         ,mathtools
	   ,url
	   ,cite
	   ,amsmath
}
\usepackage[numbers]{natbib}     % this is a better citation system
\usepackage{syntax}
\usepackage[ruled]{algorithm}
\usepackage{enumitem}
\usepackage{threeparttable}
\usepackage{pbox}
\usepackage{algpseudocode}
\usepackage{framed}
\usepackage{hhline}
\usepackage{balance}
\usepackage{color, colortbl}
\usepackage{tikz}
\usepackage{verbatim}
\usepackage{framed}
\usepackage{amsmath}

\usepackage{tikz}
\usepackage{verbatim}
\usetikzlibrary{arrows,shapes,backgrounds}
\usepackage{etoolbox}
\usepackage{tikz}
\usetikzlibrary{tikzmark}
\usetikzlibrary{calc}
\usepackage[all]{nowidow}
\usepackage[justification=centering]{caption}
\usepackage{longtable}

\tikzstyle{vertex}=[ellipse,fill=black!25,minimum size=20pt, inner sep=0pt]
\tikzstyle{edge} = [draw,thin,-]
\tikzstyle{glabel} = [text width=1cm,text centered,font=\bf]
\pgfdeclarelayer{bg}    % declare background layer
\pgfsetlayers{bg,main}  % set the order of the layers (main is the standard layer)

\usepackage{expl3}
\ExplSyntaxOn
\newcommand\latinabbrev[1]{
  \peek_meaning:NTF . {% Same as \@ifnextchar
    #1\@}%
  { \peek_catcode:NTF a {% Check whether next char has same catcode as \'a, i.e., is a letter
      #1., \@ }%
    {#1., \@}}}
\ExplSyntaxOff

\algnewcommand\algorithmicswitch{\textbf{switch}}
\algnewcommand\algorithmiccase{\textbf{case}}
\algnewcommand\algorithmicassert{\texttt{assert}}
\algnewcommand\Assert[1]{\State \algorithmicassert(#1)}%
\algdef{SE}[SWITCH]{Switch}{EndSwitch}[1]{\algorithmicswitch\ #1\ \algorithmicdo}{\algorithmicend\ \algorithmicswitch}%
\algdef{SE}[CASE]{Case}{EndCase}[1]{\algorithmiccase\ #1}{\algorithmicend\ \algorithmiccase}%
\algtext*{EndSwitch}%
\algtext*{EndCase}%

\algnewcommand{\LineComment}[1]{\State \(\triangleright\) #1}

\definecolor{LightGray}{rgb}{.9 .9 .9}

\newsavebox{\supbox}% Superscript box
\newcommand{\bsup}{\begin{lrbox}{\supbox}$\tt\scriptstyle}
\newcommand{\esup}{$\end{lrbox}{}^{\usebox{\supbox}}}% Superscript end
\def\eg{\latinabbrev{e.g}}
\def\ie{\latinabbrev{i.e}}

\definecolor{lightpurple}{rgb}{0.8,0.8,1}
\definecolor{codebg}{RGB}{255,255,255}
\definecolor{commentcolor}{RGB}{11,140,11}
\newenvironment{frshaded}{%
	\MakeFramed {\FrameRestore}}%
{\endMakeFramed}

\errorcontextlines\maxdimen

\newcommand{\ALGtikzmarkcolor}{black}% customise this, if you want
\newcommand{\ALGtikzmarkextraindent}{4pt}% customise this, if you want
\newcommand{\ALGtikzmarkverticaloffsetstart}{-.5ex}% customise this, if you want
\newcommand{\ALGtikzmarkverticaloffsetend}{-.5ex}% customise this, if you want
\makeatletter
\newcounter{ALG@tikzmark@tempcnta}

\newcommand\ALG@tikzmark@start{%
	\global\let\ALG@tikzmark@last\ALG@tikzmark@starttext%
	\expandafter\edef\csname ALG@tikzmark@\theALG@nested\endcsname{\theALG@tikzmark@tempcnta}%
	\tikzmark{ALG@tikzmark@start@\csname ALG@tikzmark@\theALG@nested\endcsname}%
	\addtocounter{ALG@tikzmark@tempcnta}{1}%
}

\def\ALG@tikzmark@starttext{start}
\newcommand\ALG@tikzmark@end{%
	\ifx\ALG@tikzmark@last\ALG@tikzmark@starttext
	\else
	\tikzmark{ALG@tikzmark@end@\csname ALG@tikzmark@\theALG@nested\endcsname}%
	\tikz[overlay,remember picture] \draw[\ALGtikzmarkcolor] let \p{S}=($(pic cs:ALG@tikzmark@start@\csname ALG@tikzmark@\theALG@nested\endcsname)+(\ALGtikzmarkextraindent,\ALGtikzmarkverticaloffsetstart)$), \p{E}=($(pic cs:ALG@tikzmark@end@\csname ALG@tikzmark@\theALG@nested\endcsname)+(\ALGtikzmarkextraindent,\ALGtikzmarkverticaloffsetend)$) in (\x{S},\y{S})--(\x{S},\y{E});%
	\fi
	\gdef\ALG@tikzmark@last{end}%
}

\apptocmd{\ALG@beginblock}{\ALG@tikzmark@start}{}{\errmessage{failed to patch}}
\pretocmd{\ALG@endblock}{\ALG@tikzmark@end}{}{\errmessage{failed to patch}}
\makeatother

\AtBeginDocument{%
	\paperwidth=\dimexpr
	1in + \oddsidemargin
	+ \textwidth
	+ 1in + \oddsidemargin
	\relax
	\paperheight=\dimexpr
	1in + \topmargin
	+ \headheight + \headsep
	+ \textheight
	+ 1in + \topmargin
	\relax
	\usepackage[pass]{geometry}\relax
}

\begin{document}

\title{The Forgotten Role of Search Queries in IR-based Bug Localization: An Empirical Study}

\author{Mohammad M. Rahman~\and Foutse Khomh~\and Shamima Yeasmin~\and Chanchal K. Roy }


\institute{Mohammad Masudur Rahman$^\dagger$,~Foutse Khomh$^\$$,~Shamima Yeasmin$^\star$,~Chanchal K. Roy$^\star$ \at Dalhousie University$^\dagger$, Polytechnique Montreal$^\$$,  University of Saskatchewan$^\star$, Canada\\
	\email{masud.rahman@dal.ca, foutse.khomh@polymtl.ca, shamima.yeasmin@usask.ca, \\
	chanchal.roy@usask.ca}           %  \\
	%           \emph{Present address:} of F. Author  %  if needed
}

\maketitle

\begin{abstract}
Being light-weight and cost-effective, IR-based approaches for bug localization have shown promise in finding software bugs. However, the accuracy of these approaches heavily depends on their used bug reports. A significant number of bug reports contain only plain natural language texts. 
%and no hints for localization (e.g., program elements). 
According to existing studies, IR-based approaches cannot perform well 
when they use these bug reports as search queries. On the other hand, there is a piece of recent evidence that suggests that even these natural language-only reports contain enough good keywords that could help localize the bugs successfully. On one hand, these findings suggest that natural language-only bug reports might be a sufficient source for good query keywords.
%Such potentially contradictory findings suggest that natural language-only bug reports might not be really poor after all. 
On the other hand, they cast serious doubt on the query selection practices in the IR-based bug localization. In this article, we attempted to clear the sky on this aspect by conducting an in-depth empirical study that critically examines the state-of-the-art query selection practices in IR-based bug localization. In particular, we use a dataset of 2,320 bug reports, employ ten existing approaches from the literature, exploit the Genetic Algorithm-based approach to construct optimal, near-optimal search queries from these bug reports, and then answer three research questions. We confirmed that the state-of-the-art query construction approaches are indeed not sufficient for constructing appropriate queries (for bug localization) from certain natural language-only bug reports. However, these bug reports indeed contain high-quality search keywords in their texts even though they might not contain explicit hints for localizing bugs (e.g., stack traces). We also demonstrate that optimal queries and non-optimal queries chosen from bug report texts are significantly different in terms of several keyword characteristics (e.g., frequency, entropy, position, part of speech). Such an analysis has led us to four actionable insights on how to choose appropriate keywords from a bug report. Furthermore, we demonstrate 27\%--34\%  improvement in the performance of non-optimal queries through the application of our actionable insights to them. Finally, we summarize our study findings with future research directions (e.g., machine intelligence in keyword selection).

%from natural language-only bug reports are very different from the ones delivered by the state-of-the-art approaches in terms of several qualitative aspects
%(e.g., , the similarity with ground truth) indicating the further potential for improvements.

\end{abstract}

%\ccsdesc[500]{Software and its engineering~Software verification and validation}
%\ccsdesc[500]{Software and its engineering~Software testing and debugging}
%\ccsdesc[300]{Software and its engineering~Software defect analysis}
%\ccsdesc[100]{Software and its engineering~Software maintenance tools}

%\printccsdesc

%\terms{Theory, Metrics, Human factors}


%\maketitle

%\IEEEpeerreviewmaketitle

\section{Introduction}
Software bugs and errors could lead to massive fatalities (\eg\ Boeing-737 MAX 8 crashes\footnote{https://goo.gl/GwXv6H}, Therac-25 accidents\footnote{https://bit.ly/2KU9IR2}) and loss of trillions of dollars every year \cite{312B,17T,tse-bug-dev-study}. Finding and fixing these bugs and errors (a.k.a., software debugging) cost about 50\% of the development time and efforts \cite{312B}. One of the most challenging and time consuming steps of software debugging is \emph{bug localization} (\ie\ locating the bugs within the code) \cite{parnin,locombined,tse-bug-dev-study}.
Over the last two decades, Information Retrieval (IR) methods have been widely adopted in bug localization \cite{buglocator,saha,versionhistory,stacktrace}.
The IR-based localization leverages the lexical similarity between bug reports (queries) and source code documents to find out the bugs \cite{buglocator}. Such a localization is reported to be light-weight, cost-effective, and also as accurate as spectra-based techniques which analyse software execution traces for the localization \cite{parninireval,raobug}. However, recent findings \cite{bias-bug-loc-lo,parninireval,icse2018} raise a major concern about the effectiveness of the IR-based localization. About 55\% of the bug reports contain plain natural language texts without providing any explicit hints for the bug localization (e.g., program elements, stack traces) \cite{icse2018,parninireval}. Based on a developer study, \citet{parninireval} report that the IR-based approaches (\eg\ BugLocator \cite{buglocator}) cannot perform well when they use these bug reports as search queries. Such \emph{natural language-only} reports are also referred to as \emph{non-localized} bug reports \cite{bias-bug-loc-lo} or bug reports \emph{without identifiable information} \cite{parninireval}.
%We use these three names interchangeably throughout the rest of the article.
The findings and conclusions of the developer study mentioned above have also been partially supported by another empirical study \cite{icse2018}.

%bug reports are used as their search queries. Poor bug reports often lack explicit \emph{localization hints} such as relevant program elements or stack traces \cite{parninireval}.
%They are generally written by the layman software users.
%As much as 55\% of the submitted bug reports of a software system could lack the explicit localization hints \cite{icse2018,parninireval}.

One of the key challenges of any IR-based solution is to select an appropriate search query that reflects the information need. In the case of bug localization, an appropriate query should capture the right keywords (\eg\ characteristics of a bug) that can pinpoint the faulty code.
Since non-localized bug reports lack explicit localization hints (\eg\ program elements), many existing approaches \cite{buglocator,saha,stacktrace,versionhistory,versionhistoryjsep} might fail to construct appropriate queries from them. Thus, the empirical and developer studies above \cite{parninireval,bias-bug-loc-lo} might have used sub-optimal search queries, which possibly led to their poor performance in the IR-based localization. However, a more recent study \cite{cmills-icsme2018} suggests that bug reports alone contain sufficient keywords (other than the localization hints) which (1) could form the appropriate search queries and (2) thus in turn could deliver optimal performance in localizing the bugs. That is, even the \emph{natural language-only} bug reports might contain high-quality search keywords. In other words, natural language-only bug reports might 
be a source for good query keywords,
%not be really poor after all, 
which in essence contradicts the findings of the past empirical studies \cite{parninireval,bias-bug-loc-lo}. 
This recent result showing that bug reports often contain optimal queries, and the fact that existing IR-based localization approaches often do not perform well prompts us to question the effectiveness of the query selection practices in these IR-based localization approaches.   % also
%However, such a contradiction also
%casts serious doubt on the query selection practices of the existing IR-based localization approaches.
A better understanding of the strengths and weaknesses of current query construction techniques and the characteristics of optimal queries
could help us build more effective, reliable, and practical tools for the IR-based bug localization.

%Interestingly, several recent findings \cite{cmills-icsme2018,oscar-duplicate,fse2018masud} suggest that the criticism about IR-based localization could not be totally fair due to two major issues.
%First, appropriate query construction is a major step of IR-based bug localization \cite{sisman} which has been largely ignored. \citet{cmills-icsme2018} demonstrate that 67\% of the bug reports contain enough search keywords other than \emph{localization hints} that can deliver the optimal  performance in bug localization with Information Retrieval methods.
%Unfortunately, the developer study above \cite{parninireval} did not use the optimal search queries, which possibly led to sub-optimal localization performances. Thus, their criticism about IR-based localization could be biased. Second, most of the existing studies \cite{buglocator,saha,stacktrace,versionhistory,versionhistoryjsep} overlook the potential of optimal (or reformulated) search queries, and use almost verbatim texts (except stop words) from the bug reports as queries for the localization. As a result, their queries are likely to be verbose \cite{verbose}, noisy \cite{parninireval,fse2018masud} or poor \cite{fse2018masud}, which do not provide the best performance. Thus, the true potential of IR-based bug localization might not be properly understood yet.
%The above findings also suggest that \emph{appropriate query construction} might have been a very important but yet less explored aspect of the IR-based bug localization.

Software developers often use a few important keywords from a bug report as a search query for bug localization \cite{sisman}. Unfortunately, even the experienced developers struggle to choose the right keywords \cite{kevic,vocaprob,sitir}. In fact, \citet{kevic} reports that developers might fail to choose the right keywords from the bug reports up to 88\% of time.
Several existing studies attempt to (1) deliver appropriate queries from bug reports \cite{kevic,saner2017masud,fse2018masud,observed} and (2) reformulate the poor queries chosen by developers \cite{refoqus,ase2016masud,sisman,gayg,ase2017masud}. A few other studies attempt to find out the best document retrieval algorithm \cite{trconfig,paniSANER2016} for a given query. Unfortunately, in the literature, there is a marked lack of research that (1) investigates the hidden variables behind the issues of \emph{natural language-only} bug reports, and (2) critically examines the state-of-the-art practices for query selection in the IR-based bug localization. By examining these important aspects, one might gain valuable insights on whether these bug reports can deliver good queries for accurately localizing bugs or not (as alluded by the past studies  \cite{parninireval,bias-bug-loc-lo,fse2018masud}).

In this article, we conduct an empirical study using 2,320 bug reports (939 natural language-only + 1,381 natural language texts and localization hints),
%, non-localized 1,546 bug reports,
and \emph{ten} existing approaches on search query construction \cite{tfidf,kevic,saner2017masud,rsv,rocchio,qsurvey,sisman,ase2017masud}, and
critically examine the state-of-the-art query construction practices in the IR-based bug localization. We perform three different analyses in our empirical study. First, we compare between two major keyword selection methodologies (\emph{frequency-based} \cite{tfidf,kevic,refoqus,sisman} and \emph{graph-based} \cite{saner2017masud,ase2017masud}) that are widely used to construct search queries for localizing bugs with Information Retrieval. Second, we employ a Genetic Algorithm-based approach \cite{cmills-icsme2018} to construct optimal, near-optimal search queries from the bug report texts. We define an \emph{optimal query} as a set of search keywords that achieve the optimal performance in bug localization, i.e., retrieves the first buggy document at the topmost position of the ranked result list. We demonstrate how optimal or near-optimal performance could have been achieved in localizing bugs if the appropriate search queries were provided.
%how optimal search queries could lead to optimal performance in IR-based localization.
Third, we contrast between optimal and non-optimal search queries constructed from the bug reports using a detailed analysis involving machine learning, data mining, and manual inspection.
%state-of-the-art queries \cite{saner2017masud} and the optimal search queries (a.k.a., the best possible queries) from the non-localized bug reports using a detailed analysis.
Our analysis demonstrates that optimal and non-optimal queries are significantly different from each other in several aspects, which indicates that there is room for improvement in query formulation techniques (e.g., application of machine intelligence in recognizing optimal keywords). %and future research opportunities.
%discuss the high potential of IR-based localization for cost-effective software debugging.
In particular, we answer three research questions in this empirical study as follows.
\begin{enumerate}
	 \setlength\itemsep{1em}
	\item \textbf{RQ$\mathbf{_1}$: How do the state-of-the-art approaches perform in identifying appropriate search keywords from software artifacts (\eg\ bug reports) for IR-based bug localization?} \\
	We investigate two major keyword selection methodologies --\emph{frequency-based} and \emph{graph-based}, and compare their strengths and weaknesses. We first divide our bug reports into four subsets based on their baseline queries and localization hints, and then extract the queries from each of them using the existing approaches. Queries from these approaches achieve a maximum of 68\% Hit@10, 43\% mean average precision and 0.43 mean reciprocal rank, which are comparable to baseline measures (Table \ref{table:freq-vs-graph-code}). Graph-based approaches were found more promising than the frequency-based ones. However, as a whole, the state-of-the-art approaches were not sufficient to make appropriate queries from the bug reports that lead to poor baseline queries. We also found that some bug reports could lead to poor search queries even if they contain
	the potential hints for localizing software bugs (e.g., program elements, stack traces) in their texts (Table \ref{table:freq-vs-graph-br-subset}), which was intriguing.
	
	\item \textbf{RQ$\mathbf{_2}$: Can optimal, near-optimal search queries be constructed from the bug reports that lack bug localization hints or simply contain natural language only texts? How do these search queries perform in the IR-based bug localization?} \\
	Optimal search queries (that deliver optimal performance) could be constructed from \textbf{50}\%--\textbf{81}\% of the bug reports that
	might neither contain any explicit localization hints (\eg\ program elements, stack traces) nor lead to any good baseline queries.
	%nor have high-quality textual contents. 
	We use Genetic Algorithm (GA) and ground truth information to construct these queries.
	It should be noted that this GA-based query construction approach cannot be used in a practical setting where the ground truth is not known beforehand.
	Our analysis also suggests that \textbf{78}\%--\textbf{93}\% Hit@10, \textbf{57}\%--\textbf{86}\% mean average precision and \textbf{0.58}--\textbf{0.85} mean reciprocal rank could be achieved by the near-optimal queries constructed from some bug reports whereas their baseline queries (i.e., preprocessed bug reports)
	%the low-quality bug reports, where the baseline queries 
	simply fail to retrieve any relevant documents within their Top-10 result positions (e.g., \textbf{0.00}\% Hit@10) (Table \ref{table:opt-vs-baseline-br}).
	
	%which are \textbf{44}\%, \textbf{109}\% and \textbf{111}\% higher respectively than the baseline measures .
	
	%We construct optimal search queries from bug reports using a GA-based approach \cite{cmills-icsme2018}, and demonstrate their strengths in the bug localization as opposed to baseline search queries.  According to our investigation, XX\% of the bug reports contain optimal queries, and they outperform the baseline queries with YY\% higher precision and ZZ\% recall.
	
	\item \textbf{RQ$\mathbf{_3}$: How optimal, near-optimal, and non-optimal search queries differ from each other in their characteristics and performances?}
		
	\noindent
	We compare optimal and near-optimal search queries with non-optimal queries collected from each of our selected bug reports. We conduct a multi-modal comparative analysis that involves machine learning, data mining, and careful manual inspection. Our analysis reveals several major insights. First, optimal search keywords are relatively less frequent than non-optimal ones in a bug report. 
	Second, they are more specific (\ie\ less ambiguous) than those in the non-optimal queries. Third, optimal keywords are more prevalent within the \emph{description} section in a bug report. Fourth, they are more likely to be nouns than of other part of speech.
	We employed these actionable insights to baseline queries and the queries from the state-of-the-art technique \cite{saner2017masud} (i.e., non-optimal queries)
	%discarded the frequent, ambiguous keywords from the description section, 
	and then significantly improve them. Although the poor baseline queries fail to retrieve any results within their Top-10 positions, our insight-based query expansion improved their Hit@10 by up to \textbf{27}\%--\textbf{34}\%, which clearly demonstrates the significant benefits of our derived insights.

	%25\% Hit@10 whereas the original Hit@10 was 0.00\% for the low-quality bug reports. %(1) often found within the \emph{description} of a bug report rather than in the \emph{title}, (2) less frequent within a system's codebase, and (3) more interestingly, similar to \emph{ground truth class names}.
	%Unfortunately, existing keyword selection methods in IR-based localization \cite{saner2017masud,ase2017masud,kevic} are not sufficient enough to identify them effectively yet. Their algorithms for term importance estimation could be limited or even outdated.

\end{enumerate}
	
	\textbf{Novelty in contribution:} Our work makes significant contributions to the literature in several aspects. First, we demonstrate that even the state-of-the-art IR-based localization approaches fail to construct the right search queries from the bug reports that lead to poor baseline queries (\textbf{RQ$_1$)}. Bug reports could provide high-quality or poor search queries irrespective of their localization hints (e.g., program elements, stack traces). Second, graph-based approaches perform consistently higher than frequency-based approaches in selecting queries both from bug reports and from source code (\textbf{RQ$_1$}). Third, although the state-of-the-art approaches fail, natural language-only bug reports actually contain high-quality search keywords in their texts, which could be identified using a Genetic Algorithm (GA) and ground truth information (\textbf{RQ$_2$}). This result strengthens the earlier finding of \citet{cmills-icsme2018}. 
	%It should be noted that this \textbf{GA-based query construction might not be applied in a real-world bug localization scenario where the ground truth is not known}.
	Fourth, although the GA-based approach might not be applied in a real-world bug localization scenario, it allows identifying insights that can be leveraged to improve keyword selection. We found that optimal search queries are significantly different from non-optimal ones in several characteristics (e.g., keyword position, frequency, ambiguity), which were previously unknown. Furthermore, these insights were also found actionable and effective according to our investigation where we improved the non-optimal baseline and state-of-the-art queries by applying our insights (\textbf{RQ$_3$}). 
	
	%the state-of-the-art queries in terms of several qualitative aspects .
	
	\textbf{Structure of the article:} The rest of the article is organized as follows -- Section \ref{sec:motivation} motivates our research challenges, Section \ref{sec:bg} provides a background overview, and Section \ref{sec:ssetup} discusses our study set up. Sections \ref{sec:rq1}, \ref{sec:rq2}, \ref{sec:rq3} answer our three research questions, and Section \ref{sec:discussion} summarizes our findings from the empirical and manual analysis. Section \ref{sec:threats} discusses threats to the validity, Section \ref{sec:related} focuses on related work, and finally Section \ref{sec:conclusion} concludes the article with future work.

\begin{table}[!t]
	\centering
	\caption{An Example Natural Language-Only Bug Report (\#229380, eclipse.jdt.debug)}
	\label{table:examplebr}
	\vspace{-.2cm}
	\normalsize
	\resizebox{4.8in}{!}{%
		\begin{threeparttable}
			\begin{tabular}{l|p{4.5in}|c}
				\hline
				\textbf{Field} & \multicolumn{2}{l}{\textbf{Content}}\\
				\hline
				\hline
				Title & \multicolumn{2}{l}{Up/Down buttons incorrectly enabled on classpath tab}\\
				\hline
				Description & \multicolumn{2}{p{4.5in}}{(1) Open a Java Launch Configuration.
					(2) Go to the classpath tab.
					(3) Ensure there is at least one bootstrap entry and one user entry. (4) Select the bottom bootstrap entry, the DOWN button is enabled, but pressing it does nothing. (5) Select the top user entry, the UP button is enabled, but pressing it does nothing. We should
					be able to update the button to reflect
					whether moving it is possible.}\\
				\hline
				
			\end{tabular}
		\end{threeparttable}
	}
	%\vspace{-.1cm}
\end{table}

\begin{table}[!t]
	\centering
	\caption{Example Queries from the Bug Report of Table \ref{table:examplebr}}
	\label{table:example-query}
	\vspace{-.2cm}
	\normalsize
	\resizebox{4.8in}{!}{%
		\begin{threeparttable}
			\begin{tabular}{l|p{3.5in}|c}
				\hline
				\textbf{Technique} & \textbf{Search Query} & \textbf{QE}\\
				\hline
				\hline
				Baseline-I & \{\emph{title}\} & 138 \\
				\hline
				Baseline-II & \{\emph{description}\} & 43 \\
				\hline
				Baseline & \{\emph{title} + \emph{description}\} & 25 \\
				\hline
				\citet{kevic} & \{\textbf{button} enabled tab entry classpath bootstrap\}  & 93 \\
				\hline
				TF-IDF \cite{tfidf} & \{\textbf{button} entry bootstrap enabled incorrectly  moving\} & 177 \\
				\hline
				\citet{saner2017masud} & \{tab classpath enabled \textbf{buttons} user entry\} & 86 \\
				\hline
				\textbf{NrOptimal$\mathbf{_{GA}}$} & \textit{\{open reflect tab bottom entry classpath\}} & \textbf{01}\\
				\hline
				\multicolumn{3}{c}{\textbf{Impact of Insight-Based Noise Filtration}} \\
				\hline
				\citeauthor{kevic} & \{enabled tab entry classpath bootstrap\} & \textbf{70} \\
				\hline
				TF-IDF & \{entry bootstrap enabled incorrectly  moving\} &  \textbf{141} \\
				\hline
				\citeauthor{saner2017masud} & \{tab classpath enabled user entry\} & \textbf{76} \\
				\hline
			\end{tabular}
			\centering
			\textbf{QE} = Rank of the first buggy result returned by a query
		\end{threeparttable}
	}
	\vspace{-.3cm}
\end{table}

\section{Motivating Example}\label{sec:motivation}
In order to contrast among baseline, state-of-the-art and optimal search queries, we provide a motivating example. Table \ref{table:examplebr} shows a natural language-only bug report that does not contain any hints for localizing bugs (\eg\ program elements, stack traces).
%While \emph{title} field summarizes an issue, the \emph{description} field explains the same issue in details.
Table \ref{table:example-query} shows multiple search queries constructed from this bug report. We see that the baseline queries from \emph{title}, \emph{description} or \emph{title} + \emph{description} fields do not perform well. They return their first buggy documents at the 138$^{th}$, 43$^{rd}$ and 25$^{th}$ positions respectively within their result lists. On the contrary, an optimal query selected by the Genetic Algorithm (NrOptimal$_{GA}$) from the bug report returns the same buggy document at the \textbf{topmost position}, which is the best possible outcome. 
%It should be noted that NrOptimal$_{GA}$ might not be applied in real-world bug localization scenario where the ground truth is not known. 
Finally, the state-of-the-art approaches in query construction \cite{saner2017masud,kevic,tfidf} achieve the best rank of 86, which is far from ideal. 

We compare optimal queries with non-optimal queries and found that the optimal search keywords are often found within the \emph{description} section of a bug report and might always not be \emph{frequent} (see details in Table \ref{table:actionale-insights}, Fig. \ref{fig:feature-distribution}). Such insights might explain why one of the frequent keywords of the bug report -- \emph{``button"} -- is not a part of the optimal query. Furthermore, when this keyword was removed from the above queries, we notice a performance improvement (\ie\ lower QE) in each of the three existing approaches-- \citet{kevic}, TF-IDF \cite{tfidf} and \citet{saner2017masud}, which suggests the benefit of this insight.

To summarize, even the state-of-the-art approach fails to provide an appropriate search query from a natural language-only bug report.
%construction of an appropriate search query is a major challenge.
Thus, existing IR-based localization approaches used in the past empirical studies \cite{parninireval,bias-bug-loc-lo} were possibly
%criticized \cite{parninireval,bias-bug-loc-lo} for the wrong reasons since they
not provided with the \emph{optimal} search queries from bug reports, which possibly led to their low performance in the bug localization. All these findings presented above clearly point out a gap in the literature (concerning search query construction) that warrants further researches and investigations.

\section{Background}\label{sec:bg}

%\subsection{Word Embeddings (WE)}
%Semantics of a word are often determined by its contexts (\ie\ surrounding words) within the texts. There have been several studies \cite{w2vec,wordsim,wembedding,ase2016masud} that define the semantics of a word by using its contexts captured from a large corpus such as Stack Overflow. \citet{w2vec} propose \emph{word2vec}, a feed-forward neural network based text mining tool that mines a corpus, and represents each word as a single point within a high-dimensional semantic space. That is, they determine the semantics of a word as stochastic approximations of its contextual words. Such approximations are generally represented as a numerical vector which is also called \emph{word embeddings} \cite{wembedding,w2vec}. \emph{word2vec} employs continuous skip-gram model to generate such a vector. In this work, we use word embeddings for determining the quality of a candidate query for our reformulation task.

\subsection{Term Weighting}
Determining the relative importance of a term within a textual document is commonly known as \emph{term weighting} \cite{masudir,sameer}. Although the underlying concepts and algorithms were introduced by the IR community, term weighting has been frequently used to develop IR-based solutions for Software Engineering problems (\eg\ code search, bug localization) \cite{refoqus,fse2018masud,sisman,sourcerer}. Two term weighting methodologies are frequently adopted in Software Engineering contexts -- \emph{frequency based} and \emph{graph-based}-- as follows.

\textbf{(a) Frequency-Based Term Weighting:} TF-IDF is a frequency-based method that is widely adopted by the literature for term weighting.
%%\citet{tfidf} introduced TF-IDF to determine importance of a term within a body of texts (\eg\ news article) couple of decades ago.
TF-IDF stands for Term Frequency (TF) $\times$ Inverse Document Frequency (IDF), and it can be calculated using the following equation.

\begin{equation}\label{eq:tfidf0}
\text{TF-IDF (t,d)}= (1+log(f_{t,d})) \times log(\frac{|D|}{n_t}+0.01)
\end{equation}
Here $f_{t,d}$ refers to the frequency of a term $t$ in the document $d$, $n_t$ refers to the number of documents containing the term $t$, and $D$ is the set of all documents in the corpus. That is, if a term is frequent within a document but not so frequent in other documents across the corpus, then this term is considered to be \emph{important} (\eg\ search keyword) with respect to the target document. TF-IDF  adopts the notion of \emph{term independence}. That is, it does not consider the impact of contexts (\eg\ surrounding terms) upon a given term in determining the importance of the term \cite{term-independence}. However, contexts play a major role in determining the term semantics and hence, in the term importance as well \cite{wordnet,wordsim}. Thus, TF-IDF could be limited in the term importance estimation \cite{rada,blanco}. However, due to its simplicity and low costs, several studies \cite{kevic,refoqus} adopt TF-IDF in term weighting for Software Engineering problems (\eg\ bug localization, concept location).
%However, idioms and phrases clearly depend on each other for their comprehensive meaning.
%For example, \emph{``search engine"} noun phrase conveys a different semantic than \emph{``search"} and \emph{``engine"} in isolation. Besides, important keywords of a document might always not be the most frequent ones which is especially observed with the source code documents \cite{devanbu-fse}. Despite this issue, many of the past studies  \cite{refoqus,qsurvey,rocchio} adopt frequency based term weights in keyword selection (and query reformulation) since they are light weight, intuitive and easy to use.
%\subsection{TextRank \& POSRank}

\textbf{(b) Graph-Based Term Weighting:} Unlike TF-IDF, several studies capture dependencies among the terms within a body of texts (\eg\ bug report) using term co-occurrences \cite{rada} and syntactic relations \cite{blanco}. First, they transform each text document into a graph where the nodes represent the distinct words and the connecting edges refer to the dependencies among the words from the document.
Second, they adapt Google's PageRank algorithm \cite{pagerank} for natural language texts, and determine the relative weight $S(v_i)$ of each of the words $v_i$ using the recursive score computation as follows:

\begin{equation}\label{eq:textrank}
S(v_{i})=(1-\phi)+\phi \times \sum_{j\epsilon V(v_{i})}\frac{S(v_{j})}{|V(v_{j})|}~~ (0 \le \phi \le1)
\end{equation}
\noindent
Here, $V(v_{j})$ refers to the list of nodes that are connected to $v_i$, and $\phi$ is the damping factor. The connecting edges among the nodes could be uni-directional or bi-directional. In the context of web link analysis, \citet{pagerank} define $\phi$ as the probability of a random surfer of staying on a web page, and $1-\phi$ as the probability of jumping off the page. PageRank relies on a voting mechanism, and computes score of each node based on the scores from connected nodes within the graph. Several studies in Software Engineering \cite{saner2017masud,ase2017masud,fse2018masud} use PageRank for term weighting and keyword selection.

In our empirical study, we compare between these two major term weighting approaches above, and analyse their relative strengths and weaknesses in the query construction for IR-based bug localization.

%(Chapter \ref{chap:s1}). Term weight based on word co-occurrences is called TextRank whereas the weight computed using syntactic dependencies is called POSRank \cite{blanco}. Several of our studies

\begin{figure*}[!t]
	\centering
	%\resizebox{3.5in}{!}{%
	\begin{tikzpicture}[scale=.63, auto,swap, block/.style={
		draw,
		fill=white,
		align=center,
		rectangle,
		text width={3cm},
		inner sep=3pt,
		fill=lightgray,
		font=\small},
	mylabel/.style={
		font=\small }]]
	% Draw a 7,11 network
	% First we draw the vertices
	
	%\node at (-2,2.5)(br) {Bug Reports};
	\node[inner sep=0pt] (bf1) at (-3,2) {\includegraphics[width=.35in]{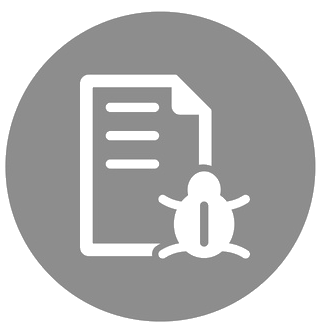}};
	\node[inner sep=0pt] (bf3) at (-2.2,2) {\includegraphics[width=.35in]{./bf1tr.png}};
	\node[inner sep=0pt] (bf2) at (-2.6,2.8) {\includegraphics[width=.35in]{./bf1tr.png}};
	
	\node at (-.2,2)(lab1) {\small Bug Reports};
	
	\node[inner sep=0pt] (com) at (-2.6,-2) {\includegraphics[width=.40in]{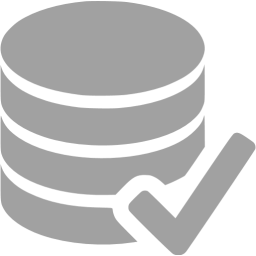}};
	\node at (-.8,-1.9) (lab2) {\small Bug-fixing};
	\node at (-.8,-2.4) (lab2) {\small Commits};
	
	\node[inner sep=0pt] (filter) at (-1,0) {\includegraphics[width=.30in]{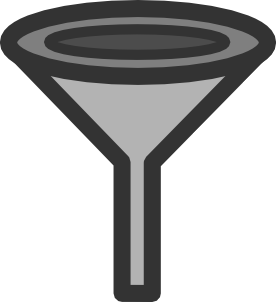}};
	\node at (-2.3,0.2) (lab3) {\small Dataset};
	\node at (-2.3,-.3) (lab3) {\small Filtration};
	
	\node[inner sep=0pt] (dsbr) at (1.2,0) {\includegraphics[width=.40in]{./bf1tr.png}};
	\node[inner sep=0pt] (gt) at (2,0) {\includegraphics[width=.40in]{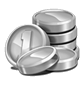}};
	\node at (3.6,.3) (lab3) {\small Refined};
	\node at (3.6,-.3) (lab3) {\small Dataset};

	%\node[inner sep=0pt] (se) at (3.6,0) %{\includegraphics[width=.35in]{./sysdiag/sengine.png}};
	
	\node[block](rq1) at (7.5,4) {\begin{tabular}{c}
	\small Frequency vs. \\
	\small Graph-based \\
	\small Keyword Selection\\
	\end{tabular}};

	\node at (4.8,4)[ellipse, draw=black, fill=gray!20] (rq10) {RQ$_1$};
	
	\node[inner sep=0pt] (scale1) at (10.7,4) {\includegraphics[width=.40in]{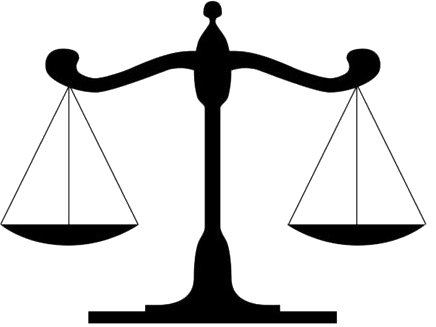}};

	\node[block](rq2) at (7.5,-4) {\begin{tabular}{c}
		\small Optimal vs. \\
		\small Baseline Queries\\
		\end{tabular}};
	\node at (4.8,-4)[ellipse, draw=black, fill=gray!20] (rq20) {RQ$_2$};
	\node[inner sep=0pt] (scale2) at (10.7,-4) {\includegraphics[width=.40in]{./scale-tr.png}};
	
	\node[block](rq3) at (9.5,0) {\begin{tabular}{c}
		\small Optimal vs. \\
		\small Non-Optimal\\
		\small Search Queries\\
		\end{tabular}};
	
	\node at (7,0)[ellipse, draw=black, fill=gray!20] (rq30) {RQ$_3$};
	\node[inner sep=0pt] (scale3) at (12.7,0) {\includegraphics[width=.40in]{./scale-tr.png}};

	\node[inner sep=0pt] (summary) at (15,0) {\includegraphics[width=.40in]{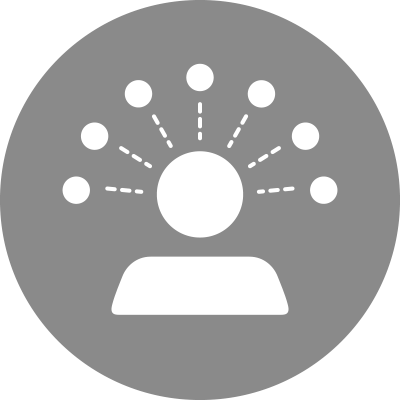}};
	\node at (14,-1.2) (lab3) {\small Finding};
	\node at (14,-1.7) (lab3) {\small Summary};

	\begin{pgfonlayer}{bg}    % select the background layer
	%\draw (foo) -- (baz);
%	\node[inner sep=0pt] (ga) at (1,0)
%	{\includegraphics[width=.38in]{./sysdiag/evolve.png}};
%	
%	\node[inner sep=0pt] (ds) at (1,1.2)
%	{\includegraphics[width=.30in]{./sysdiag/documents.png}};
%	
%	\node[inner sep=0pt] (summary) at (4.5,0)
%	{\includegraphics[width=.35in]{./sysdiag/summary.png}};
		
	\draw[->,thick] (bf3) -- (filter);
	\draw[->,thick] (com) -- (filter);
	\draw[->,thick] (filter) -- (dsbr);
	%\draw[->,thick] (gt) -- (se);
	\draw[->,thick] (gt) -- (rq10);
	\draw[->,thick] (gt) -- (rq20);
	\draw[->,thick] (gt) -- (rq30);
	
	\draw[->,thick] (rq1) -- (rq3);
	\draw[->,thick] (rq2) -- (rq3);
	
	\draw[->,thick] (scale1) -| (summary);
	\draw[->,thick] (scale2) -| (summary);
	\draw[->,thick] (scale3) -- (summary);

	\end{pgfonlayer}
	
	\end{tikzpicture}
	%}
	%\vspace{-.2cm}
	%\small
	\centering
	\caption{Schematic diagram of our empirical study}
	\label{fig:schematic-diag}
	\vspace{-.4cm}
\end{figure*}

\subsection{Genetic Algorithm (GA)}
Genetic Algorithms are a class of evolutionary search that is inspired by biological operations such as \emph{natural selection}, \emph{cross-over} and \emph{mutation}. These algorithms are widely used to solve complex optimization problems in various research domains including Software Engineering (\eg\ automatic patch generation \cite{genprog}). The whole texts (\ie\ \emph{title} + \emph{description}) of a bug report are often verbose and non-optimal as a search query \cite{verbose}. Thus, carefully chosen keywords from the report texts could form an optimal or near-optimal search query for IR-based bug localization \cite{cmills-icsme2018}. In other words, query construction could be considered as a kind of optimization problem. Genetic Algorithms equipped with an appropriate fitness function are a great fit for this optimization task. In our empirical study, we employ Genetic Algorithm with Query Effectiveness (QE) as a fitness function \cite{cmills-icsme2018}, and construct near-optimal search queries from the bug reports.

\begin{table}[!tb]
	\centering
	\caption{Study Dataset (Subject Systems \& Bug Reports)}\label{table:expds}
	%\vspace{-.2cm}
	\resizebox{4.8in}{!}{%
		\begin{threeparttable}
			\begin{tabular}{l|c|c|c|c|c}
				\hline
				\textbf{Subject System} & \textbf{\#BR$\mathbf{_{H-}}$} & \textbf{\#BR$\mathbf{_{H+}}$} & \textbf{\#Files}& \textbf{\#Methods} & \textbf{\#AGT}\\
				\hline
				\hline
				\texttt{ecf} & 58 & 141 & 2,802 & 21.5K & 1.74\\
				\hline
				\texttt{eclipse.jdt.core} & 53 & 95 & 5,908 & 66.3K & 2.03 \\
				\hline
				\texttt{eclipse.jdt.debug} & 141 & 186 & 1,532 & 15.7K & 1.76\\
				\hline
				\texttt{eclipse.jdt.ui} & 293 & 286 & 10,927 & 57.4K  & 1.78\\
				\hline
				\texttt{eclipse.pde.ui} & 332 & 188 & 5,334 & 31.8K & 1.97\\
				\hline
				\texttt{tomcat70} & 62 & 485 & 1,841 & 23.8K & 1.27\\
				\hline
				\hline
				& \textbf{939} & \textbf{1,381} &  &  & \\
				\hline
				\multicolumn{6}{c}{\textbf{Total Bug Reports: 2,320}} \\
				\hline
			\end{tabular}
		   \centering
		   \textbf{BR$\mathbf{_{H+}}$}=Bug reports with hints, \textbf{BR$\mathbf{_{H-}}$}=Bug reports without hints, \textbf{AGT}=Average number of ground truth files
		\end{threeparttable}
	}
	\vspace{-.4cm}
\end{table}

\section{Study Setup}\label{sec:ssetup}

Fig. \ref{fig:schematic-diag} shows the schematic diagram of our empirical study. We collect dataset from an existing benchmark \cite{fse2018masud}, refine it by discarding the noisy items (\eg\ tangled commits, false-positive bug reports), and then answer three research questions through comparative analyses. In this section, we discuss our dataset construction, noise filtration, and study setup as follows.

\subsection{Dataset Preparation}\label{sec:dataset}
\textbf{Bug Report Selection:} Table \ref{table:expds} shows the dataset for our empirical study. We first collect a total of 1,546 bug reports from six open source, Java-based subject systems for our study. They are taken from a publicly available \emph{benchmark dataset} \cite{fse2018masud,icse2018,fse2018data}. Several steps were taken by the benchmark authors to construct this dataset.
First, all the bug reports marked as RESOLVED were extracted from the bug repository (\eg\ BugZilla, JIRA) of each system. Second, they consulted with the version control history of each system at GitHub, and collected the bug-fixing commits (\ie\ commits solving the bugs) using a set of traditional heuristics \cite{bugid,buglocator}. In particular, a set of regular expressions 
(e.g.,\texttt{(B$|$b)ug\textbackslash s+\textbackslash d+$|$\textbackslash=\textbackslash d+$|<$Repo$>$-\textbackslash d+}) were employed for identifying these commits. For example, \texttt{(B$|$b)ug\textbackslash s+} recognizes this commit\footnote{https://bit.ly/2RnIAPK} as the bug-fixing commit of the showcase bug report in Table \ref{table:examplebr} (e.g., Bug 229380).  
Then only such bug reports were retained that have corresponding bug-fixing commits.
Third, they also detected the presence of structured entities namely \emph{bug localization hints} (\eg\ stack traces, method invocations) using regular expressions in the report texts \cite{fse2018masud,stacktrace}, and retained such bug reports that contain only unstructured regular texts and no structured entities.
\FrameSep.3em
\begin{framed}
	\noindent
	\textbf{Regular expression for stack traces}:
	\begin{verbatim}
	(.*)?(.+)\.(.+)(\((.+)\.java:\d+\)|\(Unknown Source\)|\(Native Method\))
	\end{verbatim}
	\textbf{Regular expression for method invocations}:
	\begin{verbatim}
	((\w+)?\.[\s\n\r]*[\w]+)[\s\n\r]*(?=\(.*\))|([A-Z][a-z0-9]+){2,}
	\end{verbatim}
\end{framed}
\noindent
According to \citet{parninireval}, developers often struggle to find appropriate search keywords from the reports containing only regular texts and no structured entities.
Fourth, they also discarded such bug reports (1) for which no source code documents (\eg\ Java classes) were changed, or (2) that their changed documents were missing in the current snapshot of the subject system.

\textbf{Construction of Ground Truth:} We collect the changed source documents (a.k.a., \emph{change set}) from each of the bug-fixing commits, and use them as the \emph{ground truth} for corresponding bug reports. When multiple commits are made for a single bug report, their changed document lists are merged together to construct the final ground truth. Thus, all three items -- bug reports, their corresponding ground truths and the subject systems -- are collected from the benchmark dataset.

\textbf{Dataset Cleansing with Manual Analysis:}
Recent findings \cite{its-not-a-bug,bias-bug-loc-lo} suggest that existing dataset used for IR-based bug localization could be noisy due to misclassified bug reports (a.k.a., feature requests). In order to mitigate such a threat, we conduct an extensive manual analysis and perform further data cleansing and noise removal. First, two authors individually go through \emph{title} and \emph{description} of each bug report, and annotate whether it indicates a software bug or a new feature \cite{its-not-a-bug,bias-bug-loc-lo}.
Second, we perform an agreement analysis between the two authors for all six subject systems. We reached an average agreement level of 74\%. While 26\% (100\%-74\%) is a significant disagreement between the two authors, such disagreement in our classification might also be explained. Bug reports collected from the benchmark mostly contain natural language texts rather than structured elements (e.g., stack traces). Both authors carefully analyse the textual contents to understand their hidden semantics and then classify them as either bug reports or feature requests. Since the same words/texts could be explained differently by different people based on their subjective perceptions, such disagreement might be expected. In fact, similar disagreements among the human experts were also observed in various other text processing tasks such as search keyword selection \cite{vocaprob} or textual summary generation \cite{pyramid,haiduc-summarization}. We also performed retrospective analysis and resolved the disagreements between the two authors through cross-examination, further manual analysis, and extensive discussions. During this analysis, we toggled the class labels of 192 bug reports/feature requests. Finally, we retained only such documents that were labelled as \emph{bug reports} by both authors, and discarded the rest from our dataset.

%\Foutse{it is not really clear which amount of data remained ambiguous and what happened to it...did you removed them from the dataset?}

%The class labels of 192 bug reports/feature requests were toggled during this retrospective analysis, which left us with a total of \textbf{939} bug reports.
%These bug reports are then used for our empirical study. It should be noted that we extended our dataset with 1,381 more bug reports that contain both natural language texts and bug localization hints. Thus, our final dataset contains a total of 2,320 bug reports.
%Second, these annotated reports are then randomly reviewed by the other author for agreement analysis. We achieved 86\% agreement between the first author and the second author for \texttt{tomcat70} system.
%, the disagreements were resolved through direct consultation.
%Finally, we retain only such reports that were annotated as bugs by the authors, and discard the feature requests from the dataset.
%\Foutse{the only may not make sense anymore since you analyse commits with less than 5 changes later!} 

According to recent findings \cite{relink, tangled-zeller}, the ground truth of bug reports could be bloated due to \emph{tangled commits}. In particular, the bug-fixing commits might contain changes irrelevant to the fixed bugs. 
Such bug-fixing commits are called \emph{tangled commits}.
In order to mitigate such a threat, we conduct an extensive manual analysis on the collected ground truth. First, we manually analyse only the large commits containing more than \emph{five} changed source documents.
%and consider the smaller commits as potentially noise-free. 
Out of 1,546 bug reports, we manually analysed the bug-fixing commits of 333 (22\%) bug reports.
%since each of these commits touched more than five source code documents.
In particular, we manually check for (1) changes irrelevant to the bug and (2) trivial changes (\eg\ updating code comments), and discard them from the dataset.
We identified 78 commits as \emph{``tangled"} which were discarded from further analysis. In order to mitigate the threat in the remaining commits containing five or less changed source documents, we conducted further analysis.
After discarding false-positive bug reports (i.e., feature request), we identified 1,149 bug reports with five or less ground truth documents. Since manually analyzing all of them could be very costly, we (1) selected 10\% bug reports from each of the six subject systems using stratified random sampling, and (2) manually analyzed 112 bug-fixing commits. Out of these 112 cases, we found only 5 tangled commits which are only $\approx$5\%. Thus, the impact of tangled commits on our findings might be negligible. An earlier study \cite{bias-bug-loc-lo} has also reported negligible impact of the tangled commits on bug localization performance.
Furthermore, to assess the potential impact of tangled commits on our results, we replicated the analysis on a manually validated dataset, and the detailed analysis results can be found in Table \ref{table:freq-vs-graph-br-all}.

Once both the filtration steps above were completed, we ended up with a refined dataset of \textbf{939} bug reports and their ground truths for our study. We spent \textbf{60+} man-hours in conducting our manual analysis.

\textbf{Dataset Extension:} Although the above analysis provides a total of 939 natural language-only bug reports, we further extend our dataset with such bug reports that contain explicit localization hints (e.g., stack traces, method invocations). We use the same benchmark dataset \cite{fse2018masud} as above.
Our goal was to extend our experiments and generalize our findings.
In particular, we collect such \emph{localized bug reports} (\ie\ containing localization hints) that have a single ground truth source file.  
We believe that commits modifying only one source file are less likely to contain \emph{irrelevant changes}. Our idea was to avoid the threat of \emph{tangled} commits.
Thus, we extend our dataset with an additional 1,381 bug reports from the six subject systems, which provides a total of \textbf{2,320} bug reports for the study.

\textbf{Replication Package:} Our dataset, replicated techniques and other associated materials are \emph{publicly available} \cite{emse2019-rep} for the replication and third-party reuse.

\subsection{Selection of Baseline Search Queries}\label{sec:baseline}
Developers often use \emph{title} and \emph{description} fields from a bug report as ad hoc search queries for locating the buggy code within a software system. Existing approaches \cite{refoqus,stacktrace,fse2018masud,sisman} also employ them as the baseline queries for their evaluation and validation. We thus construct three baseline queries using these fields -- \textbf{Baseline-I} (\ie\ \emph{title}), \textbf{Baseline-II} (\ie\ \emph{description}) and \textbf{Baseline} (\ie\ \emph{title} + \emph{description}) -- for our empirical study. In particular, we perform standard natural language preprocessing (\eg\ removal of stop words and punctuations, token splitting, camel case splitting) on these fields, and use their preprocessed versions as the baseline search queries for our study. We avoid stemming due to its mixed findings, as reported in the literature \cite{stemming,kevic}. 
Stemming might improve recall but often reduces the precision of search queries, which is important in our case. It should be noted that the same pre-processing steps were used for both bug reports and the source code documents in the corpus. Table \ref{table:baseline-query-stats} shows the statistics on keyword counts in the baseline queries.

\subsection{Code Search Engine} \label{sec:sengine}
In order to evaluate the performance of our search queries, we use \emph{Apache Lucene} \cite{lucene}, a popular document search engine that has been widely adopted by both academia \cite{refoqus,saner2017masud,stacktrace,trconfig} and industry (\eg\ ElasticSearch, Stack Overflow). We capture each source code document of a subject system, perform limited natural language preprocessing (\eg\ removal of stop words, keywords and punctuations, token splitting, camel case splitting) on these documents, and then use them to construct our corpus. Lucene then develops a document index  against each subject system by analysing its preprocessed corpus documents. Once a search query is issued, Lucene (1) first selects a list of candidate results using Boolean Search Model, and then (2) delivers a ranked list of relevant results using a Vector Space Model (VSM)-based search algorithm (\eg\ BM25 \cite{bm25}).

\subsection{Performance Metrics}\label{sec:pmetrics}
We employ four state-of-the-art performance metrics for evaluating the search queries in our empirical study. Since these metrics have been widely adopted earlier by relevant literature on IR-based bug localization \cite{stacktrace,saha,fse2018masud} and query reformulation \cite{refoqus,saner2017masud,ase2017masud}, their use is also justified for this study. We define the four performance metrics as follows.

\textbf{Hit@K}: It calculates the fraction of search queries (\eg\ bug reports) for each of which at least one ground truth is retrieved within the Top-K results. The higher the Hit@K value is, the better the search queries are. It is also called Recall @Top K in the literature \cite{saha}.

\textbf{Query Effectiveness (QE)}: It approximates a developer's effort in locating the reported bug within the source code of a software system \cite{stacktrace}. In practice, the metric returns the rank of the first result that matches with the ground truth, within the ranked list. The underlying idea is to provide an accurate starting point to the developer that deals with the bug discussed in the search query. The lower the effectiveness value is, the better a query is since the developer then can locate the buggy source code more quickly with less manual efforts.

\textbf{Mean Reciprocal Rank (MRR)}: Reciprocal Rank (RR) refers to the multiplicative inverse of the rank of the first buggy source code document correctly returned within the ranked result list.
In practice, only Top-K results (\eg\ $K=10$) are often analysed for each query.
Mean Reciprocal Rank (MRR) averages the RR measures for all search queries ($Q$) of a subject system. It can be defined as follows:

\begin{equation*}
\text{MRR(Q)} =\frac{1}{|Q|}\sum_{q\in Q}{\frac{1}{firstRank(q)}}
\end{equation*}
Here, $firstRank(q)$ provides the rank of the first correctly retrieved buggy document. MRR can take a maximum value of 1.00 when the correct result is found at the top-most position and a minimum value of 0.00 when no correct results are found within the whole result list.  The bigger the MRR value is, the better the search query is.

\textbf{Mean Average Precision (MAP)}: Precision@K calculates the precision at the occurrence of every single buggy source document within the ranked list. Average Precision@K (AP) averages the Precision@K for all buggy documents in the list for a search query. Thus, Mean Average Precision (MAP) is calculated from the mean of Average Precision@K (AP) for all the queries ($Q$) of a system as follows:

\begin{equation*}\label{eq:avep}
\text{AP}=\frac{\sum_{k=1}^{D}P_{k}\times buggy(k)}{|S|},~ \text{MAP}=\frac{\sum_{q\epsilon Q}\text{AP(q)}}{|Q|}
\end{equation*}
Here, $buggy(k)$ determines whether $k^{th}$ result in the ranked list is buggy (\ie\ matches ground truth) or not (\ie\ does not match ground truth). It returns either 0 (false positive) or 1 (true positive). $P_{k}$ denotes the precision calculated at the $k^{th}$ position, and $D$ refers to the number of total results. $S$ refers to the true positive instances retrieved by a query, and $Q$ is the set of all queries (\eg\ bug reports).

\subsection{Bug Report Clustering}\label{sec:br-cluster}
Our dataset contains a total of 2,320 bug reports where 939 of them are natural language only and 1,381 of them contain explicit bug localization hints (e.g., program elements, stack traces). Several studies \cite{parninireval,bias-bug-loc-lo} suggest a significant impact of localization hints in the bug reports upon IR-based bug localization methods. For our empirical study, we further divide them into subsets based on the baseline queries they produce through the pre-processing steps (Section \ref{sec:baseline}). In particular, if a baseline query (e.g., \textbf{Baseline}) achieves the query effectiveness (QE) between 1 and 10, then the query is already good. The corresponding bug report is considered as \emph{bug report leading to high-quality queries} (\textbf{HQ}). On the other hand, a bug report with a poor baseline query (i.e., QE$>$10) 
is considered as \emph{bug report leading to poor queries} (\textbf{LQ}).
%to be of \emph{low-quality}. 
Thus, based on two different dimensions -- baseline query performance and presence of localization hints (e.g., method invocations, stack traces), we divide our bug report collection into four different subsets. Dividing the bug reports into smaller categories like these can help us zoom in the issues of IR-based bug localization. Our idea was to check how the existing approaches perform with these four subsets and to derive further insights on how to make appropriate search queries from bug reports. Table \ref{table:br-cluster} shows the statistics on bug report from each subset.

\begin{table}[!t]
	\centering
	\caption{Four Subsets of Bug Report Collection}
	\label{table:br-cluster}
	\resizebox{4.4in}{!}{%
	\begin{threeparttable}
		\begin{tabular}{p{3.4in}|c}
			\hline
			\textbf{Bug Report Category} & \textbf{\#BR}\\
			\hline
			\hline
			Bug reports with high-quality baseline queries and without localization hints (HQ$\mathbf{_{H-}}$) & 567 \\
			\hline
			Bug reports with high-quality baseline queries and with localization hints (HQ$\mathbf{_{H+}}$) & 954 \\
			\hline
			Bug reports with low-quality baseline queries and without localization hints (LQ$\mathbf{_{H-}}$) & 372 \\
			\hline
			Bug reports with low-quality baseline queries and with localization hints (LQ$\mathbf{_{H+}}$) & 427 \\
			\hline
			\textbf{Total} & \textbf{2,320} \\
			\hline
		\end{tabular}
		\centering
		\textbf{BR} = Bug reports
	\end{threeparttable}	
}
\vspace{-.3cm}	
\end{table}

\section{Answering RQ$_1$: Frequency-Based vs. Graph-Based Search Keyword Selection for IR-Based Bug Localization}\label{sec:rq1}

\subsection{Determination of State-of-the-Art on Keyword Selection}
\label{sec:soa-selection}
The underlying idea of any query construction/reformulation task is to choose appropriate search keywords from the suitable information sources (\eg\ bug report, source code). According to existing literature \cite{gayg,refoqus,saner2017masud,ase2017masud,sisman}, two keyword selection
methodologies --\emph{frequency-based} and \emph{graph-based}-- have been widely adopted for query construction in the context of concept/bug localization. We choose ten different approaches from these two major methodologies for our study as follows.

\textbf{Frequency-Based Keyword Selection Methods:} TF-IDF \cite{tfidf} has been a popular term weighting method in Information Retrieval for the last five decades.
%It not only captures a term's local contribution within a document (\eg\ frequency) but also its global influence within the corpus \cite{sameer}. TF-IDF
It is also widely adopted in Software Engineering contexts. Several existing studies \cite{kevic,refoqus,kevicdict,trconfig,cmills-icsme2018} employ TF-IDF and its variants to identify important keywords from a body of texts (\eg\ bug report, source code). \citet{kevic} use TF-IDF and three heuristics (\eg\ POS tags, notation, position) to identify the important search terms from a bug report. \citet{gayg} employ Rocchio's expansion \cite{rocchio} where they use TF-IDF in reformulating search queries for concept location. \citet{refoqus} later employ three frequency-based term weighting methods --Rocchio \cite{rocchio}, RSV \cite{rsv} and Dice \cite{qsurvey}, and deliver the best performing query keywords from the source code for concept location using machine learning. \citet{sisman} leverage \emph{spatial proximity} between query keywords and candidate keywords within the source code (a.k.a., spatial code proximity (SCP)), and return such candidates that frequently co-occur with the query. To the best of our knowledge, these are the state-of-the-art studies that adopt \emph{frequency-based} keyword selection for IR-based concept/bug localization.

The above studies employ two different sources --\emph{bug report} and \emph{source code document}-- for query construction. Thus, we use \citet{kevic} to select keywords from the bug reports, and \citet{refoqus} and \citet{sisman} for selecting keywords from the source code. It should be noted that \citet{refoqus} combine three term weighting approaches --Rocchio, RSV and Dice. Thus, we select a total of \emph{eight} frequency-based approaches (Tables \ref{table:freq-vs-graph-br-all}, \ref{table:freq-vs-graph-code}) in our study that use TF-IDF and its variants for keyword selection. Since the authors' implementations were not publicly available, we carefully implement each of these approaches in our working environment with their recommended settings and the best performing parameters (\eg\ regression coefficients \cite{kevic}). Please check our replication package for further details on the used settings and parameters.

\textbf{Graph-Based Keyword Selection Methods:} Although TF-IDF has been a widely adopted method, it fails to capture a term's contexts which is a major limitation \cite{rada,blanco}. Several existing studies \cite{saner2017masud,ase2017masud,fse2018masud} attempt to overcome this issue by transforming a document into a graph representation.
In particular, they represent the individual terms and the dependencies among them using nodes and connecting edges respectively within a graph. \citet{saner2017masud} capture semantic and syntactic dependencies among the words, employ PageRank algorithm \cite{pagerank} on the constructed graph, and then deliver appropriate search keywords from a bug report for concept location. While their approach (STRICT) is suitable for regular texts, bug reports could also be noisy; containing too many structured elements (\eg\ stack traces, test cases). Thus, keywords from bug reports should be chosen carefully.
Recently, \citeauthor{fse2018masud} leverage reporting quality dynamics, graph-based term weighting, and deliver search keywords even from the noisy bug reports \cite{fse2018masud}. Despite these attempts, constructing appropriate queries could be challenging since the bug reports might always not contain the necessary information. Thus, \citeauthor{ase2017masud} also make use of source document structures and their contexts, employ graph-based term weighting, and then suggest appropriate search keywords from the relevant source code \cite{ase2017masud}.
To the best of our knowledge, these are the state-of-the-art studies that adopt \emph{graph-based} keyword selection for IR-based concept location and IR-based bug localization.

Since we consider two keyword sources in our study, we use STRICT \cite{saner2017masud} to select keywords from bug reports and ACER \cite{ase2017masud} for selecting keywords from the source code.
%We use them \cite{saner2017masud,ase2017masud} for selecting keywords from bug reports and source code in our study.
The authors' implementations for these approaches were publicly available. We thus use their prototypes for our empirical study.

%\subsection{Bug Report Categorization}\label{sec:subsets} Bug reports

\begin{table}[!t]
	\centering
	\caption{Frequency-Based vs. Graph-Based Keyword Selection from Bug Reports (Using Top-10 Results Only)}\label{table:freq-vs-graph-br-all}
	\vspace{-.2cm}
	\resizebox{4.6in}{!}{%
		\begin{threeparttable}
			\begin{tabular}{l|c|c|c|c|c|c}
				\hline
				\textbf{Technique} & \textbf{Genre} & \textbf{Hit@1} & \textbf{Hit@5} & \textbf{Hit@10} & \textbf{MRR} & \textbf{MAP} \\
				\hline
				\multicolumn{7}{c}{\textbf{Retrieval performance using all bug reports (2,320)}} \\
				\hline
				\textbf{Baseline} &  \multirow{3}{*}{--} & \textbf{31.98}\% & \textbf{57.96}\% & \textbf{66.50}\% & \textbf{0.43} & \textbf{42.69}\% \\
				\hhline{-~-----}
				Baseline-I &  & 22.37\% & 46.73\% & 57.58\% & 0.33 & 32.82\% \\
				\hhline{-~-----}
				Baseline-II &  & 29.01\% & 51.16\% & 60.43\% & 0.38 & 38.21\% \\
				\hline			
				%\hline
				%\multicolumn{6}{c}{Frequency-Based Keyword Selection} \\
				%\hline
				TF & \multirow{4}{*}{Frequency}  &  24.39\% & 46.23\% & 55.58\%
				& 0.34 & 33.94\%\\
				\hhline{-~-----}
				IDF &  & 24.77\% & 43.91\% & 52.79\% & 0.33 & 32.94\% \\
				\hhline{-~-----}
				\textbf{TF-IDF} &  & \textbf{27.02}\% & 49.92\% & 59.80\% & \textbf{0.37} & 36.76\% \\
				\hhline{-~-----}
				\citeauthor{kevic} & & 23.36\% & 44.03\% & 52.92\% & 0.32
				& 31.98\% \\
				%\hline
				%\multicolumn{6}{c}{Graph-Based Keyword Selection} \\
				\hline
				\textbf{STRICT} & Graph & 25.82\% & \textbf{52.28}\% & \textbf{63.02}\% & \textbf{0.37} & \textbf{37.31}\% \\
				\hline
				\multicolumn{7}{c}{\textbf{Retrieval performance using manually analysed bug reports (175)}} \\
				\hline
				Baseline & \multirow{3}{*}{--} & \textbf{29.65}\% & \textbf{58.77}\% & \textbf{63.77}\% & \textbf{0.41} & \textbf{38.08}\% \\
				\hhline{-~-----}
				Baseline-I & & 21.45\% & 48.65\% & 60.00\% & 0.34 & 32.87\% \\
				\hhline{-~-----}
				Baseline-II &  & 23.62\% & 45.79\% & 56.07\% & 0.33 & 30.40\% \\
				\hline
				TF & \multirow{4}{*}{Frequency} & 20.75\% & 44.45\% & 56.13\% & 0.32 & 28.72\% \\
				\hhline{-~-----}
				IDF & & 17.61\% & 38.90\% & 45.93\% & 0.26 & 24.63\% \\
				\hhline{-~-----}
				TF-IDF & & 24.37\% & 47.33\% & 55.98\% & 0.35 & 31.57\% \\
				\hhline{-~-----}
				\citeauthor{kevic} & & 26.17\% & 49.03\% & \textbf{58.40}\% & 0.36 & 31.89\% \\
				\hline
				STRICT & Graph & \textbf{26.60}\% & \textbf{50.87}\% & 58.01\% & \textbf{0.37} & \textbf{34.28}\% \\ 
				\hline
				\multicolumn{7}{c}{\textbf{Retrieval performance using single-release bug reports (360)}} \\
				\hline
				Baseline & \multirow{3}{*}{--} & \textbf{36.99}\% & \textbf{58.55}\% & \textbf{66.66}\% & \textbf{0.46} & \textbf{46.29}\% \\
				\hhline{-~-----}
				Baseline-I & & 27.42\% & 48.67\% & 59.56\% & 0.37 & 36.40\% \\
				\hhline{-~-----} 
				Baseline-II & & 35.28\% & 50.77\% & 57.00\% & 0.42 & 42.05\% \\
				\hline
				TF & \multirow{4}{*}{Frequency} & 24.89\% & 50.74\% & 58.03\% & 0.35 & 34.60\% \\
				\hhline{-~-----} 
				IDF & & 29.56\% & 48.41\% & 54.26\% & 0.38 & 37.79\% \\
				\hhline{-~-----} 
				TF-IDF & & \textbf{30.06}\% & \textbf{55.12}\% & 62.97\% & \textbf{0.41} & \textbf{40.68}\% \\
				\hhline{-~-----}
				\citeauthor{kevic} & & 22.37\% & 44.46\% & 52.39\% & 0.31 & 29.82\% \\
				\hline
				STRICT & Graph & 28.54\% & 50.31\% & \textbf{66.18}\% & 0.39 & 39.14\% \\
				\hline
				
			\end{tabular}
		\centering
		\textbf{Emboldened}=Baseline and state-of-the-art performance measures
		\end{threeparttable}
	}
	\vspace{-.3cm}
\end{table}

\subsection{Search Keyword Selection from Bug Reports}
\label{sec:ks-br}
\begin{figure}[!t]
	\centering
	\includegraphics[width=4.8in]{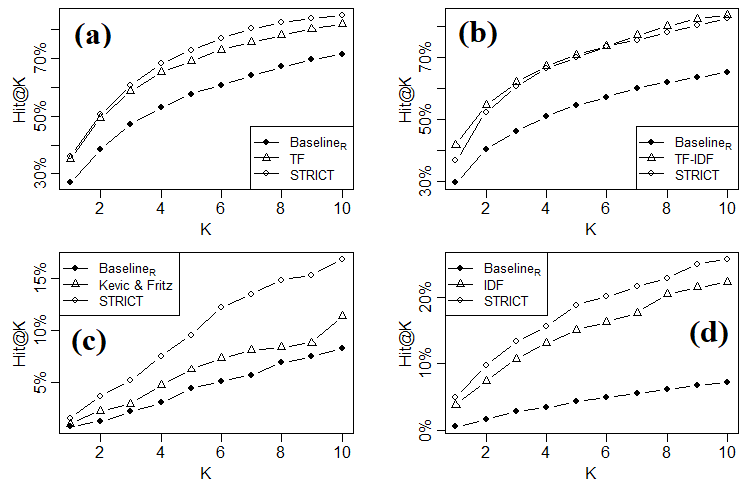}
	%\vspace{-.5cm}
	\caption{Comparison between existing techniques and random baseline (Baseline$_{R}$) for (a) bug reports with good baseline queries and no localization hints (HQ$_{H-}$), (b) bug reports with good baseline queries and with localization hints (HQ$_{H+}$), (c) bug reports with poor baseline queries and no localization hints (LQ$_{H-}$), and (d) bug reports with poor baseline queries and with localization hints (LQ$_{H+}$)}
	\label{fig:freq-vs-graph-br-random}
	\vspace{-.3cm}
\end{figure}

Tables \ref{table:freq-vs-graph-br-all}, \ref{table:freq-vs-graph-br-subset}, \ref{table:baseline-query-stats}, \ref{table:freq-vs-graph-improved}, and Fig. \ref{fig:freq-vs-graph-br-random} summarize our comparative analysis between frequency-based keyword selection and graph-based keyword selection from the bug reports.
Table \ref{table:freq-vs-graph-br-all} shows the bug localization performance of various queries constructed from a total of 2,320 bug reports.
We see that the existing approaches are not sufficient enough for delivering appropriate keywords from the bug report texts.
%that lack bug localization hints (\eg\ program elements)
Simple TF-based keywords achieve 56\% Hit@10 with 34\% mean average precision and a mean reciprocal rank of 0.34. That is, term frequency (TF) might work only half of the time in returning a relevant result within the top 10 positions.
TF-IDF-based search queries achieve 60\% Hit@10, 37\% precision and 0.37 reciprocal rank, which are marginally higher. However, they still cannot outperform the best performing baseline measures (e.g., 67\% Hit@10, 43\% MAP, and 0.43 MRR).
\citeauthor{kevic} construct a regression model for keyword selection using TF-IDF and three popular heuristics concerning the position, notation and part of speech of the keywords. Unfortunately, their model also fails to deliver the right keywords and thus performs poorly. On the contrary, \citet{saner2017masud} propose a graph-based technique namely STRICT where they leverage co-occurrences and syntactic dependencies among the keywords for search keyword selection.
%STRICT \cite{saner2017masud} adopts a graph-based term weighting approach by exploiting various dependencies among the terms.
Queries delivered by STRICT achieve 13\% higher Hit@10 and 5\% higher Hit@10 than that of the TF and TF-IDF approaches respectively.
%8\% higher Hit@10 than those of TF.
Thus, graph-based keyword selection is marginally better than frequency-based keyword selection according to our experiment.
Unfortunately, none of these existing approaches above offers a better alternative than the baseline query (\ie\ pre-processed version of a bug report), which is unexpected. It should be noted that, in our study, a baseline query contains almost all keywords (except stop words) from a bug report whereas a search query suggested by the existing approaches contains only Top-K (e.g., $K=10$) chosen keywords.

\begin{table}[!t]
	\centering
	\caption{Impact of Bug Report Quality and Localization Hints on Search Keyword Selection (Using Top-10 Results Only)}\label{table:freq-vs-graph-br-subset}
	\vspace{-.2cm}
	\resizebox{4.8in}{!}{%
		\begin{threeparttable}
			\begin{tabular}{l|c|c|c|c|c|c}
				\hline
				\textbf{Technique} & \textbf{Genre} & \textbf{Hit@1} & \textbf{Hit@5} & \textbf{Hit@10} & \textbf{MRR} & \textbf{MAP} \\
				\hline
				\hline
				\multicolumn{7}{c}{Bug reports with good baseline queries and no localization hints (\textbf{567}) (HQ$_{H-}$)}\\
				\hline
				\textbf{Baseline} & -- & \textbf{41.96}\% & \textbf{86.18}\% & \textbf{100.00}\% & \textbf{0.60} & \textbf{58.15}\% \\
				\hline
				TF & \multirow{4}{*}{Frequency}  &  35.08\% & 68.87\% & 81.63\%
				& 0.49 & 48.61\%\\
				\hhline{-~-----}
				IDF &  & 26.59\% & 53.17\% & 62.89\% & 0.38 & 35.99\% \\
				\hhline{-~-----}
				TF-IDF &  & 33.84\% & 68.00\% & 80.67\% & 0.48 & 46.67\% \\
				\hhline{-~-----}
				\citeauthor{kevic} & & 35.22\% & 67.67\% & 79.82\% & 0.49
				& 46.71\% \\
				\hline
				\textbf{STRICT} & Graph & \textbf{36.14}\% & \textbf{72.55}\% & \textbf{84.83}\% & \textbf{0.51} & \textbf{50.29}\% \\
				\hline
				\multicolumn{7}{c}{Bug reports with good baseline queries and with localization hints (\textbf{954}) (HQ$_{H+}$)}\\
				\hline
				\textbf{Baseline} & -- & \textbf{50.01}\% & \textbf{86.75}\% & \textbf{100.00}\% & \textbf{0.66} & \textbf{66.63}\% \\
				\hline
				TF & \multirow{4}{*}{Frequency} & 35.37\% & 66.61\% & 76.87\% & 0.49 & 49.53\% \\
				\hhline{-~-----}
				IDF & & 38.22\% & 63.12\% & 74.68\% & 0.49 & 50.05\% \\
				\hhline{-~-----}
				\textbf{TF-IDF} & & \textbf{41.59}\% & \textbf{70.80}\% & \textbf{83.55}\% & \textbf{0.54} & \textbf{55.15}\% \\
				\hhline{-~-----}
				\citeauthor{kevic} & & 33.19\% & 58.93\% & 69.14\% & 0.44 & 44.91\% \\
				\hline
				STRICT & Graph & 36.53\% & 69.86\% & 82.41\% & 0.51 & 51.85\% \\
				\hline
				\multicolumn{7}{c}{Bug reports with poor baseline queries and no localization hints (\textbf{372}) (LQ$_{H-}$)}\\
				\hline
				\textbf{Baseline} & -- & \textbf{0.00}\% & \textbf{0.00}\% & \textbf{0.00}\% & \textbf{0.00} & \textbf{0.00}\% \\
				\hline
				TF & \multirow{4}{*}{Frequency} & 1.22\% & 2.99\% & 7.97\% & 0.02 & 2.31\% \\
				\hhline{-~-----}
				IDF & & 1.38\% & 4.49\% & 6.94\% & 0.03 & 2.52\% \\
				\hhline{-~-----}
				TF-IDF & & 1.36\% & 6.00\% & 9.70\% & 0.03 & 3.18\%\\
				\hhline{-~-----}
				\citeauthor{kevic} &  & 1.00\% & 6.24\% & 11.42\% & 0.03 & 3.23\% \\
				\hline
				\textbf{STRICT} & Graph & \textbf{1.56}\% & \textbf{9.54}\% & \textbf{16.89}\% & \textbf{0.05} & \textbf{5.12}\% \\
				\hline
				\multicolumn{7}{c}{Bug reports with poor baseline queries and with localization hints (\textbf{427}) (LQ$_{H+}$)}\\
				\hline
				\textbf{Baseline} & -- & \textbf{0.00}\% & \textbf{0.00}\% & \textbf{0.00}\% & \textbf{0.00} & \textbf{0.00}\% \\
				\hline
				 TF & \multirow{4}{*}{Frequency} & 0.38\% & 2.86\% & 7.02\% & 0.02 & 1.64\% \\
				\hhline{-~-----}
				IDF &  & 3.86\% & 15.11\% & 22.28\% & 0.09 & 8.74\% \\
				\hhline{-~-----}
				TF-IDF & & 1.56\% & 9.31\% & 16.80\% & 0.05 & 5.11\% \\
				\hhline{-~-----}
				\citeauthor{kevic} & & 1.12\% & 8.02\% & 12.93\% & 0.04 & 4.01\% \\
				\hline
				\textbf{STRICT} & Graph & \textbf{4.91}\% & \textbf{18.86}\% & \textbf{25.77}\% & \textbf{0.11} & \textbf{10.66}\% \\
				\hline
			\end{tabular}
		\centering
		\textbf{Emboldened}=Baseline and state-of-the-art performance measures
		\end{threeparttable}
	}
	%\vspace{-.2cm}
\end{table}

\begin{table}[!t]
	\centering
	\caption{Statistics on Keyword Counts from the Search Queries}
	\vspace{-.2cm}
	\label{table:baseline-query-stats}
	\resizebox{4.2in}{!}{%
		\begin{threeparttable}
			\begin{tabular}{l|l|c|c|c|c}
				\hline
				\textbf{Query} & \textbf{Keyword Source} & \textbf{Min} &  \textbf{Median} & \textbf{Mean} & \textbf{Max} \\
				\hline
				\hline
				Baseline & \{\emph{title}+\emph{description}\} & 03 & 32 & 49 & 406 \\
				\hline
				Baseline-I & \{\emph{title}\} & 01 & 07 & 07 & 24 \\
				\hline
				Baseline-II & \{\emph{description}\} & 01 & 28 & 46 & 404 \\
				\hline
				TF-IDF & \{\emph{title}+\emph{description}\} & 03 & 10 & 10 & 10 \\
				\hline
				STRICT & \{\emph{title}+\emph{description}\} & 03 & 10 & 10 & 10 \\
				\hline
				\multirow{2}{*}{SCP} & \{\emph{title}+\emph{description}+ & \multirow{2}{*}{03} & \multirow{2}{*}{40} & \multirow{2}{*}{56} &  \multirow{2}{*}{406} \\
				& \emph{source code}\} & & & & \\
				\hline
				\multirow{2}{*}{ACER} & \{\emph{title}+\emph{description}+ & \multirow{2}{*}{13} & \multirow{2}{*}{40} & \multirow{2}{*}{56} & \multirow{2}{*}{375} \\
				& \emph{source code}\} & & & & \\
				\hline
			\end{tabular}
		\end{threeparttable}	
	}
	\vspace{-.3cm}	
\end{table}

Since the existing approaches cannot outperform the baseline, we conduct further investigation to gain more insights. In particular, we divide our bug report collection into four different subsets based on the quality of their baseline queries and the presence of localization hints (e.g., stack traces) in the report texts (check Section \ref{sec:br-cluster} for details). Table \ref{table:freq-vs-graph-br-subset} shows the performance of existing approaches for these four different subsets. We see that some bug reports could make good queries regardless of whether they contain any localization hints (e.g., HQ$_{H+}$) or not (e.g., HQ$_{H-}$). Baseline queries from these bug reports achieve 42\%--50\% Hit@1 and 100\% Hit@10, which are highly promising and interesting.
That is, in this case, the preprocessed version of a bug report (a.k.a., Baseline) is already a good search query for the bug localization. The existing approaches choose only Top-10 keywords from each bug report, and their performance is comparatively lower than the baseline. However, they also achieve $\approx$85\% Hit@10 using their top ten search keywords, which is a high performance. 

From Table \ref{table:freq-vs-graph-br-subset}, we also see another interesting finding. Our dataset contains a total of 799 (372 LQ$_{H-}$+427 LQ$_{H+}$) bug reports with their baseline queries that fail to retrieve any buggy source document within their Top-10 results. That is, they deliver 0.00\% Hit@10 in the bug localization.
It should be noted that 427 of these bug reports contain explicit localization hints (e.g., stack traces, program elements). Unfortunately, in this case, the baseline search queries were clearly not sufficient. On the contrary, the existing approaches achieve up to 17\% Hit@10 and 26\% Hit@10 for these bug reports with no localization hints (i.e., LQ$_{H-}$) and with localization hints (i.e., LQ$_{H+}$).
Although 26\% Hit@10 could still be considered as low, it is clearly a better alternative than the baseline for these 799 bug reports which constitute 34\% of our dataset. We also note that graph-based approaches (e.g., STRICT) perform better than frequency-based approaches (e.g., TF-IDF) in delivering appropriate search keywords from these types of bug reports.

According to our analysis in Table \ref{table:freq-vs-graph-br-all}, baseline approach outperforms each of the five existing techniques on keyword-selection (e.g., TF, STRICT). Further investigation in Table \ref{table:freq-vs-graph-br-subset} demonstrates that the baseline method can only outperform the existing approaches when they are provided with the bug reports containing good baseline queries. Besides this corner case, the above findings clearly question the benefits of existing sophisticated techniques. However, it should be noted that our baseline queries (\emph{title} + \emph{description}) contain a total of 49 keywords on average (median: 32), whereas each query from the existing techniques contains at most 10 keywords (details in Table \ref{table:baseline-query-stats}).
%The title contains a median of 6 keywords.
Thus, the comparison between them was not head-to-head using same settings.
%less than fair.
 We thus further wanted to see whether the suggested keywords are any better than 10 randomly selected keywords from the bug report texts (a.k.a., random baseline, Baseline$_R$), which could show the benefits of the existing keyword selection algorithms (e.g., TF, TF-IDF, STRICT). 
 We thus repeat the process of random keyword selection five times, evaluate their performance in finding buggy documents, and then report their average performance.
 Fig. \ref{fig:freq-vs-graph-br-random} demonstrates the comparison between random baseline (Baseline$_{R}$) and the best performing existing techniques for each of the four subsets of bug reports. Interestingly, we found that the suggested keywords (by existing techniques) are indeed better than the randomly chosen baseline keywords. They achieve higher Hit@K than that of the random baseline method in all circumstances. In particular, STRICT, the graph-based approach, performs consistently higher than not only the random baseline but also the competing frequency-based techniques. We conducted non-parametric statistical tests (\eg\ \emph{Wilcoxon Signed Rank}, \emph{Cliff's delta}), and found that STRICT outperforms the random baseline (Baseline$_R$) with significant margin (\emph{p-value}$\le$0.01) and large effect size (0.54$\le\delta\le$0.90) with each of the four subsets of bug reports (Fig. \ref{fig:freq-vs-graph-br-random}-(a), (b), (c), (d)).

To determine the potential impact of \emph{tangled commits} in our ground truth (e.g., $\approx$5\%, details in Section \ref{sec:dataset}), we conduct a limited experiment using 175 bug reports. We randomly choose these bug reports and ensure that their ground truths do not contain any tangled commits through a careful manual analysis. From Table \ref{table:freq-vs-graph-br-all}, we see that the performance of baseline queries and the queries from the existing approaches reduce marginally with this limited dataset. However, the overall findings did not change. The queries from state-of-the-art approaches still cannot outperform the baseline queries in bug localization according to multiple performance metrics (e.g., Hit@K, MAP, MRR).

According to our investigation, the dataset might have contained bug reports from multiple versions of a software system, which could pose a threat to our reported findings. However, to mitigate this threat, we conducted additional experiments. In particular, we identified the project version at our disposal and selected all the bug reports from the dataset that targeted this version, which left us with 360 bug reports from six subject systems. 
The associated artifacts are included in the replication package.
From our experiments in Tables \ref{table:freq-vs-graph-br-all}, \ref{table:freq-vs-graph-improved}, we see that our major findings did not change. The performance of the search queries from each existing method increases to some extent, which aligns to an earlier finding in the similar scenario \cite{bench4bl}. However, the queries from state-of-the-art approaches \cite{saner2017masud,ase2017masud} still cannot outperform the baseline search queries, which was one of our major findings from RQ1. Furthermore, as shown in Table \ref{table:freq-vs-graph-improved}, graph-based approaches perform consistently higher than frequency-based approaches in improving the baseline queries for IR-based bug localization. Thus, our major findings remain the same even with the bug reports from a single software release.

%The associated artifacts are included in the replication package. From our experiments in Table \ref{table:freq-vs-graph-br-all}, we see that our major findings did not change. The performance of search queries from each existing method increases to some extent, which supports an earlier finding for the similar scenario \cite{bench4bl}. However, the queries from state-of-the-art approaches still cannot outperform the baseline search queries. 

%(\ie\ \emph{p-value}$\le$0.01, $\delta$=0.32 (small)) and the TF (\ie\ \emph{p-value}$\le$0.01, $\delta$=0.36 (medium)) with significant margins and small to medium effect sizes.

%the frequency-based approaches in suggesting appropriate search queries. However, as stated above, baseline queries consistently achieve the highest Hit@K. This finding  questions the benefits of the existing researches. We thus further contrast between the suggested queries and the baseline queries in a more granular level. Fig. \ref{fig:freq-vs-graph-br}-(b) demonstrates the comparison between Top-K keywords from each suggested query and first K keywords from each baseline query. We see that STRICT clearly delivers higher quality search keywords than that of baseline, and underscores the benefits of existing literature. STRICT also performs significantly higher than TF, the highest performing frequency-based approach.

\begin{figure}[!t]
	\centering
	\includegraphics[width=4.8in]{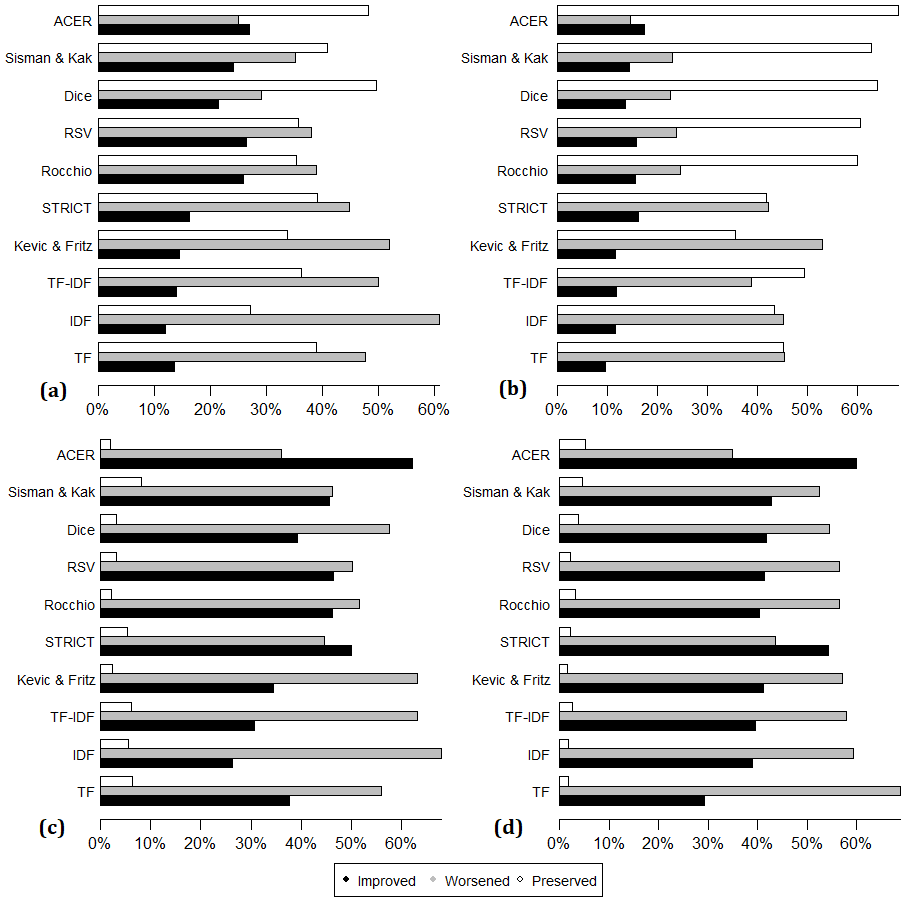}
	\caption{Improvement and worsening of baseline queries by existing techniques for (a) bug reports with good baseline queries and no localization hints, (b) bug reports with good baseline queries and with localization hints, (c) bug reports with poor baseline queries and no localization hints, and (d) bug reports with poor baseline queries and with localization hints}
	\label{fig:freq-vs-graph-br-improved}
	\vspace{-.3cm}
\end{figure}

While the above analyses considered only Top-10 results, we further contrast between frequency-based and graph-based approaches by analysing all the results retrieved by their queries. In particular, we compare each suggested query with its baseline counterpart in terms of their Query Effectiveness (\ie\ position of the first ground truth, defined in Section \ref{sec:pmetrics}). If the suggested query achieves a higher rank than the baseline, existing literature \cite{trconfig,refoqus} defines it as \emph{query improvement}, otherwise it is considered to be %and the vice versa as 
\emph{query worsening}. If both queries achieve the same rank, then it is called \emph{query preserving}.
From Table \ref{table:freq-vs-graph-improved}, we see that TF-IDF performs the best among the four frequency-based approaches. It improves 21\% and preserves 31\% of the baseline queries. Unfortunately, TF-IDF based approach also worsens 49\% of the queries. On the contrary, STRICT, a graph-based approach, improves 29\% and worsens 44\% of the queries which are 40\% higher and 11\% lower respectively. 
When this experiment is repeated using the 175 manually selected bug reports and 360 single-release bug reports above, our findings also remain aligned (Table \ref{table:freq-vs-graph-improved}). 
However, as shown above, none of these existing techniques improves more baseline queries than it worsens. 
We thus further investigate this issue using different subsets of bug reports. From Fig. \ref{fig:freq-vs-graph-br-improved}, we see that STRICT can actually improve more baseline queries than it worsens when the technique is provided with the bug reports containing poor baseline queries (LQ$_{H-}$, LQ$_{H+}$). Unfortunately, it fails to do so when the bug reports contain good baseline queries (i.e., Query Effectiveness $\le$10) in their texts.

\begin{table}[!t]
	\centering
	\caption{Improvement \& Worsening of Baseline Queries by Existing Techniques}\label{table:freq-vs-graph-improved}
	%\vspace{-.2cm}
	\resizebox{4.8in}{!}{%
		\begin{threeparttable}
			\begin{tabular}{l|c|c|c|c}
				\hline
				\textbf{Query} & \textbf{Genre} & \textbf{Improvement} & \textbf{Worsening} & \textbf{Preserving}\\
				\hline
				\hline
				\multicolumn{5}{c}{\textbf{Query Construction using Bug Reports Only (2,320)}} \\
				\hline
				TF & \multirow{4}{*}{Frequency} & 434 (18.71\%) &  1,204 (51.90\%) & 682 (29.40\%)\\
				\hhline{-~---}
				IDF & & 443 (19.10\%) & 1,281 (55.22\%) & 596 (25.69\%) \\
				\hhline{-~---}
				\textbf{TF-IDF} & & 475 (20.47\%) & 1,135 (48.92\%) & 710 (30.60\%) \\
				\hhline{-~---}
				\citeauthor{kevic} & & 496 (21.38\%) & 1,278 (55.09\%) & 546 (23.53\%) \\
				\hline
				\textbf{STRICT} & Graph & \textbf{664 (28.62\%)} & \textbf{1,008 (43.45\%)} & \textbf{648 (27.93\%)} \\
				\hline
				\multicolumn{5}{c}{\textbf{Query Construction using Bug Reports (2,320) + Source Code Documents}} \\
				\hline
				Rocchio & \multirow{4}{*}{Frequency} & 639 (27.54\%) & 887 (38.23\%) & 794 (34.22\%) \\
				\hhline{-~---}
				RSV &  & 650 (28.02\%) & 869 (37.46\%) & 801 (34.53\%) \\
				\hhline{-~---}
				Dice &  & 575 (24.78\%) & 827 (35.65\%) & 918 (39.57\%) \\
				\hhline{-~---}
				SCP &  & 626 (26.98\%) & 814 (35.09\%) & 880 (37.93\%) \\
				\hline
				\textbf{ACER} & Graph & \textbf{805 (34.70\%)} & \textbf{563 (24.27\%)} & \textbf{952 (41.03\%)} \\
				\hline
				\multicolumn{5}{c}{\textbf{Query construction using manually selected bug reports (175)}}\\
				\hline
				TF & \multirow{4}{*}{Frequency} & 34 (19.43\%) & 81 (46.29\%) & 60 (34.29\%) \\
			    \hhline{-~---}
			    IDF & & 27 (15.43\%) & 113 (64.57\%) & 35 (20.00\%) \\
			    \hhline{-~---}
			    TF-IDF &  & 36 (20.57\%) & 92 (52.57\%) & 47 (26.86\%) \\
			    \hhline{-~---}
			    \citeauthor{kevic} & & 36 (20.57\%) & 93 (53.14\%) & 46 (26.29\%) \\
			    \hline
			    \textbf{STRICT} & \multirow{2}{*} {Graph} & \textbf{47 (26.86\%)} & \textbf{75 (42.86\%)} & \textbf{53 (30.29\%)} \\
			    \hhline{-~---}
			    \textbf{ACER} & & \textbf{57 (32.57\%)} & \textbf{55 (31.43\%)} & \textbf{63 (36.00\%)} \\
				\hline
				\multicolumn{5}{c}{\textbf{Query construction using single-release bug reports (360)}}\\
				\hline
				TF & \multirow{4}{*}{Frequency}  & 65 (18.06\%) & 199 (55.28\%) & 96 (26.67\%) \\
				\hhline{-~---}
				IDF & & 75 (20.83\%) & 203 (56.39\%) & 82 (22.78\%) \\
				\hhline{-~---}
				TF-IDF &  & 81 (22.50\%) & 181 (50.28\%) & 98 (27.22\%) \\
				\hhline{-~---}
				\citeauthor{kevic} & & 76 (21.11\%) & 213 (59.17\%) & 71 (19.72\%) \\
				\hline
				\textbf{STRICT} & \multirow{2}{*} {Graph} & \textbf{106 (29.44\%)} & \textbf{168 (46.67\%)} & \textbf{86 (23.89\%)} \\
				\hhline{-~---}
				\textbf{ACER} & & \textbf{133 (36.94\%)} & \textbf{87 (24.17\%)} & \textbf{140 (38.89\%)} \\
				\hline
			\end{tabular}
		\centering
		\textbf{Emboldened}=State-of-the-art performance measures
		\end{threeparttable}
	}
	\vspace{-.2cm}
\end{table}

\begin{table}[!t]
	\centering
	\caption{Frequency-Based vs. Graph-Based Keyword Selection from Source Code Document (Using Top-10 Results Only)}\label{table:freq-vs-graph-code}
	\vspace{-.2cm}
	\resizebox{4.6in}{!}{%
		\begin{threeparttable}
			\begin{tabular}{l|c|c|c|c|c|c}
				\hline
				\textbf{Technique} & \textbf{Genre} & \textbf{Hit@1} & \textbf{Hit@5} & \textbf{Hit@10} & \textbf{MRR} & \textbf{MAP} \\
				\hline
				\hline
				Baseline & \multirow{3}{*}{--} & 31.98\% & 57.96\% & 66.50\% & 0.43 & 42.69\% \\
				\hhline{-~-----}
				Baseline-I &  & 22.37\% & 46.73\% & 57.58\% & 0.33 & 32.82\% \\
				\hhline{-~-----}
				Baseline-II &  & 29.01\% & 51.16\% & 60.43\% & 0.38 & 38.21\% \\				
				\hline
				Rocchio & \multirow{4}{*}{Frequency} &  30.17\% & 55.19\% & 65.03\%
				& 0.41 & 40.76\%\\
				\hhline{-~-----}
				RSV &  & 30.37\% & 55.75\% & 65.00\% & 0.41 & 41.19\%\\
				\hhline{-~-----}
				Dice &  &  31.15\% & 56.99\% & 66.10\% & 0.42 & 41.86\% \\
				\hhline{-~-----}
				SCP & & 31.34\% & 56.03\% & 66.20\% & 0.42 & 41.95\% \\
				\hline
				\textbf{ACER} & Graph & \textbf{32.21}\% & \textbf{58.88}\% & \textbf{67.45}\% & \textbf{0.43} & \textbf{43.34}\% \\
				\hline
			\end{tabular}
		\centering
		\textbf{Emboldened}=State-of-the-art performance measures
		\end{threeparttable}
	}
	\vspace{-.4cm}
\end{table}

\begin{table}[!t]
	\centering
	\caption{Impact of Bug Report Quality and Localization Hints on Query Expansion (Using Top-10 Results Only)}\label{table:freq-vs-graph-code-subset}
	%\vspace{-.2cm}
	\resizebox{4.8in}{!}{%
		\begin{threeparttable}
			\begin{tabular}{l|c|c|c|c|c|c}
				\hline
				\textbf{Technique} & \textbf{Genre} & \textbf{Hit@1} & \textbf{Hit@5} & \textbf{Hit@10} & \textbf{MRR} & \textbf{MAP} \\
				\hline
				\multicolumn{7}{c}{Bug reports with good baseline queries and no localization hints (\textbf{567}) (HQ$_{H-}$)}\\
				\hline
				Baseline & -- & \textbf{41.96}\% & \textbf{86.18}\% & \textbf{100.00}\% & \textbf{0.60} & \textbf{58.15}\% \\				
				\hline
				Rocchio & \multirow{4}{*}{Frequency} &  38.92\% & 79.32\% & 92.20\%
				& 0.56 & 54.45\%\\
				\hhline{-~-----}
				RSV &  & 38.24\% & 80.81\% & 92.56\% & 0.56 & 55.29\%\\
				\hhline{-~-----}
				Dice &  & \textbf{43.32}\% & 84.13\% & \textbf{95.74}\% & 0.60 & 58.17\% \\
				\hhline{-~-----}
				SCP & & 41.36\% & 81.09\% & 92.68\% & 0.58 & 56.46\% \\
				\hline
				\textbf{ACER} & Graph & 42.55\% & \textbf{85.62}\% & 94.87\% & \textbf{0.61} & \textbf{58.90}\% \\
				\hline
				\multicolumn{7}{c}{Bug reports with good baseline queries and with localization hints (\textbf{954}) (HQ$_{H+}$)}\\
				\hline
				Baseline & -- & \textbf{50.01}\% & \textbf{86.75}\% & \textbf{100.00}\% & \textbf{0.66} & \textbf{66.63}\% \\
				\hline
				Rocchio & \multirow{4}{*}{Frequency} & 47.92\% & 83.69\% & 94.46\% & 0.63 & 63.71\% \\
				\hhline{-~-----}
				RSV & & 48.61\% & 84.04\% & 94.58\% & 0.63 & 64.40\% \\
				\hhline{-~-----}
				Dice & & 47.88\% & 85.47\% & 96.08\% & 0.63 & 64.46\% \\
				\hhline{-~-----}
				SCP & & 48.02\% & 83.71\% & 95.80\% & 0.63 & 64.22\% \\
				\hline
				\textbf{ACER} & Graph & \textbf{50.20}\% & \textbf{88.95}\% & \textbf{97.43}\% & \textbf{0.66} & \textbf{67.06}\% \\
				\hline
				\multicolumn{7}{c}{Bug reports with poor baseline queries and no localization hints (\textbf{372}) (LQ$_{H-}$)}\\
				\hline
				Baseline & -- & \textbf{0.00}\% & \textbf{0.00}\% & \textbf{0.00}\% & \textbf{0.00} & \textbf{0.00}\% \\
				\hline
				Rocchio & & 0.00\% & 2.68\% & 9.99\% & 0.02 & 1.51\% \\
				\hhline{-~-----}
				RSV & & 0.00\% & 2.20\% & 10.71\% & 0.02 & 1.62\% \\
				\hhline{-~-----}
				Dice &  & 0.00\% & 0.00\% & 4.77\% & 0.01 & 1.00\% \\
				\hhline{-~-----}
				SCP & Graph & 0.00\% & 1.92\% & 11.39\% & 2.05\% & 2.09\% \\
				\hline
				\textbf{ACER} & & 0.00\% & 2.36\% & \textbf{11.78}\% & 1.76\% & 1.81\% \\
				\hline
				\multicolumn{7}{c}{Bug reports with poor baseline queries and with localization hints (\textbf{427}) (LQ$_{H+}$)}\\
				\hline
				Baseline & -- & \textbf{0.00}\% & \textbf{0.00}\% & \textbf{0.00}\% & \textbf{0.00} & \textbf{0.00}\% \\
				\hline
				Rocchio &  & 0.00\% & 1.00\% & 5.44\% & 0.01 & 1.00\% \\
				\hhline{-~-----}
				RSV &  & 0.00\% & 1.00\% & 4.91\% & 0.01 & 1.00\% \\
				\hhline{-~-----}
				Dice &  & 0.00\% & 1.01\% & 6.65\% & 0.01 & 1.00\% \\
				\hhline{-~-----}
				SCP &  & 0.00\% & 1.82\% & 5.83\% & 0.01 & 1.09\% \\
				\hline
				\textbf{ACER} & Graph &  0.00\% & 1.00\% & \textbf{8.78}\% & 0.01 & 1.15\% \\
				\hline
			\end{tabular}
		\centering
		\textbf{Emboldened}=Baseline and state-of-the-art performance measures
		\end{threeparttable}
	}
	\vspace{-.5cm}
\end{table}

\subsection{Search Keyword Selection from Source Code Document}
Our analyses show that (1) bug reports could make poor baseline queries and (2) existing techniques for selecting keywords from them might not be sufficient. Thus, several approaches \cite{rocchio,rsv,sisman,qsurvey,ase2017masud} attempt to complement the baseline queries with appropriate keywords chosen from apparently relevant source code documents. That is, each search query from these approaches is a combination of baseline query (from the bug report) and suggested keywords from the source code.
Tables \ref{table:freq-vs-graph-improved}, \ref{table:freq-vs-graph-code}, \ref{table:freq-vs-graph-code-subset}, and Fig. \ref{fig:freq-vs-graph-br-improved} summarize our comparative analysis between frequency-based and graph-based keyword selection from source code documents. From Table \ref{table:freq-vs-graph-code}, we see that all four frequency-based approaches -- Rocchio \cite{rocchio}, RSV \cite{rsv}, Dice \cite{refoqus} and SCP \cite{sisman} -- perform almost equally. They achieve a maximum of 66\% Hit@10, 42\% precision and a reciprocal rank of  0.42 which are marginally lower than the baseline measures. Thus, TF-IDF and its variants might not be enough to select keywords neither from the bug reports (e.g., Table \ref{table:freq-vs-graph-br-subset}) nor from the source code. On the other hand, ACER, a graph-based approach, achieves 68\% Hit@10 with 43\% precision and 0.43 reciprocal rank which are marginally higher than the baseline measures. We also further investigate the performances of baseline and existing techniques using four subsets of bug reports. From Table \ref{table:freq-vs-graph-code-subset}, we see that the keywords collected from the source code improve the baseline queries marginally when they are already good (e.g., HQ$_H-$, HQ$_{H+}$). However, these keywords from source code can significantly improve the baseline queries when they are poor (e.g., LQ$_{H-}$, LQ$_{H+}$). In particular, the best performing approach--ACER-- achieves 9\%--12\% Hit@10 for the bug reports with poor baseline queries, which is 0.00\% for the baseline method. It should be noted that these accuracy measures (Tables \ref{table:freq-vs-graph-code}, \ref{table:freq-vs-graph-code-subset}) are based on Top-10 results only. We also note that the graph-based approaches are more effective than their frequency-based counterparts in delivering appropriate search keywords from the source code documents.

We also compare between graph-based and frequency-based approaches based on how much they can improve the baseline queries through query expansion. From Table \ref{table:freq-vs-graph-improved}, we see that SCP \cite{sisman} was found more effective than other frequency-based approaches in delivering appropriate keywords from the source code. It improves 27\% and preserves 38\% of the baseline queries. However, this approach also worsens 35\% of the queries. On the contrary, ACER improves 35\% and worsens 24\% of the baseline queries which are 29\% higher and 31\% lower respectively. That is, the graph-based approach achieves $\approx$ 10\% \emph{net improvement} (i.e., improvement -- worsening) in the baseline queries. 
When this experiment is repeated using 175 manually selected and 360 single-release bug reports, our findings also remain aligned.
We also further investigate using four different bug report groups. From Fig. \ref{fig:freq-vs-graph-br-improved}, we see that ACER can actually improve more baseline queries than it worsens. It also should be noted that while the net improvement is marginal for the bug reports with good baseline queries, it is significantly large (e.g., 26\%) for the bug reports with poor baseline queries. All these empirical findings above suggest that graph-based approaches are a better choice for search keyword selection from the source code documents in the context of IR-based bug localization.

%are 4\%, 8\% and 9\% higher respectively. Such performance improvements against the baseline are 3\%, 8\% and 7\% respectively.
%We also conducted significance tests (\eg\ \emph{WSR}, \emph{Cliff's delta}), and found that ACER outperforms the baseline in MRR (\ie\ \emph{p-value}=0.03, $\delta$=0.61(large)) and MAP (\ie\ \emph{p-value}=0.03, $\delta$=0.39 (medium)) with significant margins. ACER also outperforms Dice, the highest performing frequency-based approach, in MRR (\ie\ \emph{p-value}=0.03, $\delta$=0.67 (large)) and MAP (\ie\ \emph{p-value}=0.03, $\delta$=0.50 (large)) with significant margins.

%Table \ref{table:freq-vs-grap-code-improved} further contrasts between frequency-based and graph-based approaches in terms of their query improvement capabilities. We see that RSV improves a maximum of 361 out of 1,065 queries while worsening 435 queries. On the contrary, ACER improves 413 queries and worsens 272 queries which are 14\% higher and 38\% lower respectively.

\vspace{-.2cm}
\FrameSep5pt
\begin{frshaded}
	\noindent
	\textbf{Summary of RQ$\mathbf{_1}$:}
	Bug reports could lead to poor search queries despite having explicit bug localization hints (e.g., program elements, stack traces) and to good search queries
	despite lacking such hints, which challenges the conventional wisdom. Queries from existing approaches achieve a maximum of 68\% Hit@10, 43\% MAP and 0.43 MRR in bug localization, which are comparable to the baseline measures. However, neither existing techniques nor baseline technique might be able to deliver appropriate search queries from a significant number of bug reports (e.g., 34\%), which warrants further investigation. According to our experiments, graph-based approaches consistently perform higher than the best performing frequency-based approaches in suggesting appropriate keywords both from bug reports and from source code documents.
	
	%Unfortunately, the queries suggested by the existing approaches might always not outperform the ad-hoc baseline queries when the bug reports lack localization hints (\eg\ relevant program elements, stack traces).
\end{frshaded}

\section{Answering RQ$_2$: Comparing Optimal, Near-Optimal Search Queries with Baseline and State-of-the-Art Queries in IR-Based Bug Localization}
\label{sec:rq2}

\subsection{Genetic Algorithm (GA)-Based Keyword Selection}\label{sec:ga}
According to \textbf{RQ$\mathbf{_1}$}, existing state-of-the-art approaches might not be enough for delivering appropriate search queries from the bug reports that lead to poor baseline queries (LQ$_{H+}$, LQ$_{H-}$). Their queries achieve a maximum of 26\% Hit@10 that outperforms baseline Hit@10 (0.00\%) when only Top-10 results are analyzed. However, this still indicates a low performance. We thus wanted to determine whether the these bug reports actually contain any meaningful keywords that could form any optimal or near-optimal search queries. We define an \emph{optimal search query} as a set of keywords that can retrieve the first faulty source document at the topmost position of the result list (Table \ref{table:nr-query-def}). On the other hand, \emph{near-optimal queries} return their first faulty document at the 2$^{nd}$ to 10$^{th}$ position of their result list. 

%More specifically, we investigate whether there exists any near optimal search query within the report that outperforms the baseline.
%know  better than the baseline or not. Thus,
%Thus,
%\Foutse{Yet, the insights generated from this analysis could.....(provide some argumentation to support the fact that although the NrOptimal$_{GA}$ might not be used in practice, it could help gather useful insights to improve other techniques for example... )}

We employ a Genetic Algorithm (GA)-based approach for identifying optimal and near-optimal search keywords from each bug report,
%We attempt to identify the optimal search keywords from each bug report using a ,
as done by an earlier study \cite{cmills-icsme2018}. Algorithms \ref{algo} and \ref{algo2} show the pseudo code of our GA-based approach, \textbf{NrOptimal$\mathbf{_{GA}}$}. It should be noted that this \textbf{NrOptimal$_{GA}$ might not be applied in a real-world bug localization scenario where the ground truth is not known} \cite{cmills-icsme2018}. Although NrOptimal$_{GA}$ could not be used in practice, it could help gather useful insights (e.g., characteristics of optimal keywords) to improve other existing techniques (e.g., Tables \ref{table:actionable-baseline}, \ref{table:actionable-strict}). 
We thus attempt to construct an optimal query from each bug report and if not possible, a near-optimal query from each bug report using five different steps inspired by biological evolution as follows.

\textbf{(a) Initial Population Generation:} The first step of a Genetic Algorithm is the initialization of solution candidates (a.k.a., \emph{population}) where each candidate (or chromosome) is characterized with a fixed set of genes (Lines 4--6, Algorithm \ref{algo}). In the context of our query construction problem, we initialize a population $pop$ of $M$ candidate queries. Each candidate comprises of $K$ terms (or genes) randomly taken from all terms (or keywords) $R$ of a bug report. It should be noted that we choose $K=$ 5, 10, 15, 20 terms as the initial length of each solution candidate. Population from each iteration of the algorithm is also called a \emph{generation}.

\textbf{(b) Fitness Evaluation:} Since Genetic Algorithms attempt to find out an optimal (or near-optimal) solution to a problem, each solution candidate from each generation is evaluated with a \emph{fitness function}. In this study, we design our GA-based approach with Query Effectiveness (Section \ref{sec:pmetrics}) as the fitness function. That is, NrOptimal$_{GA}$ attempts to find out a candidate query that retrieves the ground truth as its topmost search result (\ie\ QE=1) (Lines 8--9, 17, Algorithm \ref{algo}). Although the Query Effectiveness (QE) has been primarily used in our algorithm, other performance metrics (e.g., MAP) could be easily employed as the fitness function. Fitness function is frequently invoked during the selection step.

\textbf{(c) Selection:} The third step of a Genetic Algorithm is the population selection. The idea behind natural selection is to select the fittest individuals and let them pass their genes to the next generation. In the context our query construction problem, we select two fittest individuals (i.e., high performing candidate queries) from a population (generation) as parents and send them forward for further evolution. In particular, we arrange two tournaments by randomly selecting two subsets ($N$ candidates each) of a population $pop$, and identify the two fittest candidates as parents (Lines 9--12, Algorithm \ref{algo2}).

\begin{algorithm}[!t]
	\caption{Query Construction from a Bug Report using Genetic Algorithm}
	\label{algo}
	\normalsize
	\begin{algorithmic}[1]
		\Procedure{NrOptimal$_{GA}$}{$R$}
		\LineComment{$R$: all terms from a bug report}
		\LineComment{$Q$: near optimal search query}
		\LineComment{population selection}
		\State $pop \gets$generatePopulation($R$, $M$, $K$)
		\State $generation$++
		\While {$pop$.getFittest().fitness$<$MAX\_FITNESS}
		\LineComment{fitness evaluation}
		\State Individual $fittest\gets pop$.getFittest()
		\LineComment{crossover and mutation}
		\State $pop\gets$evolvePopulation($fittest$, $pop$)
		\If {$generation$==MAX\_GENERATION}
		\State \textbf{break}
		\EndIf
		\EndWhile
		\LineComment{suggesting the near-optimal query}
		\State $Q\gets pop$.getFittest().genes
		\State \textbf{return} $Q$
		\EndProcedure
	\end{algorithmic}
\end{algorithm}

\begin{algorithm}[!t]
	\caption{Evolve the Population of Candidate Queries}
	\label{algo2}
	\normalsize
	\begin{algorithmic}[1]
		\Procedure{evolvePopulation}{$pop$,$fittest$}
		\LineComment{$pop$: current population of candidate queries}
		\LineComment{$fittest$: fittest individual}
		\State $newPop\gets \{\}$
		\LineComment{retaining the fittest individual}
		\If {$elitismEnabled$}
		\State  $newPop[0]\gets$$fittest$
		\EndIf
		\LineComment{evolve through selection and crossover operations}
		\For {$i\in pop$.index$-1$}
		\State $indiv1\gets$tournamentSelection($pop$)
		\State $indiv2\gets$tournamentSelection($pop$)
		\State $newIndiv\gets$crossover($indiv1$, $indiv2$)
		\State $newPop[i]\gets$$newIndiv$
		\EndFor
		\LineComment{evolve through mutation operation}
		\For {$i\in newPop$.index}
		\State $indiv\gets$$newPop[i]$
		\State $muIndiv\gets$mutate($indiv$)
		\State $newPop[i]\gets$$muIndiv$
		\EndFor
		\LineComment{returning the evolved population}
		\State \textbf{return} $newPop$
		\EndProcedure
	\end{algorithmic}
\end{algorithm}

\textbf{(d) Crossover:} Chromosomal \emph{crossover} involves reconstruction of a new chromosome from two given chromosomes through gene overlapping. In our problem context, we choose two fittest candidate queries from the selection step above and generate a new candidate by randomly swapping their keywords (or genes). We adopt a \emph{uniform crossover} strategy for generating the new chromosome. That is, for each gene index within the chromosome, the new gene (keyword) is taken from either parent chromosome (i.e., candidate query) with an equal probability (i.e., 0.5). 
%The crossover operation might sometimes lead to \emph{duplicate} keywords (genes), which are discarded as they do not improve a candidate's fitness according to our investigation. 
We repeat the process $M-1$ times and generate a new population $newPop$ using crossover and elitism strategy (Lines 9--15, Algorithm \ref{algo2}). Elitism ensures that the fittest candidate ($fittest$) from a previous generation joins the new generation without any crossover operation \cite{genprog} (Lines 5--8, Algorithm \ref{algo2}).

\begin{table}[!t]
	\centering
	\caption{A Working Example of Query Construction with NrOptimal$_{GA}$}\label{table:example-workflow}
	\vspace{-.2cm}
	\resizebox{4.8in}{!}{%
		\begin{threeparttable}
			\begin{tabular}{l|p{3in}|c}
				\hline
				\textbf{Step} & \textbf{Candidate Queries} & \textbf{QE} \\	
				\hline
				\hline
				All keywords & \{bug, change, dialogs, primitive, object, accessible,	command, variables, expressions, view, contexts\} & 105 \\
				\hline
				& C$_1$: \{variables, command, variables, variables, contexts\}  & $\infty$ \\
				& C$_2$: \{variables, dialogs, change, object, object\} & 167 \\
				& C$_3$: \{expressions, change, command, command, primitive\} & 233 \\
				& C$_4$: \{dialogs, accessible, accessible, expressions, change\} & 121 \\
				Initialize & C$_5$: \{dialogs, contexts, object, primitive, expressions\} & 13\\
				population & C$_6$: \{contexts, view, view, dialogs, command\} & 07 \\
				& C$_7$: \{dialogs, dialogs, contexts, Bug, accessible\} & 26 \\
				& ................. & - \\
				& ................. & - \\
				& C$_{M}$: \{contexts, command, object, accessible, contexts\} & 619\\
				\hline
				\multirow{2}{*}{Selection} & P$_1$: \{contexts, view, view, dialogs, command\} & 07 \\
				& P$_2$: \{dialogs, contexts, object, primitive, expressions\} & 13 \\
				\hline
				Crossover & Child: \{dialogs, view, view, primitive, command\} & \textbf{05}\\
				\hline
				Mutation & MC: \{view, primitive, command, dialogs\} & \textbf{04} \\
				\hline
				\multicolumn{3}{c}{\textbf{After Three Generation of Evolutions}} \\
				\hline
				\multicolumn{3}{c}{Duplicate keywords removal} \\
				\hline
				\textbf{Solution} & \textbf{\{primitive, dialogs\}} & \textbf{01} \\
				\hline
			\end{tabular}
			\centering
			\textbf{QE} = Query Effectiveness, \textbf{C$_i$} = Chromosome, \textbf{P$_i$} = Parent chromosome, \textbf{MC} = Mutated chromosome
		\end{threeparttable}
	}
	\vspace{-.6cm}
\end{table}

\textbf{(e) Mutation:} Once a new population is created, \emph{mutation} modifies each candidate (chromosome) by randomly \emph{flipping} genes in random positions within a chromosome \cite{genprog}. Similarly, we randomly \emph{replace} the keywords with other keywords within each candidate query and mutate all candidates of a population (Lines 16--21, Algorithm \ref{algo2}). In particular, we use a non-uniform, single point mutation strategy with a low mutation rate where each point of mutation is chosen with a small probability of 0.015. We implement a restrictive mutation policy to preserve the diversity of the population that was created in the crossover step above. Once completed, we finalize the new population for the next round of evolution.

Four genetic operations above -- selection, fitness calculation, crossover, mutation -- are performed iteratively on each generation until a candidate query with the \emph{maximum fitness} (\eg\ QE = 1, ground truth as the topmost result) is found or the generation count reaches the maximum threshold (\eg\ 100). 
The crossover and mutation operations above might sometimes lead to \emph{duplicate} keywords (genes) in a candidate query, which are discarded as they do not improve a candidate's fitness according to our investigation. 
We initialize each of the candidate queries (chromosome) with $K$ unique keywords where $5\le$$K\le20$.
Further justification on this can be found in Table \ref{table:ga-config}.
%We construct each candidate query using $K$=10 terms from the bug report, as Top-K keywords are suggested by the existing approaches \cite{saner2017masud,refoqus}.
Since Genetic Algorithms involve randomness, we execute our approach \emph{three} times on our dataset and report the performance for a combined set of optimal and near-optimal search queries.

\subsection{An Example Workflow of NrOptimal$_{GA}$ Algorithm} \label{sec:example-workflow}
Table \ref{table:example-workflow} shows a working example of query construction from a given bug report (e.g., Bug \#189012, \texttt{eclipse.jdt.debug}). We first collect all unique keywords from the \emph{title} and \emph{description} of the report and initialize a population of $M$ solution candidates (or chromosomes). As shown in the Table \ref{table:example-workflow}, each candidate consists of $K = 5$ search keywords that determine the candidate's fitness to be a solution. We select the two fittest candidates (e.g., P$_1$ and P$_2$) from this population and generate a new candidate query from them using crossover operation (i.e., keyword overlapping). The operation also led to duplicate keywords, which were discarded. It should be noted that the crossover step improves the candidate's fitness. We repeat crossover operation $M$ times and generate a new population of solution candidates. 
Then each of these candidates goes through a mutation step that helps them further improve their fitness. After three generation of evolution (selection, crossover and mutation), our algorithm, NrOptimal$_{GA}$ delivers the an optimal search query --\emph{\{primitive, dialogs\}}--that achieves the best possible outcome in the IR-based bug localization (i.e., \textbf{QE=1}).

\begin{table}[!t]
	\centering
	\caption{Query Definitions}\label{table:nr-query-def}
	\vspace{-.2cm}
	\resizebox{4.8in}{!}{%
		\begin{threeparttable}
			\begin{tabular}{l|p{3.2in}}
				\hline
				\textbf{Query} & \textbf{Description}\\
				\hline
				\hline
				\emph{Baseline query} & The pre-processed version of the title and description texts from a bug report. \\
				\hline
				\emph{Optimal query} & The search query that returns the first faulty source document at the \emph{topmost} position of its result list. \\
				\hline
				\emph{Near-optimal query} & The search query that returns the first faulty source document at the 2$^{nd}$ to 10$^{th}$ position of its result list. \\
				\hline
				\emph{Non-optimal query} & The search query that returns the first faulty source document below the 10$^{th}$ position of its result list. \\
				\hline
			\end{tabular}
		\end{threeparttable}
	}
	\vspace{-.1cm}
\end{table}

\begin{table}[!t]
	\centering
	\caption{Baseline vs. Near-Optimal Search Queries from Bug Reports (Using Top-10 Results Only)}\label{table:opt-vs-baseline-br}
	%\vspace{-.2cm}
	\resizebox{4.8in}{!}{%
		\begin{threeparttable}
			\begin{tabular}{l|c|c|c|c|c}
				\hline
				\textbf{Query} & \textbf{Hit@1} & \textbf{Hit@5} & \textbf{Hit@10} & \textbf{MRR} & \textbf{MAP} \\
				\hline
				\hline
				\multicolumn{6}{c}{Performance with all bug reports (\textbf{2,320})} \\
				\hline
				\textbf{Baseline}  & \textbf{31.98}\% & \textbf{57.96}\% & \textbf{66.50}\% & \textbf{0.43} & \textbf{42.69}\% \\
				\hhline{------}
				Baseline-I &   22.37\% & 46.73\% & 57.58\% & 0.33 & 32.82\% \\
				\hhline{------}
				Baseline-II &   29.01\% & 51.16\% & 60.43\% & 0.38 & 38.21\% \\
				\hline
				\textbf{NrOptimal$\mathbf{_{GA}}$} &  \textbf{87.41\%}\% & \textbf{93.94}\% & \textbf{95.74}\%
				& \textbf{0.90} & \textbf{90.00}\%\\
				\hline
				\multicolumn{6}{c}{Bug reports with good baseline queries and no localization hints (\textbf{567}) (HQ$_{H-}$)}\\
				\hline
				Baseline & 41.96\% & 86.18\% & 100.00\% & 0.60 & 58.15\% \\
				\hline
				Baseline-I & 32.26\% & 64.94\% & 79.72\% & 0.46 & 45.57\% \\
				\hline
				Baseline-II & 35.10\% & 69.96\% & 81.96\% & 0.50 & 47.65\% \\
				\hline
				\textbf{NrOptimal$\mathbf{_{GA}}$} & \textbf{95.93}\% & \textbf{100.00}\% & \textbf{100.00}\% & \textbf{0.98} & \textbf{93.51}\% \\
				\hline
				\multicolumn{6}{c}{Bug reports with good baseline queries and with localization hints (\textbf{954})(HQ$_{H+}$)}\\
				\hline
				Baseline & 50.01\% & 86.75\% & 100.00\% & 0.66 & 66.63\% \\
				\hline
				Baseline-I & 29.97\% & 61.20\% & 71.16\% & 0.43 & 43.79\% \\
				\hline
				Baseline-II & 46.11\% & 79.18\% & 92.07\% & 0.60 & 60.90\% \\
				\hline
				\textbf{NrOptimal$\mathbf{_{GA}}$} & \textbf{98.04}\% & \textbf{99.91}\% & \textbf{100.00}\% & \textbf{0.99} & \textbf{100.00}\% \\
				\hline
				\multicolumn{6}{c}{Bug reports with poor baseline queries and no localization hints (\textbf{372})(LQ$_{H-}$)}\\
				\hline
				Baseline & 0.00\% & 0.00\% & 0.00\% & 0.00 & 0.00\% \\
				\hline
				Baseline-I & 2.44\% & 8.40\% & 16.17\% & 0.05 & 5.05\% \\
				\hline
				Baseline-II & 0.00\% & 1.00\% & 5.55\% & 0.01 & 1.00\% \\
				\hline
				\textbf{NrOptimal$\mathbf{_{GA}}$} & \textbf{50.04}\% & \textbf{69.68}\% & \textbf{77.96}\% & \textbf{0.58} & \textbf{56.47}\% \\
				\hline
				\multicolumn{6}{c}{Bug reports with poor baseline queries and with localization hints (\textbf{427})(LQ$_{H+}$)}\\
				\hline
				Baseline & 0.00\% & 0.00\% & 0.00\% & 0.00 & 0.00\% \\
				\hline
				Baseline-I & 5.01\% & 18.99\% & 28.20\% & 0.11 & 10.98\% \\
				\hline
				Baseline-II & 0.00\% & 0.00\% & 0.02\% & 0.00 & 0.00\% \\
				\hline
				\textbf{NrOptimal$\mathbf{_{GA}}$} & \textbf{80.70}\% & \textbf{91.00}\% & \textbf{93.37}\% &
				\textbf{0.85} & \textbf{86.19}\% \\
				\hline
			\end{tabular}
		\centering
		\textbf{Emboldened}=Baseline and NrOptimal$_{GA}$ performance measures
		
		\end{threeparttable}

	}
	\vspace{-.6cm}
\end{table}

\subsection{Comparison between Optimal or Near-Optimal Query and Baseline Query}
\begin{figure}[!t]
	\centering
	\includegraphics[width=4.6in]{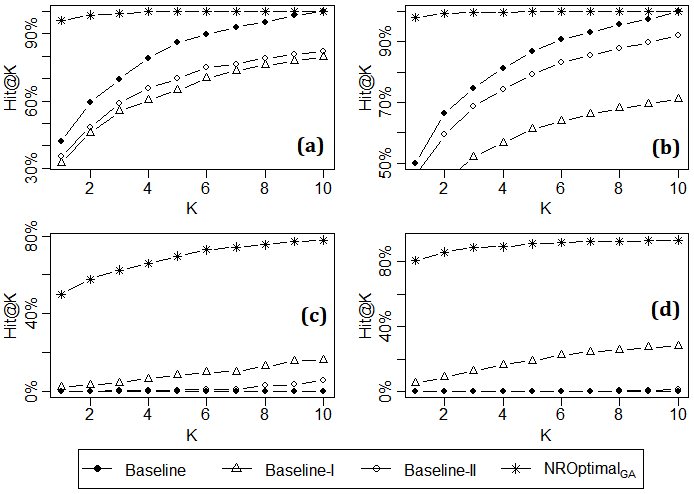}
	%\vspace{-.2cm}
	\caption{Comparison of Hit@K between near-optimal and baseline queries from (a) bug reports with good baseline queries and no localization hints, (b) bug reports with good baseline queries and with localization hints, (c) bug reports with poor baseline queries and no localization hints, and (d) bug reports with poor baseline queries and with localization hints}
	\label{fig:opt-vs-baseline}
	\vspace{-.5cm}
\end{figure}

\begin{figure}[!t]
	\centering
	\includegraphics[width=4.6in]{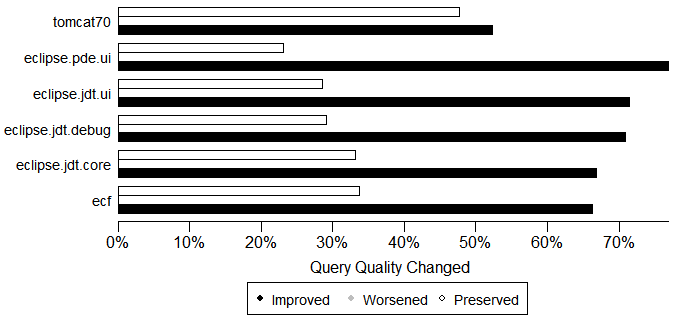}
	%\vspace{-.6cm}
	\caption{Query improvement upon the Baseline alternatives by near-optimal search queries from the bug reports}
	\label{fig:opt-vs-baseline-system}
	\vspace{-.5cm}
\end{figure}

%Both baseline and state-of-the-art approaches perform poorly in locating appropriate search keywords from the bug reports (RQ$_1$). However, we wanted to investigate whether
Once optimal or near-optimal search queries are constructed from each of the bug reports, we compare them with baseline queries to determine their potential in IR-based bug localization.
Tables \ref{table:opt-vs-baseline-br}, \ref{table:optimal-vs-baseline-improved} and Figures \ref{fig:opt-vs-baseline}, \ref{fig:opt-vs-baseline-system} summarize our comparative analyses. From Table \ref{table:opt-vs-baseline-br}, we see that the baseline queries (defined in Section \ref{sec:baseline}) achieve a maximum of 67\% Hit@10 with 43\% mean average precision and a mean reciprocal rank of 0.43. On the contrary, the near-optimal search queries (NrOptimal$_{GA}$) achieve 96\% Hit@10 with 90\% MAP and 0.90 MRR, which are 44\%, 111\% and 109\% higher respectively. These queries also achieve 94\% Hit@5 which is 62\% higher than the baseline counterpart.
Furthermore, 87\% of the bug reports contain optimal queries that can deliver the ground truth results at the topmost position of their result list.
%they achieve 87\% Hit@1 and 94\% Hit@5, which are 173\% and 62\% higher respectively than the baseline measures.
%That is, the near-optimal queries can deliver the ground truth results at the topmost position for 87\% of bug reports and within the Top-5 positions for 94\% of the bug reports. 
Thus, all the above findings clearly suggest that appropriate search keywords indeed exist in the bug report texts and they are highly effective for localizing bugs using Information Retrieval.

Although the above findings are promising, we further investigate whether they hold for all four subsets of bug reports. 
%in how the quality of bug reports or the presence of localization hints (e.g., program elements) in them could affect the near-optimal search queries. 
From Table \ref{table:opt-vs-baseline-br}, we see that search queries from NrOptimal$_{GA}$ achieve 98\% Hit@1 and 100\% Hit@5 when bug reports leading to good baseline queries (HQ$_{H-}$, HQ$_{H+}$) are considered. That is, the GA-based approach can always deliver optimal or near-optimal queries when the baseline queries are already good, which is expected and explainable. However, we also notice an interesting phenomenon for the bug reports leading to poor baseline queries (LQ$_{H-}$, LQ$_{H+}$).
Baseline queries from these bug reports fail to retrieve any ground truth documents within their Top-10 results (i.e., 0.00\% Hit@10). On the contrary, near-optimal queries constructed (by NrOptimal$_{GA}$) from these bug reports achieve 78\%--94\% Hit@10, which is both surprising and interesting. More interestingly, these queries can return the buggy source documents at the topmost position for 
50\%--81\% of cases, which is highly promising.
%localizing the buggy source documents. 
Thus, optimal and near-optimal query keywords are present even in the bug reports that make poor baseline queries (i.e., QE$>$10). Unfortunately, given our experimental findings (\eg\ Tables \ref{table:freq-vs-graph-br-subset}, \ref{table:opt-vs-baseline-br}), the baseline and existing approaches from literature \cite{kevic,saner2017masud,tfidf} are simply not effective enough to identify these keywords accurately.

%where the baseline or state-of-the-art approaches often

%could actually exist in the bug report texts even though they might not contain the explicit localization hints (\eg\ program elements, stack traces).

We also contrast between near-optimal and baseline queries by analysing their top 1 to 10 search results. From Fig. \ref{fig:opt-vs-baseline}, we see that near-optimal queries outperform the best performing baseline (\ie\ Baseline) with a significant margin (\ie\ \emph{p-value}$\le$0.01, 0.82$\le$$\delta$$\le$1.00) for each of four subsets of bug reports. Similar conclusions can also be drawn for their precision and reciprocal rank measures.

%We also contrast between optimal and baseline queries by analysing their top 1 to 10 keywords. Fig. \ref{fig:opt-vs-baseline}-(b) demonstrates that optimal query keywords perform significantly higher (\ie\ \emph{p-value}$\le$0.01, $\delta$=0.42 (medium)) than the baseline keywords.

While the above analyses use Top-10 results only, we further contrast between near-optimal and baseline search queries using all the results retrieved by each query. In particular, we determine how many near-optimal queries improve upon the baseline alternatives. From Table \ref{table:optimal-vs-baseline-improved},
we see that near-optimal queries improve upon the baseline for 67\%--77\% of the queries and preserve 23\%--33\% of the queries, which are highly promising according to existing literature \cite{refoqus,trconfig}. That is, $\approx$72\% of the baseline queries could have been improved either by choosing appropriate keywords from them or by discarding the noisy ones from them. Fig. \ref{fig:opt-vs-baseline-system} further demonstrates how near-optimal queries could improve upon $\approx$72\% of the baseline queries from each of the six subject systems. It should be noted that NrOptimal$_{GA}$ does not worsen any of the baseline queries. We also investigate whether this finding holds for all four subsets of bug reports.
%the quality of a bug report has any impact on improvement and worsening of the baseline queries by NrOptimal$_{GA}$.

\begin{table}[!t]
	\centering
	\caption{Query Improvement \& Worsening against Baseline (Keywords from Bug Report)}\label{table:optimal-vs-baseline-improved}
	\vspace{-.2cm}
	\resizebox{4.8in}{!}{%
		\begin{threeparttable}
			\begin{tabular}{l|c|c|c}
				\hline
				\textbf{Query Pair} & \textbf{Improvement} & \textbf{Worsening} & \textbf{Preserving}\\
				\hline
				\multicolumn{4}{c}{Performance for all Bug Reports (\textbf{2,320})} \\
				\hline
				NrOptimal$_{\text{GA}}$ - Baseline-I & 1,789 (77.11\%) & 0 (0.00\%) & 531 (22.89\%)\\
				\hline
				NrOptimal$_{\text{GA}}$ - Baseline-II & 1,643 (70.82\%) & 0 (0.00\%) & 677 (29.18\%)\\
				\hline
				\textbf{NrOptimal$_{\textbf{GA}}$ - Baseline} & \textbf{1,563 (67.37\%)} & 0 (0.00\%) & \textbf{757 (32.63\%)} \\
				\hline
				\multicolumn{4}{c}{Bug reports with good baseline queries and no localization hints (\textbf{567})}\\
				\hline
				NrOptimal$_{\text{GA}}$ - Baseline-I & 398 (70.19\%) & 0 (0.00\%) & 169 (29.81\%) \\
				\hline
				NrOptimal$_{\text{GA}}$ - Baseline-II & 373 (65.78\%) & 0 (0.00\%) & 194 (34.22\%) \\
				\hline
				\textbf{NrOptimal$_{\text{GA}}$ - Baseline} & \textbf{331 (58.38\%)} & 0 (0.00\%) & \textbf{236 (41.62\%)}\\
				\hline
				\multicolumn{4}{c}{Bug reports with good baseline queries and with localization hints (\textbf{954})}\\
				\hline
				NrOptimal$_{\text{GA}}$ - Baseline-I & 630 (66.04\%) & 0 (0.00\%) & 324 (33.96\%) \\
				\hline
				NrOptimal$_{\text{GA}}$ - Baseline-II & 478 (50.11\%) & 0 (0.00\%) & 476 (49.90\%)\\
				\hline
				\textbf{NrOptimal$_{\text{GA}}$ - Baseline} & \textbf{439 (46.02\%)} & 0 (0.00\%) & \textbf{515 (53.98\%)} \\
				\hline
				\multicolumn{4}{c}{Bug reports with poor baseline queries and no localization hints (\textbf{372})}\\
				\hline
				NrOptimal$_{\text{GA}}$ - Baseline-I & 361 (97.04\%) & 0 (0.00\%) & 11 (2.96\%) \\
				\hline
				NrOptimal$_{\text{GA}}$ - Baseline-II & 368 (98.93\%) & 0 (0.00\%) & 4 (1.08\%)\\
				\hline
				\textbf{NrOptimal$_{\text{GA}}$ - Baseline} & \textbf{369 (99.19\%)} & 0 (0.00\%) & \textbf{3 (0.01\%)} \\
				\hline
				\multicolumn{4}{c}{Bug reports with poor baseline queries and with localization hints (\textbf{427})}\\
				\hline
				NrOptimal$_{\text{GA}}$ - Baseline-I & 400 (93.68\%) & 0 (0.00\%) & 27 (6.32\%) \\
				\hline
				NrOptimal$_{\text{GA}}$ - Baseline-II & 424 (99.30\%) & 0 (0.00\%) & 3 (0.01\%) \\
				\hline
				\textbf{NrOptimal$_{\text{GA}}$ - Baseline} & \textbf{424 (99.30\%)} & 0 (0.00\%) & \textbf{3 (0.01\%)} \\
				\hline
				\multicolumn{4}{c}{Manually selected bug reports (\textbf{175})} \\
				\hline
				NrOptimal$_{\text{GA}}$ - Baseline-I& 137 (78.29\%) & 0 (0.00\%) & 38 (21.71\%) \\
				\hline
				NrOptimal$_{\text{GA}}$ - Baseline-II& 126 (72.00\%) & 0 (0.00\%) & 49 (28.00\%) \\
				\hline
				\textbf{NrOptimal$_{\text{GA}}$ - Baseline} & \textbf{117 (66.86\%)} & 0 (0.00\%) & \textbf{58 (33.14\%)} \\
				\hline
				\multicolumn{4}{c}{Single-release bug reports (\textbf{360})} \\
				\hline
				NrOptimal$_{\text{GA}}$ - Baseline-I& 283 (78.61\%) & 0 (0.00\%) & 77 (21.39\%) \\
				\hline
				NrOptimal$_{\text{GA}}$ - Baseline-II& 261 (72.50\%) & 0 (0.00\%) & 99 (27.50\%) \\
				\hline
				\textbf{NrOptimal$_{\text{GA}}$ - Baseline} & \textbf{248 (68.89\%)} & 0 (0.00\%) & \textbf{112 (31.11\%)} \\
				\hline
			\end{tabular}
		\centering
		\textbf{Emboldened}=Improvement and preserving of baseline search queries by the NrOptimal$_{GA}$ approach
		\end{threeparttable}
	}
	\vspace{-.3cm}
\end{table}

\begin{table}[!t]
	\centering
	\caption{Impact of Candidate Query Length and Fitness Function on Near-Optimal Queries (Using Top-10 Results Only)}
	\label{table:ga-config}
	\vspace{-.2cm}
	\resizebox{4.8in}{!}{%
		\begin{threeparttable}
			\begin{tabular}{l|c|c|c|c|c|c}
				\hline
				\textbf{Search Query} & & \textbf{Hit@1} & \textbf{Hit@5} & \textbf{Hit@10} & \textbf{MRR} & \textbf{MAP}\\
				\hline
				\hline
				\multicolumn{7}{c}{\textbf{Impact of Candidate Search Query Length}} \\
				\hline
				NrOptimal$\mathbf{_{GA}}$ & 5-Keywords & 85.93\% & 93.74\% & 95.65\% & 0.89 & 89.16\% \\
				\hline
				NrOptimal$\mathbf{_{GA}}$ & 10-Keywords & 86.89\% & 93.73\% & 95.34\% & 0.90 & 89.70\% \\
				\hline
				NrOptimal$\mathbf{_{GA}}$ & 15-Keywords & 87.26\% & 93.69\% & 95.34\% & 0.90 & 89.57\% \\
				\hline
				NrOptimal$\mathbf{_{GA}}$ & 20-Keywords & 86.91\% & 93.65\% & 95.20\% & 0.90 & 89.43\% \\
				\hline
				\multicolumn{7}{c}{\textbf{Impact of Fitness Function}} \\
				\hline
				NrOptimal$\mathbf{_{GA}}$ & QE & 86.89\% & 93.73\% & 95.34\% & 0.90 & 89.70\% \\
				\hline
				NrOptimal$\mathbf{_{GA}}$ & MAP & 86.83\% & 93.49\% & 95.12\% & 0.90 & 91.06\% \\
				\hline
			\end{tabular}
		\end{threeparttable}
	}
\vspace{-.3cm}
\end{table}

\begin{table}[!t]
	\centering
	\caption{State-of-the-Art vs. Near-Optimal Search Queries from Bug Reports (Using Top-10 Results Only)}\label{table:sota-vs-optimal}
	\vspace{-.2cm}
	\resizebox{4.8in}{!}{%
		\begin{threeparttable}
			\begin{tabular}{l|c|c|c|c|c|c}
				\hline
				\textbf{Technique} & \textbf{Genre} & \textbf{Hit@1} & \textbf{Hit@5} & \textbf{Hit@10} & \textbf{MRR} & \textbf{MAP} \\
				\hline
				\hline
				\multicolumn{7}{c}{Bug reports with good baseline queries and no localization hints (\textbf{567}) \textbf{(HQ$_{H-}$)}}\\
				\hline
				Baseline & -- & 41.96\% & 86.18\% & 100.00\% & 0.60 & 58.15\% \\
				\hline
				STRICT & Graph  & 36.14\% & 72.55\% & 84.83\% & 0.51 & 50.29\% \\
				\hline
				ACER & Graph & 42.55\% & 85.62\% & 94.87\% & 0.61 & 58.90\% \\
				\hline
				\textbf{NrOptimal$\mathbf{_{GA}}$} & GA & 95.93\% & 100.00\% & 100.00\% & 0.98 & 93.51\% \\
				\hline
				\multicolumn{7}{c}{Bug reports with good baseline queries and with localization hints (\textbf{954}) \textbf{(HQ$_{H+}$)}}\\
				\hline
				Baseline & -- & 50.01\% & 86.75\% & 100.00\% & 0.66 & 66.63\% \\
				\hline
				TF-IDF & Frequency & 41.59\% & 70.80\% & 83.55\% & 0.54 & 55.15\% \\
				\hline
				ACER  & Graph & 50.20\% & 88.95\% & 97.43\% & 0.66 & 67.06\% \\
				\hline
				\textbf{NrOptimal$\mathbf{_{GA}}$} & GA & 98.04\% & 99.91\% & 100.00\% & 0.99 & 100.00\% \\
				\hline
				\multicolumn{7}{c}{Bug reports with poor baseline queries and no localization hints (\textbf{372}) \textbf{(LQ$_{H-}$)}}\\
				\hline
				Baseline & -- & 0.00\% & 0.00\% & 0.00\% & 0.00 & 0.00\% \\
				\hline
				STRICT & Graph & 1.56\% & 9.54\% & 16.89\% & 0.05 & 5.12\% \\
				\hline
				\textbf{NrOptimal$\mathbf{_{GA}}$} & GA & 50.04\% & 69.68\% & 77.96\% & 0.58 & 56.47\% \\
				\hline
				\multicolumn{7}{c}{Bug reports with poor baseline queries and with localization hints (\textbf{427}) \textbf{(LQ$_{H+}$)}}\\
				\hline
				Baseline & -- & 0.00\% & 0.00\% & 0.00\% & 0.00 & 0.00\% \\
				\hline
				STRICT & Graph & 4.91\% & 18.86\% & 25.77\% & 0.11 & 10.66\% \\
				\hline
				\textbf{NrOptimal$\mathbf{_{GA}}$} & GA & 80.70\% & 91.00\% & 93.37\% &
				0.85 & 86.19\%\\
				\hline
			\end{tabular}
		\end{threeparttable}
	}
	\vspace{-.5cm}
\end{table}

From Table \ref{table:optimal-vs-baseline-improved}, we see that near-optimal queries are better than 46\%--58\% of 1,521 (567 + 954) baseline queries that were constructed from the bug reports leading to good baseline queries (HQ$_{H-}$, HQ$_{H+}$). On the contrary, NrOptimal$_{GA}$ can deliver better queries than 99\% of 799 (372 + 427) baseline queries that were constructed from the bug reports leading to poor baseline queries (LQ$_{H-}$, LQ$_{H+}$). 
We also repeat our experiments using 175 manually selected and 360 single-release bug reports (see details in Table \ref{table:optimal-vs-baseline-improved}), and our findings remain aligned with the above (e.g., $\approx$70\% query improvement).
All these empirical findings above clearly suggest that optimal and near-optimal queries exist in bug report texts and they are a far better alternative than the baselines.

We also investigate the impact of two important parameters --\emph{candidate query length} and \emph{fitness function} -- upon the GA-based algorithm (NrOptimal$_{GA}$) used in our study. Table \ref{table:ga-config} summarizes our investigation details. In NrOptimal$_{GA}$, we initialize our solution candidates (chromosomes) with four different query lengths (e.g., 5, 10, 15, 20) and collect the optimal and near-optimal search queries generated from them. However, we did not find any significant difference in the performance of search queries generated from these four query lengths. According to our analysis, although the solution candidates started with many keywords (e.g., 20), the algorithm, NrOptimal$_{GA}$, eventually found out a subset of these keywords with the maximum fitness (e.g., QE = 1) through evolution and removal of duplicate keywords. However, construction of optimal and near-optimal queries from the long candidate queries is more time-consuming than with the short ones. We also conduct experiments using two different fitness functions -- Query Effectiveness (QE) and Mean Average Precision (MAP) (defined in Section \ref{sec:pmetrics}), and collect the optimal and near-optimal queries for them. We did not find any significant difference in the performance of search queries generated using these two fitness functions (Table \ref{table:ga-config}). However, we found that MAP-based query construction is more time-consuming than QE-based one.

\subsection{Comparison between Optimal or Near-Optimal Search Query and State-of-the-art Search Query}
We compare optimal and near-optimal search queries with the queries from existing approaches \cite{saner2017masud,ase2017masud,tfidf}.
%queries constructed from four subsets of bug reports. 
From Table \ref{table:sota-vs-optimal}, we see that both GA-based approach and existing approaches were able to provide good search queries (e.g., 84\%--97\% Hit@10) from the bug reports that already make good baseline queries (i.e., HQ$_{H-}$ and HQ$_{H+}$). 
%both near-optimal and state-of-the-art queries perform satisfactorily ( when the bug reports contain high-quality textual contents 
However, the best-performing existing approach (STRICT \cite{saner2017masud} according to Table \ref{table:freq-vs-graph-br-subset}) often struggles to provide appropriate queries from the bug reports that make poor baseline queries (i.e., LQ$_{H-}$ and LQ$_{H+}$).
%contain low-quality contents (i.e., LQ$_{H-}$ and LQ$_{H+}$). 
It should be noted that bug reports could make poor baseline queries despite having 
%be of low-quality as search queries despite having 
explicit hints for bug localization (e.g., stack traces). 
From Table \ref{table:sota-vs-optimal}, we see that the queries from STRICT, the state-of-the-art technique, achieve 17\%--26\% Hit@10 for 799 (372 + 427) bug reports with poor baseline.
On the contrary, search queries from the GA-based approach (NrOptimal$_{GA}$) for the same dataset achieve 78\%--93\% Hit@10, which are three to five times higher. Similar conclusions can also be drawn for other performance metrics (e.g., MAP, MRR).
Furthermore, this approach was able to deliver optimal search queries from 50\%--81\% of the bug reports containing poor baseline queries (\ie\ 50\%--81\% Hit@1).

According to our investigation, about 65.56\% (567 HQ$_H-$ + 954 HQ$_H+$) bug reports inherently contain more useful keywords for localizing the bugs than others (372 LQ$_H-$ + 427 LQ$_H+$). We annotate these bug reports as the bug reports leading to good baseline queries and the bug reports leading to poor baseline queries. Besides, different existing approaches have their own strengths and weaknesses. Given these two dimensions, we see significant differences in techniques’ performance across the bug report clusters (Table \ref{table:freq-vs-graph-br-subset}).

%extract near-optimal queries from the same set of bug reports. More interestingly, these queries achieve 78\%--93\% Hit@10, which are three to five times higher than the state-of-the-art Hit@10. 

One can argue about the novelty of our findings in RQ$_2$ compared to that of \citet{cmills-icsme2018}. However, we not only strengthen the earlier finding but also  make further contributions in this work. First, we conduct a more extensive experiment using 2,320 bug reports (as opposed to 620 by \citeauthor{cmills-icsme2018}) that are categorized into four different subsets and demonstrate that optimal search queries are present even in the bug reports that make poor baseline queries.
%low-quality, natural language-only bug reports. 
Second, we show how effective the optimal and near-optimal search queries are compared to the baseline and state-of-the-art queries. Third, we demonstrate the impact of candidate query length and fitness function on the optimal or near-optimal search queries constructed by the genetic algorithm. Fourth, we also provide a detailed pseudo-code (Algorithms \ref{algo}, \ref{algo2}) and an example workflow of the GA-based query construction, which were not provided by \citet{cmills-icsme2018}.

%However, the evolution of long chromosomes is costlier than those of short chromosome contain
%we did not find any significant difference among these different candidate query lengths in terms of the performance of their

%\vspace{-.2cm}
\begin{frshaded}
	\noindent
	\textbf{Summary of RQ$\mathbf{_2}$:} Bug reports could lead to poor baseline queries or could miss the explicit hints for localizing bugs (\eg\ program elements, stack traces) in their texts. However, they generally \textbf{contain sufficient keywords} to construct optimal or near-optimal search queries, as demonstrated by the GA-based approach. These queries can deliver optimal performance in bug localization for \textbf{50}\%--\textbf{81}\% of bug reports. They also \textbf{outperform} the baseline and state-of-the-art search queries in all \emph{four} performance measures (e.g., Hit@K, MAP, MRR, QE) by a \textbf{large margin}.
\end{frshaded}

\section{Answering RQ$_3$: Optimal vs. Non-Optimal Search Queries for IR-Based Bug Localization}
\label{sec:rq3}
According to RQ$_1$, state-of-the-art approaches \cite{saner2017masud,kevic} might not be able to deliver appropriate search queries from many bug reports despite their texts containing explicit hints for bug localization (e.g., program elements) (Table \ref{table:freq-vs-graph-br-subset}).
However, these bug reports actually contain optimal and near-optimal search queries in their texts as shown in RQ$_2$.
Table \ref{table:sota-vs-optimal} also demonstrates that near-optimal search queries from these bug reports perform at least \emph{three} to \emph{five} times higher than the state-of-the-art queries (STRICT \cite{saner2017masud}) in terms of Hit@10 and eight to ten times higher in terms of MAP and MRR. All these findings above beg an obvious question -- \emph{``What are the differences between optimal and non-optimal search queries from a bug report?"} A comprehensive understanding of these differences is essential since they can be leveraged to improve query reformulation approaches.
%We thus further investigate how near-optimal search queries are different from the non-optimal queries.
%the search queries suggested by the state-of-the-art \cite{saner2017masud}.
We thus conduct a multi-modal analysis involving machine learning, data mining and manual inspection to answer this important question. Tables \ref{table:sota-vs-optimal}, \ref{table:opt-study-dataset}, \ref{table:model-performance} and Figures \ref{fig:feature-importance},  \ref{fig:feature-distribution}, \ref{fig:boxplot-feature-distribution} summarize our analysis details.

%In particular, we conduct a detailed comparative analysis (Table \ref{table:sota-vs-optimal-dimension}) between them using 12 widely used query attributes across four different dimensions as follows.

\subsection{Optimal, Near-Optimal and Non-Optimal Search Queries} \label{sec:optimal-dataset}
Since we wanted to differentiate among optimal, near-optimal and non-optimal search queries, we construct a dataset of 13,914 queries from 2,320 bug reports.
Table \ref{table:opt-study-dataset} shows detailed dataset for our comparative analysis.
We first collect queries from five existing approaches (TF, IDF, TF-IDF \cite{tfidf}, \citet{kevic} and STRICT \cite{saner2017masud}) and the GA-based approach, NrOptimal$_{GA}$, and then carefully categorize them into optimal, near-optimal and non-optimal queries based on their effectiveness (Table \ref{table:nr-query-def}). If a search query returns any ground truth result at the topmost position of its result list (i.e., QE=1), we consider it as an \emph{optimal query}. If the query returns the ground truth between the 2$^{nd}$ and 10$^{th}$ position, we consider it as a \emph{near-optimal query}.
Developers often check only top 10 results before making another search with a reformulated query. Existing studies \cite{buglocator,saha,refoqus} often also analyze top 10 results for a query, which justifies our definition of near-optimal query. On the other hand, if the search query returns the ground truth result below the 10$^{th}$ position, it is considered as a \emph{non-optimal query}. 
%and otherwise as a non-optimal query. Although this restriction works well with the queries from high-quality bug reports (HQ$_{H-}$, HQ$_{H+}$), it leaves us with a very small number of near-optimal queries from low-quality bug reports (e.g., 619). We thus relax QE-based restriction for the queries from low-quality bug reports, consider each query with 1$\le$QE$\le$10 as near-optimal and otherwise as non-optimal query. 
Thus, as shown in Table \ref{table:opt-study-dataset}, we end up with 
4,893 optimal, 3,857 near-optimal, and 5,164 non-optimal search queries in our dataset.

%four sets of queries (Q-HQ$_{H-}$, Q-HQ$_{H+}$, Q-LQ, Q-All) where each set contains both near-optimal and non-optimal search queries.

\subsection{Search Query Characteristics}\label{sec:q-chars}
We select a total of 31 traditional metrics (Table \ref{table:query-features}) from five different dimensions to characterize each of the search queries. Our goal was to differentiate among optimal, near-optimal and non-optimal search queries using these metrics and then derive meaningful insights to improve the non-optimal queries.
In particular, we choose the following five dimensions in our comparative analysis.

\textbf{Frequency} has been a popular proxy of keyword importance for decades \cite{tfidf}.
It had been regularly used by the Information Retrieval community for the last 50 years.
The underlying idea is that if a word frequently occurs within a document, it could serve as a potential query keyword for retrieving the document from a corpus.
Since our optimal and non-optimal queries differ significantly in their performance (e.g., Query Effectiveness), it would be interesting to see whether they also differ significantly in terms of their frequency-based statistics. We thus use three frequency-based metrics (e.g., Term Frequency (TF), Inverse Document Frequency (IDF), Term Frequency $\times$ Inverse Document Frequency (TF-IDF)) in our comparative analysis. It should be noted that these metrics are often adopted by relevant literature \cite{refoqus,qperf-mills-tosem}.
As shown in the Table \ref{table:query-features}, we capture four descriptive statistics (e.g., average, median, maximum, standard deviation) of each metric from each query where TF, IDF and TF-IDF are calculated for each query keyword. In short, we collect 12 statistics to characterize each search query using their frequency-related properties.

%TF, IDF and TF-IDF, and contrast between state-of-the-art \cite{saner2017masud} and optimal queries using frequency statistics.
%From Table \ref{table:sota-vs-optimal-dimension}, we see that keywords from the optimal queries (a.k.a., optimal keywords) are less frequent than that of the state-of-the-art within a bug report. More interestingly, they are also less frequent across the project's codebase, which is demonstrated by their higher IDF measures. Finally, we see that optimal keywords have less TF-IDF (\ie\ TF$\times$ IDF) than the keywords from STRICT. Thus, frequency statistics might not be able to explain the differences between optimal and state-of-the-art queries.

\begin{table}[!t]
	\centering
	\caption{Optimal, Near-Optimal \& Non-Optimal Search Queries}\label{table:opt-study-dataset}
	\vspace{-.2cm}
	\resizebox{4.8in}{!}{%
		\begin{threeparttable}
			\begin{tabular}{l|l|c|c|c|c}
				\hline
				\textbf{Techniques} & \textbf{QS} & \textbf{Optimal} & \textbf{Near-Optimal} & \textbf{Non-Optimal} & \textbf{TQ} \\
				\hline
				\hline
				TF, IDF, TF-IDF & Q-HQ$_{H-}$ & 1,415 & 1,313 & 674 & 3,402 \\
				\hhline{~-----}
				\citeauthor{kevic}, & Q-HQ$_{H+}$ & 2,859 & 1,885 & 980 & 5,724 \\
				\hhline{~-----}
				STRICT, & Q-LQ$_{H-}$ & 212 & 277 & 1,740 & 2,229 \\
				\hhline{~-----}
				NrOptimal$_{GA}$ & Q-LQ$_{H+}$ & 407 & 382 & 1,770 & 2,559 \\
			    \hhline{~-----}
				 & \textbf{Q-All} & \textbf{4,893} & \textbf{3,857} & \textbf{5,164} & \textbf{13,914} \\
				\hline
			\end{tabular}
			\centering
			\textbf{QS} = Query set, \textbf{HQ}$_{H-}$ = Bug reports without hints leading to good baseline queries, \textbf{HQ}$_{H+}$ = Bug reports with hints leading to good baseline queries, \textbf{LQ}$_{H-}$ = Bug reports without hints leading to poor baseline queries, \textbf{LQ}$_{H+}$ = Bug reports with hints leading to poor baseline queries, \textbf{QE}=Rank of the first buggy document within the result list, \textbf{TQ} = Total queries.
		\end{threeparttable}
	}
\vspace{-.3cm}
\end{table}

\textbf{Entropy} has been a useful proxy for determining ambiguity or specificity of a given textual entity \cite{carmel,qsurvey}. It has been inspired by the concept of entropy from Information Theory domain \cite{shannon-entropy}.
That is, if a query contains keywords with high entropies (or ambiguities), the query could be hard to answer and thus might fail to retrieve any relevant documents. We were interested to see whether optimal queries are more specific (i.e., less ambiguous)  than non-optimal queries. Such an insight could be helpful for selecting appropriate keywords from a textual entity (e.g., bug report). Thus, we use three entropy-based metrics from the existing literature -- Term Entropy \cite{msrch2015masud,qperf-mills-tosem}, Jensen Shannon Divergence (JSD) \cite{carmel}, and Query Specificity Index (QSI) \cite{specificity}-- in our investigation. The entropy dimension provides six statistics for each query.

\textbf{Mutual Information} is another probabilistic metric that could be useful to differentiate between optimal and non-optimal search queries. It approximates the semantic dependency between two words based on their co-occurrences across multiple documents. That is, presence of keywords that share a lot of mutual information indicates the cohesiveness of a search query. We were interested to see whether the optimal queries are more \emph{cohesive} than the non-optimal ones. We thus calculate Point-wise Mutual Information (PMI) for each of the keyword-pairs from a search query. Then we capture four descriptive statistics of this probabilistic metric and use them in our comparative analysis.

\textbf{Part of Speech (POS)} has been a popular and widely used heuristic for extracting keywords or phrases from a body of texts (\eg\ bug report, Q\&A threads) \cite{rack,saner2017masud,time-aware-term-weighting,autocomment}. In particular, nouns and verbs are preferred to other POS as keywords since they are intuitive and convey rich semantics. We were interested to see whether optimal and non-optimal queries differ from each other in the distribution of their noun or verb keywords. We thus consider four POS-based metrics (e.g., \emph{nounRatio}, \emph{verbRatio}, \emph{nounVerbRatio} and \emph{otherPOSRatio}) for our comparison.

%From Table \ref{table:sota-vs-optimal-dimension}, we see that optimal queries comprise of more verbs than STRICT queries. Earlier investigation \cite{shepherd} also suggest the importance of verbs as search keywords (\eg\ Verb-DO). Besides, the method signatures often contain both verbs and nouns, which might have helped the optimal queries perform higher.

\textbf{Keyword Position} could be another important dimension to compare
optimal queries with non-optimal queries. 
Each bug report has two textual fields -- \emph{title} and \emph{description}.
We were interested to see how the keywords from optimal and non-optimal queries are distributed across these two fields. 
%that form either an optimal, a near-optimal or a non-optimal query. 
In particular, we wanted to see whether the optimal keywords are extracted 
from only title, only description or from both fields in a bug report.
%from title only, description only or from both fields of a report. 
We thus use three position-related metrics (e.g., \emph{titleKeywordRatio}, \emph{bodyKeywordRatio}, \emph{titleBodyKWRatio}) in our comparative analysis 
to differentiate between optimal and non-optimal queries.

We also select two ad-hoc metrics (e.g., \emph{groundTruthTermRatio}, \emph{uniqueKeywords}) through a reverse engineering process. We wanted to see whether the keywords from optimal queries match with the ones collected from the ground truth file names. The idea is that if there exists a strong connection between
these two keyword lists (i.e., high \emph{groundTruthTermRatio}), one can focus on identifying the potential solution documents to make a better search query.
%query construction algorithms could emphasize on potential solution files to construct better search queries. 
We also consider the number of unique keywords in each query and attempt to understand 
the difference between optimal and non-optimal queries in this aspect.

Several earlier studies \cite{refoqus,ase2017masud,qperf-mills-tosem} employ dozens of pre-retrieval and post-retrieval query difficulty metrics to classify the \emph{difficult} and \emph{easy} queries. Although several of our adopted metrics overlap with theirs, we avoid several of them (e.g., Spatial Auto-Correlation, Query Scope). Our goals are to demonstrate how optimal queries are different from the non-optimal queries in terms of simple, intuitive, popular measures and also to derive meaningful insights in the process. Unfortunately, those metrics are computationally expensive, complex, and less intuitive, which makes them less than ideal for our comparative analysis.

\begin{table}[!t]
	\centering
	\caption{Performance of Query Classification Models}\label{table:model-performance}
	\vspace{-.2cm}
	\resizebox{4.8in}{!}{%
		\begin{threeparttable}
			\begin{tabular}{l|c|c|c|c|c|c}
				\hline
				\multirow{2}{*}{\textbf{Model}} & \multirow{2}{*}{\textbf{\#Queries}} & \textbf{Precision} & \textbf{Recall} & \textbf{Precision} & \textbf{Recall} &
				\multirow{2}{*}{\textbf{Accuracy}} \\
				\hhline{~~----~}
				& & \multicolumn{2}{c|}{\textbf{Optimal}}
				& \multicolumn{2}{c|}{\textbf{Non-Optimal}}  &  \\
				\hline
				\hline
				RF-HQ & 5,928 & 87.30\% & 89.40\% & 70.70\% & 66.40\% & 82.96\% \\
				\hline
				RF-LQ & 4,129  & 83.30\% & 81.40\% & 96.70\% & 97.10\% & \textbf{94.77}\% \\
				\hline
				RF-All & 10,057 & 84.60\% & 79.80\% & 81.90\% & 86.20\% &  \textbf{83.12}\% \\
				\hline
				& & \multicolumn{2}{c|}{\textbf{Near-Optimal}} & \multicolumn{2}{c}{\textbf{Non-Optimal}} & \\
				\hline
				RF-HQ & 4,852 & 80.60\% & 68.40\% & 52.80\% & 68.30\% & 68.36\% \\
				\hline
				RF-LQ & 4,169 & 53.90\% & 43.60\% & 89.80\% & 93.00\% & 85.20\% \\
				\hline
				RF-All & 9,021 & 63.80\% & 70.70\% & 76.20\% & 70.00\% & 70.29\% \\
				\hline
			\end{tabular}
			\centering
			\textbf{RF} = Random Forest, \textbf{RF-HQ} = Random Forest model based on bug reports leading to good baseline queries, \textbf{RF-LQ} = Random Forest model based on bug reports leading to poor baseline queries.
		\end{threeparttable}
	}
	\vspace{-.3cm}
\end{table}

\subsection{Comparison between Optimal and Non-Optimal Search Queries using Machine Learning-Based Feature Importance Analysis} \label{sec:feature-importance}

Optimal and non-optimal search queries might not be linearly separable across all characteristics discussed above. 
We thus attempt to separate them using a non-linear approach -- machine learning algorithm. We use supervised machine learning to classify the search queries, identify such features that the classification algorithms found as \emph{important}, and then analyze the distinctive characteristics of optimal and non-optimal queries as follows.

First, we make use of optimal (QE=1), near-optimal (2$\le$QE$\le$10), and non-optimal search queries (QE$>$10) (Table \ref{table:opt-study-dataset}), calculate their metrics, and then train \emph{six} classification models from them (Table \ref{table:model-performance}). Since the queries were taken from various subsets of bug reports, the trained models were named accordingly.
Given the imbalanced set of queries (Table \ref{table:opt-study-dataset}), these models might suffer from over-fitting issues. We thus use RandomForest algorithm that was reported as robust against model over-fitting issues \cite{msr2017amasud,qperf-mills-tosem}, SMOTE-based oversampling \cite{smote} and 10-fold cross validation to mitigate the data-imbalance issue. Table \ref{table:model-performance} shows the performance of our trained models.
We see that these models were able to classify the optimal and non-optimal queries with 83\% to 95\% accuracy, which is promising. While optimal and non-optimal queries differ significantly in terms of their retrieval performance (e.g., query effectiveness (QE)), this high classification accuracy of the models indicates that these queries can also be different in terms of their lexical or statistical characteristics (Section \ref{sec:q-chars}).
That means, meaningful insights could be derived from these characteristics to improve poor or non-optimal search queries. However, as shown in the Table \ref{table:model-performance}, near-optimal and non-optimal queries cannot be classified with a high accuracy. That is, they might not be very different in terms of their lexical or statistical characteristics.   

%Table \ref{table:model-performance} shows the performance of our models trained with 31 numerical features (Table \ref{table:query-features}) and one class label (e.g., optimal, near-optimal, non-optimal). 
%We construct four machine learning models based on the four query collections (Q-HQ$_{H-}$, Q-HQ$_{H+}$, Q-LQ and Q-All, Table \ref{table:opt-study-dataset}) where each model is trained with 31 numerical features (Table \ref{table:query-features}) and one class label (e.g., near-optimal, non-optimal). 
%Since the query collections were imbalanced, we also use . Table \ref{table:model-performance} shows the performance details of each model in classifying the near-optimal and non-optimal search queries. Overall, our models achieve 73\%--83\% classification accuracy, which is promising.

\begin{figure}[!t]
	\centering
	\includegraphics[width=4.8in]{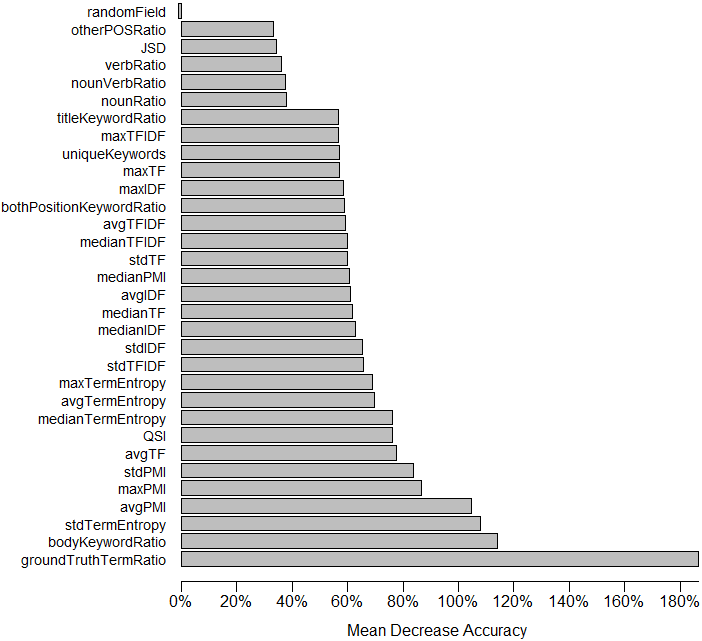}
	\vspace{-.2cm}
	\caption{Feature importance in the RandomForest model (RF-All)}
	\label{fig:feature-importance}
	\vspace{-.4cm}
\end{figure}

\begin{figure}[!t]
	\centering
	\includegraphics[width=4.8in]{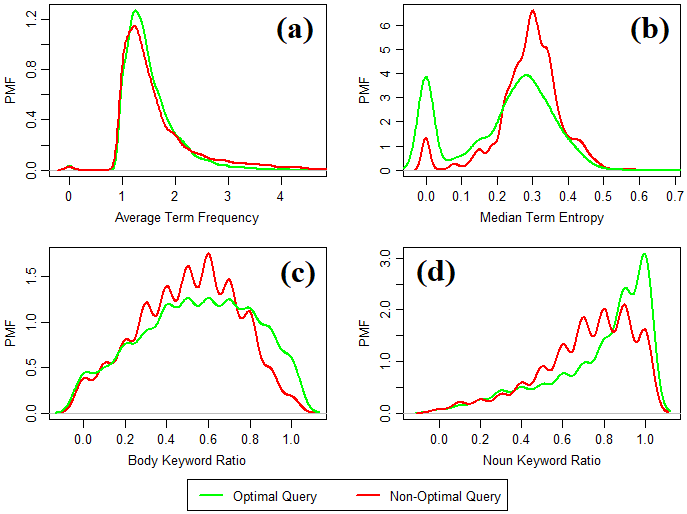}
	\vspace{-.3cm}
	\caption{Comparison between optimal and non-optimal keywords (from all bug reports) using their distributions (e.g., PMF=probability mass function) of (a) frequency, (b) entropy, (c) percentage from the body/description section of a bug report, and (d) percentage of nouns}
	\label{fig:feature-distribution}
	\vspace{-.3cm}
\end{figure}

Second, although non-linear, machine learning approach was able to separate optimal queries from non-optimal ones, the underlying decision trees could be difficult to interpret due to their sheer sizes. We thus rely on feature importance analysis and identify such features that the training algorithms found more important than others in separating the optimal queries from the non-optimal ones. Fig. \ref{fig:feature-importance} shows the relative importance of all features in terms of mean decrease accuracy, which indicates how much accuracy a model might lose if a certain feature is discarded from the model.
We select the top 10 most important features including \emph{groundTruthTermRatio}, \emph{bodyKeywordRatio}, \emph{stdTermEntropy}, \emph{avgPMI} and \emph{avgTF}, collect their corresponding numerical values, and then compare their distribution for optimal and non-optimal search queries. Then we shortlist four features
for which the optimal and non-optimal queries demonstrate \emph{significant differences} (i.e., \emph{p-value}$\le$0.05) in terms of  non-parametric statistical tests (e.g., Mann-Whitney Wilcoxon, cliff's delta).
%added for revision.
In particular, we notice significant differences in the \emph{distribution} of their term frequency, median term entropy, body keyword ratio, and noun keyword ratio (check Figures \ref{fig:feature-distribution}, \ref{fig:boxplot-feature-distribution}, \ref{fig:feature-distribution-lq}, \ref{fig:boxplot-feature-distribution-lq} for details).
It should be noted that \emph{randomField}, a randomly generated feature, has been found as the least important feature for classification, which is expected and thus instills confidence in our feature importance estimation.
Fig. \ref{fig:feature-distribution} shows the probability distribution of the four features -- average term frequency, median term entropy, percentage of keywords from body section, and percentage of noun keywords as follows.

%\begin{figure}[!t]
%	\centering
%	\includegraphics[width=4.8in]{reqs/rq3/hq-nl-mda}
%	\vspace{-.3cm}
%	\caption{Feature importance in the RandomForest model (RF-HQ$_H-$) based on the search queries constructed from bug reports with good baseline queries and no localization hints (Q-HQ$_{H-}$)}
%	\label{fig:hq-nl-mda}
%	%\vspace{-.2cm}
%\end{figure}

%Each of the RandomForest models above learns the relative importance (e.g., Mean Decrease Accuracy, Mean Decrease Gini) of their features as a part of the training. 
%We thus capture the feature importance plot from each of our models and examine them carefully. Fig. \ref{fig:all-dataset-mda} shows the Top-10 important features for identifying a near-optimal query where the model is trained on the whole collection of 13,914 queries (Q-All). We see that eight traditional features related to \emph{frequency} and \emph{entropy} do not help much. They are more important for identifying a non-optimal search query.

From Fig. \ref{fig:feature-distribution}-(a), we see that the frequency of keywords in both optimal and non-optimal search queries has a lognormal distribution. 
Both queries have a high probability mass on a lower frequency (e.g., 1).
However, non-optimal queries contain slightly more frequent keywords, especially the keywords that occur more than twice in a bug report. Box plots in Fig. \ref{fig:boxplot-feature-distribution}-(a) also show a higher variance in keyword frequency for non-optimal search queries. We also repeated this experiment using bug reports that lead to poor baseline queries (LQ$_{H-}$, LQ$_{H+}$) and found that median frequency of optimal keywords is less than that of non-optimal keywords (Fig. \ref{fig:boxplot-feature-distribution-lq}-(a)). Traditionally, frequency has been considered as a popular proxy of keyword importance, which does not apply well to optimal search queries. One can argue about using TF-IDF as a proxy of keyword importance. However, according our investigation, non-optimal queries also contain keywords with significantly higher TF-IDF measures. From Fig. \ref{fig:feature-distribution}-(b), we also see that entropy measures of keywords from both queries have a mixed distribution. However, non-optimal search queries contain more keywords of high entropy. That is, optimal search queries contain less ambiguous keywords than non-optimal queries do, which aligns to conventional wisdom \cite{carmel,qperf}. Our statistical tests (e.g., Mann-Whitney Wilcoxon) also report a p-value of $<$0.001 with a \emph{small} effect size (\ie\ $\delta$=0.30). Furthermore, this finding is confirmed by the box plots in Fig. \ref{fig:boxplot-feature-distribution}-(b) where median entropy of optimal queries is lower (i.e., less ambiguous) than that of non-optimal queries. We also calculate the percentage of keywords in queries that were extracted from the \emph{description} section (or body) of a bug report, and Fig. \ref{fig:feature-distribution}-(c) compares their distribution between two query types.
We see that optimal queries are more likely to contain keywords from the description section than non-optimal ones. Although our statistical test reported a significant \emph{p-value}, the effect size is negligible. Finally, Fig. \ref{fig:feature-distribution}-(d) shows the distribution of noun keywords in both optimal and non-optimal queries. While both distributions are irregular, we see that optimal queries are more likely to contain 100\% noun keywords. This observation is further strengthened by the box plot analysis in Fig. \ref{fig:boxplot-feature-distribution}-(d). We see that optimal queries are more likely to contain nouns as keywords. Our statistical tests also report a \emph{p-value} of $\le$0.001 with a small effect size (i.e., 0.20), which strengthens this finding. 
We also repeated each of these analysis above (e.g., Figures \ref{fig:feature-distribution}, \ref{fig:boxplot-feature-distribution}) using optimal and non-optimal queries from the bug reports that lead to poor baseline queries (LQ$_{H-}$, LQ$_{H+}$). We were able to reproduce each finding as demonstrated in the Figures \ref{fig:feature-distribution-lq}, \ref{fig:boxplot-feature-distribution-lq}, which further strengthens our findings on the four query features.

We also perform regression analysis between the four query features discussed above and query performance. Table \ref{table:logistic} summarizes our analysis using Logistic regression. We see that all four features are found statistically significant, which provides confidence in their coefficients. That is, for every unit increase in \emph{nounRatio} feature, the odd of being a query to be \emph{optimal} (i.e., QE=1) increases by 1.81. Similarly, every unit increase in \emph{bodyKeywordRatio} increases the same odd ratio by 1.25. In short, nouns and keywords from body section are likely to improve the performance of a search query. On the other hand, every unit increase in \emph{medianTF} and \emph{medianTermEntropy} features changes the odd of being a query to be optimal by only 0.01 and 0.75 respectively. That is, frequent, ambiguous keywords are not likely to improve a search query. Thus, a Logistic regression-based analysis also leads us to the same conclusion regarding the four query features.

\subsection{Actionable Insights} \label{sec:actionable}
\begin{figure}[!t]
	\centering
	\includegraphics[width=4.8in]{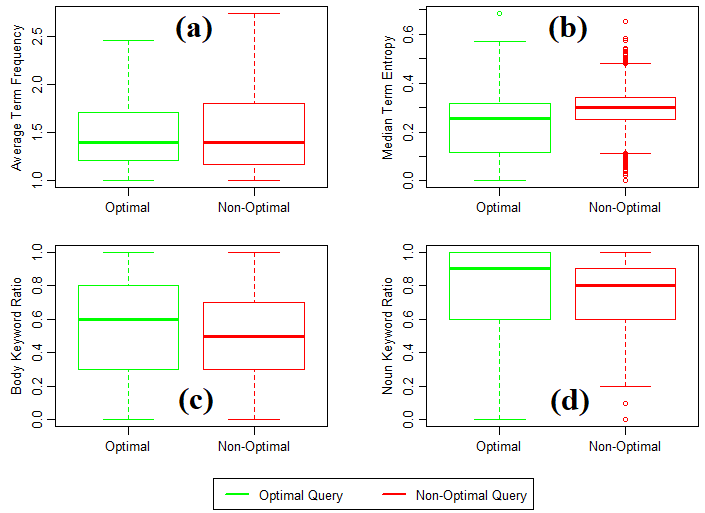}
	\vspace{-.3cm}
	\caption{Comparison between optimal and non-optimal keywords (from all bug reports) using their box plot of (a) frequency, (b) entropy, (c) percentage from the body/description section of a bug report, and (d) percentage of nouns}
	\label{fig:boxplot-feature-distribution}
	%\vspace{-.2cm}
\end{figure}

The above comparative analysis (Section \ref{sec:feature-importance}, Figures \ref{fig:feature-distribution}, \ref{fig:boxplot-feature-distribution}) has reported several interesting findings, which can be turned into actionable insights. Table \ref{table:actionale-insights} shows four actionable insights regarding frequency, ambiguity, position and part of speech of query keywords as follows. 

\textbf{I$_1$:} Frequency might not be a reliable proxy of keyword importance and highly frequent keywords should not be included in a query. This insight contradicts the conventional wisdom from existing literature \cite{kevic,refoqus}. To implement this insight, we determine \emph{median} frequency of keywords in a bug report and select the ones with less than median frequency for query expansion.

\textbf{I$_2$:} Term entropy is an important proxy of keyword importance and the keywords with less entropy, i.e., less ambiguous, should be included in a query. This insight is mostly aligned to earlier literature on query difficulty \cite{carmel,qperf-mills-tosem,qperf}. To implement this, we determine \emph{median} entropy of keywords in a bug report and then select the ones with less than median term entropy for query expansion.
 
\textbf{I$_3$}: Keywords from the \emph{description} section are more prevalent in optimal queries and thus they should be prioritized over others during query formulation.
We determine the position of each keyword within a bug report and select the keywords that are only found in the description section for expanding a query.

\textbf{I$_4$:} Optimal queries contain more nouns than non-optimal queries and thus, noun keywords should be prioritized during query formulation. We determine the part of speech of each keyword in a bug report using Stanford POS tagger \cite{postagger} and select the nouns for expanding a given query.

From Fig. \ref{fig:feature-importance}, we also notice several ad hoc features (e.g., \emph{uniqueKeywords}, \emph{ground-TruthTermRatio}) that were found important for separating the optimal queries from the non-optimal ones.
%In fact, we see them consistently in the Top-10 features of multiple models above.
Although they are strong features for classification,
they might not be actionable for keyword selection since they were derived through reverse-engineering. However, they could still help us better understand the optimal and non-optimal search queries.

%near-optimal keywords are more likely to be extracted from the \emph{description} section only from a bug report (i.e., \emph{bodyKeywordRatio}), which is interesting and unexpected. When we check the remaining three plots (Figures \ref{fig:hq-nl-mda}, \ref{fig:hq-pest-mda}, \ref{fig:lq-nl-pest-mda}), we find that the same observation is true for at least two cases (Figures \ref{fig:hq-pest-mda}, \ref{fig:lq-nl-pest-mda}). Traditionally, keywords from the \emph{title} section or the keywords found both in \emph{title} \cite{saner2017masud} and in \emph{description} sections are considered to be more important than others \cite{kevic}. Thus, our finding on the position of near-optimal keywords challenges the conventional wisdom, which indicates a new insight. We also see that the frequency-based features (e.g., avgTF, stdTF) are consistently less important across the four models in identifying the near-optimal queries. In other words, according to our investigation, the near-optimal keywords are less frequent than the non-optimal ones within a bug report. However, this finding also challenges the conventional wisdom about the keyword importance.

%\begin{figure}[!t]
%	\centering
%	\includegraphics[width=4.8in]{reqs/rq3/hq-pest-mda}
%	\vspace{-.3cm}
%	\caption{Feature importance in the RandomForest model (RF-HQ$_H+$) based on the search queries constructed from bug reports with good baseline queries and with localization hints (Q-HQ$_{H+}$)}
%	\label{fig:hq-pest-mda}
%	\vspace{-.3cm}
%\end{figure}

\begin{table}[!t]
	\centering
	\caption{Regression Analysis between Query Features and Query Performance}\label{table:logistic}
	\vspace{-.2cm}
	\resizebox{4.8in}{!}{%
		\begin{threeparttable}
			\begin{tabular}{l|c|c|c|c|c}
				\hline
				\textbf{Coefficient} & \textbf{Estimate} & \textbf{OR} & \textbf{SE} & \textbf{z-value} & \textbf{p-value} \\
				\hline
				\hline
				Intercept & 1.13 & 3.10 & 0.10 & -11.06 & $<$2e-16*** \\ 
				\hline
				\textit{medianTF} & -0.29 & 0.75 & 0.03 & 8.68 & $<$2e-16*** \\
				\hline
				\textit{medianTermEntropy} & -5.14 & 0.01 & 0.20 & 26.31 & $<$2e-16*** \\
				\hline
				\textit{bodyKeywordRatio} & 0.22 & \textbf{1.25} & 0.09 & -2.44 & 0.02* \\
				\hline
				\textit{nounRatio} & 0.59 & \textbf{1.81} & 0.10 & -6.07 & 1.28e-09*** \\
				\hline
				\multicolumn{6}{c}{\textbf{OR}=Odd Ratio, \textbf{SE}=Std. Error, \textbf{***}=Strongly significant, \textbf{*}=Significant}\\
				\multicolumn{6}{c}{\textbf{Emboldened}=Odd ratios of strong features from the optimal search queries}\\
				\hline
			\end{tabular}
		\end{threeparttable}
	}
	\vspace{-.1cm}
\end{table}

%text quality of a bug report often determines the performance of the search queries constructed from it.

\begin{table}[!t]
	\centering
	\caption{Actionable Insights from Feature Importance Analysis}\label{table:actionale-insights}
	\vspace{-.2cm}
	\resizebox{4.5in}{!}{%
		\begin{threeparttable}
			\begin{tabular}{l|p{3.6in}}
				\hline
				\textbf{No.} & \textbf{Insight} \\
				\hline
				\hline
				I$_1$ & Optimal search keywords are less frequent within a bug report than the non-optimal ones. \\
				\hline
				I$_2$ & Optimal search keywords are less ambiguous (i.e., have less entropy) than the non-optimal ones. \\
				\hline
				I$_3$ & Optimal search keywords are more likely to be found in the \emph{description} section of a bug report than the non-optimal ones. \\
				\hline
				I$_4$ & Optimal search keywords are more likely to be \emph{noun} than the non-optimal ones.\\
				\hline
			\end{tabular}
		\end{threeparttable}
	}
%\vspace{-.3cm}
\end{table}

%\begin{table}[!t]
%	\centering
%	\caption{Optimal vs. Non-Optimal Queries using their Features}\label{table:feature-difference}
%	\vspace{-.2cm}
%	\resizebox{4.5in}{!}{%
%		\begin{threeparttable}
%			\begin{tabular}{l|c|c|c|c}
%				\hline
%				\textbf{Feature} & \textbf{Optimal (Median)} & \textbf{Non-Optimal (Median)} & \textbf{p-value} & \textbf{Cliff's} $\delta$ \\
%				\hline
%				\hline
%				avgTF & 1.40 & 1.40 & 0.21 & 
%				
%				\hline
%			\end{tabular}
%		\end{threeparttable}
%	}
%	\vspace{-.3cm}
%\end{table}

\begin{table}[!t]
	\centering
	\caption{Improvement of Baseline Queries with Actionable Insights (Using Top-10 Results Only)}\label{table:actionable-baseline}
	\vspace{-.2cm}
	\resizebox{4.8in}{!}{%
		\begin{threeparttable}
			\begin{tabular}{l|c|c|c|c|c|c}
				\hline
				\textbf{Technique} & \textbf{Insight} & \textbf{Hit@1} & \textbf{Hit@5} & \textbf{Hit@10} & \textbf{MRR} & \textbf{MAP} \\
				\hline
				\multicolumn{7}{c}{Performance with all bug reports (\textbf{2,320})}\\
				\hline
				Baseline & & \textbf{31.98}\% & \textbf{57.96}\% & \textbf{66.50}\% & \textbf{0.43} & \textbf{42.69}\% \\
				\hline
				Baseline & I$_1$ & 32.63\% & 58.33\% & 67.52\% & 0.44 & 43.76\% \\
				\hline
				Baseline & I$_2$ & 28.82\% & 54.32\% & 64.00\% & 0.40 & 39.65\% \\
				\hline
				Baseline & I$_3$ & 33.15\% & 56.99\% & 66.73\% & 0.43 & 43.31\% \\
				\hline
				Baseline & I$_4$ & 28.95\% & 54.02\% & 63.43\% & 0.40 & 39.57\% \\
				\hline
				Baseline & \textbf{I$_1$+I$_2$} & 32.82\% & 59.38\% & 68.69\% & 0.44 & 44.09\%\\
				\hline
				Baseline & I$_1$+I$_3$ & 32.49\% & 56.32\% & 65.54\% & 0.43 & 42.69\% \\
				\hline
				Baseline & I$_2$+I$_3$ & 32.16\% & 57.32\% & 66.45\% & 0.43 & 42.65\% \\
				\hline
				Baseline & I$_1$+I$_4$ & 33.16\% & 58.81\% & 68.27\% & 0.44 & 43.97\% \\
				\hline
				Baseline & I$_1$+I$_2$+I$_3$ & 32.94\% & 57.98\% & 66.31\% & 0.44 & 43.38\% \\
				\hline
				Baseline & I$_2$+I$_3$+I$_4$ & 33.46\% & 59.07\% & 67.38\% & 0.44 & 43.88\% \\
				\hline
				Baseline & I$_3$+I$_4$+I$_1$ & 33.39\% & 57.34\% & 67.13\% & 0.44 & 43.52\% \\
				\hline
				Baseline & \textbf{I$_1$+I$_2$+I$_4$} & \textbf{34.42}\% & \textbf{60.30}\% & \textbf{69.08}\% & \textbf{0.46} & \textbf{45.44}\% \\
				\hline
				Baseline & all & 34.02\% & 58.86\% & 67.08\% & 0.46 & 44.44\% \\				
				\hline
				\multicolumn{7}{c}{Bug reports with good baseline queries and no localization hints (\textbf{567}) \textbf{(HQ$_{H-}$)}}\\
				\hline
				Baseline & -- & 41.96\% & 86.18\% & 100.00\% & 0.60 & 58.15\% \\
				\hline
				Baseline & I$_1$+I$_2$ & 38.12\% & 76.51\% & 87.27\% & 0.54 & 51.68\%\\ 
				\hline
				Baseline & I$_1$+I$_4$ & 41.35\% & 78.92\% & 89.64\% & 0.57 & 55.07\% \\
				\hline
				Baseline & I$_1$+I$_2$+I$_4$ & 43.24\% & 79.39\% & 90.12\% & 0.58 & 55.71\% \\
				\hline
				\multicolumn{7}{c}{Bug reports with good baseline queries and with localization hints (\textbf{954}) \textbf{(HQ$_{H+}$)}}\\
				\hline
				Baseline & -- & 50.01\% & 86.75\% & 100.00\% & 0.66 & 66.63\% \\
				\hline
				Baseline & I$_1$+I$_2$ & 50.44\% & 83.02\% & 92.01\% & 0.65 & 65.67\% \\
				\hline
				Baseline & I$_1$+I$_4$ & 50.63\% & 81.29\% & 90.53\% & 0.64 & 64.84\% \\
				\hline
				Baseline & I$_1$+I$_2$+I$_4$ & 52.71\% & 83.83\% & 92.21\% & 0.66 & 67.61\% \\ 
				\hline
				\multicolumn{7}{c}{Bug reports with poor baseline queries and no localization hints (\textbf{372}) \textbf{(LQ$_{H-}$)}}\\
				\hline
				Baseline & -- & 0.00\% & 0.00\% & 0.00\% & 0.00 & 0.00\% \\
				\hline
				Baseline & I$_1$+I$_2$ & 2.05\% & 7.74\% & 15.51\% & 0.05 & 5.19\% \\
				\hline
				Baseline & \textbf{I$_1$+I$_4$} & 1.28\% & 7.37\% & \textbf{16.09}\% & 0.04 & 4.50\% \\
				\hline
				Baseline & I$_1$+I$_2$+I$_4$ & 2.23\% & 7.33\% & 15.01\% & 0.05 & 4.79\% \\
				\hline
				\multicolumn{7}{c}{Bug reports with poor baseline queries and with localization hints (\textbf{427}) \textbf{(LQ$_{H+}$)}}\\
				\hline
				Baseline & -- & 0.00\% & 0.00\% & 0.00\% & 0.00 & 0.00\% \\
				\hline
				Baseline & \textbf{I$_1$+I$_2$} & 6.79\% & \textbf{23.06}\% & \textbf{34.29}\% & \textbf{0.14} & \textbf{13.79}\% \\
				\hline
				Baseline & I$_1$+I$_4$ & \textbf{7.01}\% & 20.04\% & 31.25\% & 0.13 & 12.80\% \\
				\hline
				Baseline & I$_1$+I$_2$+I$_4$ & 6.96\% & 22.41\% & 32.95\% & 0.14 & 13.70\% \\
				\hline
			\end{tabular}
		\centering
		\textbf{Emboldened:} Original version of Baseline and the best-performing combinations of actionable insights\\
		\end{threeparttable}
	}
\vspace{-.5cm}
\end{table}

\begin{table}[!t]
	\centering
	\caption{Improvement of STRICT Queries with Actionable Insights (Using Top-10 Results Only)}\label{table:actionable-strict}
	\vspace{-.2cm}
	\resizebox{4.8in}{!}{%
		\begin{threeparttable}
			\begin{tabular}{l|c|c|c|c|c|c}
				\hline
				\textbf{Technique} & \textbf{Insight} & \textbf{Hit@1} & \textbf{Hit@5} & \textbf{Hit@10} & \textbf{MRR} & \textbf{MAP} \\
				\hline
				\multicolumn{7}{c}{Performance with all bug reports (\textbf{2,320})}\\
				\hline
				STRICT & -- & \textbf{25.82}\% & \textbf{52.28}\% & \textbf{63.02}\% & \textbf{0.37} & \textbf{37.31}\% \\
				\hline
				STRICT & \textbf{I$_1$} & 33.97\% & 59.56\% & 68.60\% & 0.45 & 44.85\% \\
				\hline
				STRICT & I$_2$ & 30.75\% & 56.43\% & 67.38\% & 0.42 & 42.09\% \\
				\hline
				STRICT & I$_3$ & 33.53\% & 57.74\% & 67.10\% & 0.44 & 43.86 \% \\
				\hline
				STRICT & I$_4$ & 30.35\% & 56.59\% & 65.57\% & 0.41 & 41.45\% \\
				\hline
				STRICT & I$_1$+I$_2$ & 35.40\% & 61.02\% & 69.92\% & 0.46 & 46.19\% \\
				\hline
				STRICT & I$_1$+I$_3$ & 35.06\% & 59.81\% & 68.75\% & 0.46 & 45.48\% \\
				\hline
				STRICT & I$_2$+I$_3$ & 34.46\% & 58.85\% & 67.45\% & 0.45 & 44.72\% \\
				\hline
				STRICT & \textbf{I$_1$+I$_4$} & 34.87\% & 62.02\% & 70.41\% & 0.46 & 46.04\% \\
				\hline
				STRICT & I$_2$+I$_4$ & 34.43\% & 61.06\% & 70.34\% & 0.46 & 45.70\% \\
				\hline
				STRICT & I$_3$+I$_4$ & 34.09\% & 60.11\% & 69.50\% & 0.45 & 45.12\% \\
				\hline
				STRICT & I$_1$+I$_2$+I$_3$ & \textbf{35.50}\% & 60.59\% & 69.43\% & 0.46 & 46.03\% \\
				\hline
				STRICT & I$_2$+I$_3$+I$_4$ & 34.76\% & 61.35\% & 69.99\% & 0.46 & 45.79\% \\
				\hline
				STRICT & I$_3$+I$_4$+I$_1$ & 35.04\% & 62.14\% & 70.28\% & 0.46 & 46.05\% \\
				\hline
				STRICT & \textbf{I$_1$+I$_2$+I$_4$} & 35.35\% & \textbf{62.58}\% & \textbf{70.90}\% & \textbf{0.47} & \textbf{46.52}\% \\
				\hline
				STRICT & all & 35.41\% & 62.06\% & 70.63\% & 0.46 & 46.31\% \\
				\hline
				\multicolumn{7}{c}{Bug reports with good baseline queries and no localization hints (567) \textbf{(HQ$_{H-}$)}}\\
				\hline
				STRICT & - & 36.14\% & 72.55\% & 84.83\% & 0.51 & 50.29\% \\
				\hline
				STRICT & \textbf{I$_1$+I$_2$+I$_4$} & \textbf{41.90}\% & \textbf{82.74}\% & \textbf{92.12}\% & \textbf{0.58} & \textbf{56.03}\% \\
				\hline
				\multicolumn{7}{c}{Bug reports with good baseline queries and with localization hints (954) \textbf{(HQ$_{H+}$)}}\\
				\hline
				STRICT & -- & 36.53\% & 69.86\% & 82.41\% & 0.51 & 51.85\% \\
				\hline
				STRICT & \textbf{I$_1$+I$_2$+I$_4$} & \textbf{55.17}\% & \textbf{88.00}\% & \textbf{95.26}\% & \textbf{0.69} & \textbf{70.09}\% \\
				\hline
				\multicolumn{7}{c}{Bug reports with poor baseline queries and no localization hints (372) \textbf{(LQ$_{H+}$)}}\\
				\hline
				STRICT & -- & 1.56\% & 9.54\% & 16.89\% & 0.05 & 5.12\% \\
				\hline
				STRICT & I$_1$+I$_2$+I$_4$ & 2.66\% & 7.18\% & 15.18\% & 0.05 & 5.09\% \\
				\hline
				\multicolumn{7}{c}{Bug reports with poor baseline queries and with localization hints (427) \textbf{(LQ$_{H+}$)}}\\
				\hline
				STRICT & -- & 4.91\% & 18.86\% & 25.77\% & 0.11 & 10.66\% \\
				\hline
				STRICT & \textbf{I$_1$+I$_2$+I$_4$} & \textbf{6.80}\% & \textbf{23.31}\% & \textbf{32.83}\% & \textbf{0.13} & \textbf{13.39}\% \\
				\hline
			\end{tabular}
		\centering
		\textbf{Emboldened:} Original version of STRICT and the best-performing combinations of actionable insights\\
		\end{threeparttable}
	}
\vspace{-.5cm}
\end{table}

%Thus, this finding on near-optimal keyword posi
%This insight on keyword position is interesting since
%and (2) more similar to the ground-truth file names. While these insig

%for differentiating between near-optimal and non-optimal search queries
%According to our investigation, 61\% of the keywords of the optimal queries are extracted from the \emph{description} of a bug report. On the contrary, the state-of-the-art queries choose the majority (35\%) of their keywords from the \emph{title} of a bug report.
 %\ref{fig:strict-optimal-br-fields} further shows the distribution of keywords across the two fields of a bug report.
%Given these findings,
%Thus, \emph{title} clearly does not contain all the keywords needed for constructing an optimal query. The state-of-the-art queries contain more unique keywords than the optimal queries in their Top-10 keywords. However, we found that optimal queries (from bug reports) \textbf{share more terms} with the ground truth class names than their counterparts, which possibly explains their high performance. It should be noted that Optimal$_{GA}$ employs the ground truths only for evaluating the candidate queries, but not for selecting the search keywords from them (Section \ref{sec:ga}).
%Thus, although the poor bug reports might not contain explicit localization hints (\eg\ program elements), useful hints could be hidden within their bulky texts (\eg\ \emph{description}).

\subsection{Improvement of Baseline Queries and State-of-the-art Search Queries with Actionable Insights}\label{sec:improve-by-insight}
Our feature importance analysis above (Section \ref{sec:feature-importance}) offers four actionable insights (Table \ref{table:actionale-insights}). We apply these insights to both baseline and STRICT queries (i.e., state-of-the-art), expand them using appropriate keywords and determine their benefits. Tables \ref{table:actionable-baseline}, \ref{table:actionable-strict} summarize our experimental analysis as follows. 

In Table \ref{table:actionable-baseline}, we investigate the benefits of four insights both in isolation and in combination by applying them to baseline queries.
We see that expansion of baseline queries using less frequent keywords (i.e., I$_1$) improves their performance marginally. They get further improved when less ambiguous keywords are added to them (i.e., I$_1$+I$_2$). For example, they achieve a Hit@10 of 69\% with 44\% mean average precision which are 3\% improvement over the baseline. However, a combination of three insights (I$_1$+I$_2$+I$_4$) leads to further improvement of baseline queries. That is, when less frequent, less ambiguous and noun keywords are added to 2,320 baseline queries from their corresponding bug reports, the extended queries achieve a 34\% Hit@1, 69\% Hit@10, 0.46 MRR and 45\% MAP, which are 8\%, 4\%, 7\% and 6\% higher respectively, which are promising.

We also investigate the impact of our actionable insights upon the baseline queries from different subsets of bug reports. From Table \ref{table:actionable-baseline}, we see that they fail to improve the already good baseline queries, i.e., collected from HQ$_{H-}$ and HQ$_{H+}$ bug reports. Although the query expansion based on our actionable insights did not help, the expanded queries still remained of high-quality (e.g., 90\% Hit@10). However, such an expansion improved the poor baseline queries, i.e., collected from LQ$_{H-}$ and LQ$_{H+}$ bug reports, with a significant margin. For example, the poor baseline queries fail to return any ground truth within their Top-10 results (i.e., 0.00\% Hit@10). When more keywords were added to them using two actionable insights on frequency and entropy (i.e., I$_1$+I$_2$), the expanded versions achieve up to 23\% Hit@5, 34\% Hit@10, 0.14 MRR and 14\% MAP, which are highly promising.

In Table \ref{table:actionable-strict}, we investigate the effectiveness of four insights both in isolation and in combination by applying them to the queries of STRICT technique \cite{saner2017masud}.
We see that expansion of STRICT queries with less frequent keywords (i.e., I$_1$) improve Hit@1, Hit@10, MRR and MAP by 32\%, 9\%, 22\%, and 20\% respectively. Addition of nouns from bug reports (i.e., I$_1$+I$_4$) further improves these queries and they achieve 70\% Hit@10 with 46\% MAP, which are 12\% and 24\% higher than original performance measures. However, a combination of three insights on frequency, entropy and part of speech of keywords (i.e., I$_1$+I$_2$+I$_4$) leads to the maximum improvement of 2,320 queries from STRICT -- 71\% Hit@10, 0.47 MRR and 47\% MAP. It should be noted that these performance metrics are 7\%, 9\%, and 9\% higher than their baseline counterparts (Table \ref{table:actionable-baseline}).  

We also investigate the impact of our actionable insights upon the STRICT queries from different subsets of bug reports. From Table \ref{table:actionable-strict}, we see that they improve queries from three subsets of bug reports - HQ$_{H-}$, HQ$_{H+}$ and LQ$_{H+}$. STRICT achieves 70\%--73\% Hit@5, 82\%--85\% Hit@10, 0.51 MRR and 50\%--52\% MAP when queries are constructed from the bug reports that lead to good baseline queries (HQ$_{H-}$, HQ$_{H+}$). Expansion of these queries applying the actionable insights on keyword frequency, entropy and part of speech (I$_1$+I$_2$+I$_4$) leads to 83\%--88\% Hit@5, 92\%--95\% Hit@10, 0.58--0.69 MRR, and 56\%--70\% MAP, which clearly demonstrates the positive impacts of our derived insights. We also notice improvement in query performance due to query expansion, especially for STRICT queries from the bug reports with poor baseline (LQ$_{H+}$). The original queries from STRICT technique achieves 26\% Hit@10 with 11\% MAP. Our expanded version of STRICT queries achieve 33\% Hit@10 with 13\% MAP, which are 27\% and 18\% higher respectively. All these empirical findings clearly suggest the benefits of our derived actionable insights (Table \ref{table:actionale-insights}).

According to our investigation, simple keyword removal strategies and existing techniques (e.g., STRICT) perform comparably when dealing with bug reports that lead to poor baseline queries. For example, simply removing the duplicate keywords from a baseline query can lead to 25\% Hit@10 for LQ$_{H+}$ bug reports, which is comparable to STRICT's original performance (Table \ref{table:freq-vs-graph-br-subset}). Combination of unique keywords from the title and body sections of a bug report also leads to 34\% Hit@10 (Table \ref{table:actionable-baseline}). Thus, further investigation is warranted in this context. Our queries constructed using all 15 possible combinations (i.e., $^4$C$_1$+$^4$C$_2$+$^4$C$_3$+$^4$C$_4$) of the four actionable insights can also be found in the replication package \cite{emse2019-rep} for further investigation and reuse.

\subsection{Manual Analysis of Bug Reports Leading to Poor Baseline Queries}
Given the feature importance analysis, actionable insights and query improvements above, we further analyze the bug reports that lead to poor baseline queries to gain a deeper understanding of them. Since analyzing all of them could be impractical, we choose a random subset of 120 bug reports. In particular, we randomly select 20 bug reports (10 with localization hints + 10 without localization hints) from each of the six subject systems for our analysis.
We go through the \emph{title} and \emph{description} texts of each bug report and attempt to find out important trends using Grounded Theory (GT) approach \cite{grounded-theory}. In our GT approach, we perform three levels of coding namely \emph{open coding}, \emph{axial coding} and \emph{selective coding}. During open coding, we keep an open mind and attempt to describe a bug report with a set of appropriate key phrases. During axial coding, we put these key phrases on a spreadsheet document and connect the semantically related key phrases. That is, we use the same colour to represent similar concepts. In our selective coding step, we gather the low-level categories from the axial coding and derive high-level theories and meaningful insights from them.
%, and then derive meaningful insights from the coded keywords. 
%\Foutse{did we have a single evaluator or more? could you provide more details about the tagging process?}
%This approach involves three stages of coding -- open coding, axial coding and selective coding.
The artifacts from our Grounded Theory-based investigation can be found in the replication package \cite{emse2019-rep}. We spent $\approx$20 man-hours for this manual inspection.
Based on our analysis, we make several important observations about the bug reports that lead to poor baseline queries (i.e., QE$>$10) as follows.

First, bug reports leading to poor baseline queries often contain generic, verbose, and noisy textual contents that might not be helpful to understand a reported bug. Although our Genetic Algorithm-based approach was able to extract the optimal and near-optimal queries from them, their keywords might always not be intuitive enough. However, interestingly, many of them match with the keywords from ground truth file names. Furthermore, the percentage of this match has been found as one of the most important features for separating between optimal and non-optimal queries (see Fig. \ref{fig:feature-importance} for details).

Second, many of these bug reports (LQ$_{H+}$) contain noisy stack traces with hundreds of trace lines, configuration information and complex memory dump. Although these contents could be useful for human developers with relevant experience, they are not helpful enough for constructing appropriate queries using the existing automated tool supports.

Third, many of these bug reports were submitted with screenshots, videos, stack traces and data dump in the attachment. Therefore, the submitter did not feel the necessity of writing a detailed bug report using texts. As a result, sufficient textual contents were not available to the existing techniques for constructing appropriate queries. Thus, their queries were not effective for localizing the bugs.

Fourth, several of these bug reports were submitted with patched code. Developers often identify software bugs in an application, fix themselves and then submit both the bug reports and fixed code. Since the bug is already fixed, they often either provide a very generic explanation of the bug or submit the placeholder texts. Since existing approaches mostly rely on textual contents, they fail to construct appropriate search queries from these bug reports.

Fifth, many bug reports contain simple code examples as a part of explaining their bugs. However, these examples might always not point towards actual buggy elements within the software code. Thus, despite containing apparent localization hints, 
these bug reports might not be found useful to construct appropriate queries for IR-based bug localization.

\begin{frshaded}
		\noindent
		 \textbf{Summary of RQ$\mathbf{_3}$:} Optimal search queries are significantly different from the non-optimal ones not only in performance but also in their characteristics. The optimal keywords are \emph{less frequent} than the non-optimal ones within a bug report. They are \emph{more specific} (less ambiguous) than non-optimal keywords. Optimal keywords are also more likely to be found within the \emph{description} section of the report. They are more likely to be \emph{nouns} than the non-optimal ones. When these actionable insights are applied, baseline queries and state-of-the-art queries can achieve up to \textbf{34}\% higher Hit@10 and \textbf{27}\% higher Hit@10 respectively, which are highly promising.
		 Our manual analysis with Grounded Theory approach also provides further insights on why the traditional approaches might fail to construct right queries from the bug reports that lead to poor baseline queries.
\end{frshaded}

\section{Findings Summary \& Discussions} \label{sec:discussion}
In our empirical study, we conduct three comparative analyses and answer three important research questions. In the following section, we summarize our important findings, actionable insights and also point out the potential gaps in the literature of IR-based bug localization.
\begin{enumerate}
	\setlength\itemsep{1em}
	
\item \textbf{Bug reports could produce poor baseline queries despite containing explicit localization hints.} About 34\% (799) of our collected bug reports 
lead to poor baseline queries (i.e., QE$>$10)
where 18\% (427) reports contain explicit hints for localizing the bugs (e.g., program elements, stack traces). Thus, the idea of using localization hints might not be sufficient to extract appropriate search keywords. Through the application of actionable insights, we were able to improve the queries from these bug reports significantly (e.g., 27\%--34\% improvement in Hit@10), which demonstrates the benefits of our insights.

%Baseline queries constructed from them fail to retrieve any ground-truth document within their top 10 results.
	
\item \textbf{State-of-the-art approaches are not enough.} The proxies adopted by state-of-the-art approaches for keyword importance (\eg\ TF, TF-IDF, TextRank \cite{saner2017masud}) might not be sufficient to capture appropriate keywords from the bug reports that lead to poor baseline queries. Although these reports might sometimes contain explicit hints for localizing bugs, on average, their textual contents are generic, verbose, and noisy.
%As a result, baseline queries constructed from them fail to retrieve any ground-truth documents within their top 10 results (Table \ref{table:freq-vs-graph-br-subset}).
Thus, the state-of-the-art approaches often fail to choose the right queries from these reports and as a result, could perform poorly in the IR-based bug localization (Tables \ref{table:freq-vs-graph-br-subset}, \ref{table:freq-vs-graph-code-subset}).

%These reports often contain generic, verbose and noisy textual contents.
%poor bug reports that lack explicit localization hints (RQ$_1$, RQ$_3$).

\item\textbf{Graph-based approaches have more potential for search keyword selection.} According to RQ$_1$, graph-based approaches (\eg\ STRICT, ACER) provide better search queries than frequency-based counterparts regardless of the bug report clusters (e.g., Table \ref{table:freq-vs-graph-br-subset} Section \ref{sec:br-cluster}) or the origin of their keywords (\eg\ bug reports, source code) (Section \ref{sec:rq1}). Graph-based approaches were able to extract important keywords even from the bug reports that lead to poor baseline queries, which demonstrates their promising aspect.

\item \textbf{Optimal search keywords exist even in the bug reports leading to poor baseline queries.} Contrary to the popular beliefs \cite{parninireval,bias-bug-loc-lo}, our investigation (RQ$_2$) suggests that bug reports that lead to poor baseline queries contain sufficient keywords to form optimal or near-optimal search queries.
Our finding thus strengthens a similar observation of \citet{cmills-icsme2018}.
Optimal and near-optimal queries from these bug reports achieve 78\%--93\% Hit@10, which are three to five times higher than the state-of-the-art Hit@10 (Table \ref{table:sota-vs-optimal}).

\item \textbf{Optimal queries are different from non-optimal queries.}
Optimal keywords are less frequent than the non-optimal ones within a bug report.
They are more specific (less ambiguous) than the non-optimal keywords. 
Unlike keywords chosen by existing approaches \cite{saner2017masud,kevic}, optimal keywords are more likely to be found within the \emph{description} section of a bug report.
They are also more likely to be nouns than the non-optimal keywords.
Thus, optimal keywords are significantly different from the non-optimal ones in several aspects.

\item \textbf{Optimal search keywords are limited in number.} According to our investigation, the average length of an optimal query does not vary across subject systems.
When candidate queries with 5 keywords were provided, our GA-based algorithm returned optimal queries with an average length of 5 keywords.
However, the candidates with 10, 15, 20 keywords ended up with the optimal queries containing 9, 12, 14 keywords, on average. We also performed keyword overlap analysis, and found that the overlap ratio ranged from 40\% to 60\%. That is, out of 5 optimal keywords, 2-3 keywords overlap with a 9-keywords optimal query, on average. When we repeat this process between 12-keywords and 14-keywords optimal queries, we notice an overlap of 6 keywords, which also indicates 50\% overlap (6/12). Thus, optimal queries overlap significantly although they might have different initial candidates. That is, regardless of various initial keywords (from a bug report) in candidate queries, optimal candidates are likely to converge to a small number of important keywords (a.k.a., optimal keywords) through evolution.

\item \textbf{Simple keyword removal strategies work!} According to our investigation, simple keyword removal strategies and existing techniques (e.g., STRICT) perform comparably when dealing with bug reports that lead to poor baseline queries. For example, simply removing the duplicate keywords from a baseline query can lead to 25\% Hit@10 for LQ$_{H+}$ bug reports, which is comparable to STRICT's original performance (Table \ref{table:freq-vs-graph-br-subset}). Combination of unique keywords from the title and body sections of a bug report also leads to 34\% Hit@10 (Table \ref{table:actionable-baseline}). Thus, although these simple strategies might not lead to a few important keywords (e.g., GA-based approach, STRICT), further investigation involving these strategies is warranted.

%Near-optimal queries are also likely to contain part-of-speeches other than nouns and verbs.

%They contain more verbs as keywords, and their keywords are less frequent within the project's codebase.
%More interestingly, they share more terms with the ground truth class names (RQ$_3$), which possibly explains their high performance.

\item\textbf{Machine intelligence could be effective in recognizing optimal keywords.} About 47\% of 799 bug reports that lead to poor baseline queries 
do not contain any explicit hints for localizing bugs (e.g., stack traces, program elements).
%Since low-quality bug reports lack explicit localization hints (\eg\ program elements),
According to an earlier study \cite{parninireval}, developers often fail to recognize appropriate keywords from these reports.
In this study, we perform feature importance analysis (Section \ref{sec:feature-importance}) and attempt to better separate optimal queries from non-optimal ones using machine learning (e.g., Table \ref{table:model-performance}). 
Our investigation has also resulted into a few actionable insights that were found effective for improving the queries (e.g., Tables \ref{table:actionale-insights}, \ref{table:actionable-baseline}, \ref{table:actionable-strict}).
Future studies can invest more efforts in recognizing the right search keywords from a bug report and machine learning could be a feasible option.

%incorporate appropriate machine intelligence that help the developers recognize the right keywords from these low-quality bug reports.

%Thus, more intelligent tools are warranted for automatically identifying and delivering such keywords on the fly during the IR-based bug localization.

\item\textbf{Practical fitness function for Genetic Algorithm-based query construction is warranted.} We identify optimal and near-optimal search queries from a bug report using ground truth information (RQ$_2$) which might be unknown in practice.
That is, the GA-based approach discussed in this paper might not be practical.
Thus, a fitness function is warranted that can reliably evaluate the fitness of a candidate query (Section \ref{sec:ga}) without needing the ground truth. Such a function has high potential 
for improving GA-based query construction from any textual entities (e.g., bug reports, change requests) and thus will improve several IR-based software engineering tasks such as bug localization, concept location or code search.
%of introducing a novel research topic namely search-based keyword selection from bug reports.
\end{enumerate}

\section {Threats to Validity} \label{sec:threats}
We identify a few threats to the validity of our findings. In this section, we discuss these threats and necessary steps taken to mitigate them as follows.

\subsection{Threats to Internal Validity}
Threats to \emph{internal validity} relate to experimental errors and human bias \cite{wordsim}. Construction of the dataset and replication of the existing approaches are two potential sources of these threats. However, we collect the dataset from an existing benchmark \cite{fse2018masud}, and carefully discard the tangled commits and false-positive bug reports (a.k.a., feature requests) through a manual analysis of 60+ man-hours.
We employ independent coding of the bug reports by two authors, perform agreement analysis and conflict resolution, and finalized our dataset rigorously. We also use the authors' implementation for two approaches \cite{saner2017masud,ase2017masud} and carefully re-implemented the rest (\ie\ unavailable prototypes) with their best settings and parameters (\eg\ regression coefficients \cite{kevic}).
Our dataset, implementation of existing approaches and other associated resources are available in the replication package \cite{emse2019-rep} for third party reuse.
%due to the lack of their available prototypes.
In this way, we mitigate the threats to internal validity.

According to our analysis, the dataset contains bug reports from multiple versions of a software system whereas
the corpus is made of a single snapshot of the codebase, which could be a source of threat.
%Choosing a single snapshot of the codebase while the bug reports targeting multiple versions of a subject system
%rather than a single version 
%could be a source of threat. 
That is, due to this potential misuse of software versions, the bug localization might take place against the fixed code rather than the buggy code.
However, we investigate this issue with a limited experiment using 360 bug reports and careful manual analysis (Section \ref{sec:ks-br}), and demonstrate that our core findings remain the same. While this threat has been mitigated, we nevertheless recommend a consistent use of the bug reports and their software versions for the relevant future studies.

\subsection{Threats to External Validity}
Threats to \emph{external validity} relate to the generalizability of the findings \cite{wordsim}. The findings based on single benchmark might always not generalize to others. We thus conduct a parallel experiment using 332 bug reports from another benchmark dataset (\eg\ SWT, ZXing, AspectJ) \cite{buglocator}, and was able to partially replicate our findings (RQ$_1$ and RQ$_2$). In particular, we reproduced that (1) graph-based approaches indeed provide higher quality keywords than the frequency-based ones, and (2) near-optimal performance could be achieved for 65\% of the bug reports. NrOptimal$_{GA}$ achieves 8\% higher performance on natural language-only bug reports from our dataset (Table \ref{table:opt-vs-baseline-br}, RQ$_2$) which could fall within the margin of error. Thus, the threats to external validity might also be mitigated.

\subsection{Threats to Construct Validity}
Threats to construct validity are associated with the appropriateness of the performance metrics adopted in a study. We use four performance metrics (Section \ref{sec:pmetrics}) that are widely adopted by the existing studies including the state-of-the-art on IR-based concept/bug localization \cite{buglocator,saha,versionhistoryjsep,saner2017masud,ase2017masud,fse2018masud}. Thus, such threats are also mitigated.

\subsection{Threats to Conclusion Validity}
Threats to \emph{conclusion validity} arise when conclusions about relationships between two given variables are not reasonable \cite{trigger}. We answer three research questions using 2,320 bug reports, ten existing approaches including the state-of-the-art and a GA-based approach. We also use statistical tests (\eg\ \emph{Wilcoxon Signed Rank}, \emph{Cliff's delta}) and report the test details (\eg\ \emph{p-value}, $\delta$) before drawing any conclusion. Thus, such threats are also mitigated.

\section{Related Work} \label{sec:related}
\subsection{IR-Based Bug Localization}

Automatic localization of software bugs has been an active area of research for the last five decades \cite{bl-survey-wong-tse}. While traditionally static and dynamic analyses are used \cite{bl-survey-wong-tse,bl-thesis2016}, Information Retrieval has been adopted in the bug localization for more than a decade. There have been several attempts using complex models such as Latent Semantic Indexing \cite{irmarcus,lsiPoshyvanyk} or Latent Dirichlet Allocation (LDA) \cite{raobug}. However, existing findings \cite{raobug,bavota} suggest that simpler models (\eg\ VSM) often perform higher than the complex ones in bug localization. Towards this end, there have been a number of IR-based approaches \cite{buglocator,saha,brtracer,versionhistory,blia,locus,versionhistoryjsep,kak-version-history}. \citet{buglocator} first use revised Vector Space Model (rVSM) and past bug reports for bug localization using Information Retrieval. \citet{saha} leverage the structures of both bug reports and source code documents to improve the bug localization. Several studies \cite{kak-version-history,versionhistory,versionhistoryjsep,locus} make use of code change history, and combine them with Information Retrieval. A few studies \cite{brtracer,fse2018masud,stacktrace} leverage the structured items (\eg\ stack traces) found in the bug report to boost up the localization performance. However, recent findings \cite{parninireval,icse2018,bl-misoo,bias-bug-loc-lo} suggest that IR-based localization could be limited due to poor queries from bug reports. In particular, the existing approaches might not perform well without the presence of localization hints (\eg\ program elements) in the bug reports (queries). We \emph{investigate} this \emph{limitation} of IR-based localization through an empirical study, and attempt to better understand the underlying causes. Appropriate query construction is a major step of any IR-based text retrieval \cite{refoqus,sisman}. Our investigation suggests that appropriate queries might not have been used by many of the existing approaches.

 \subsection{Query Reformulation in IR-Based Bug Localization}
  There have been a number of studies on query construction/reformulation that support the developers during concept location \cite{saner2017masud,rocchio,ase2017masud,ase2016masud,refoqus,kevic,gayg}, feature/concern location \cite{flt-model-qr,time-aware-term-weighting,hillicse09,shepherd}  and bug localization \cite{sisman,fse2018masud,observed,verbose}.
 One key challenge of any automated query construction task is -- \emph{appropriate keyword selection}.
 Existing studies can be classified into two broad categories based on their keyword selection -- \emph{frequency based} and \emph{graph-based}. TF-IDF \cite{tfidf} has been a popular frequency-based term weighting method for the last five decades. Several existing studies \cite{kevic,refoqus,kevicdict,trconfig,cmills-icsme2018} employ TF-IDF and its variants to identify the important keywords from a body of texts (\eg\ bug report, source code). \citet{kevic} use TF-IDF and three heuristics to identify the important keywords from a bug report.
 \citet{gayg} employ Rocchio's expansion \cite{rocchio} where they use TF-IDF for keyword selection from the source code.
 \citet{refoqus} later employ three frequency-based term weighting methods --Rocchio \cite{rocchio}, RSV \cite{rsv} and Dice \cite{qsurvey}, and deliver the best performing query keywords from the source code using machine learning. \citet{sisman} leverage \emph{spatial code proximity}, and suggest such terms that frequently co-occur with the query keywords within the source code. We select a total of \emph{eight} frequency-based approaches (including six above \cite{tfidf,kevic,rocchio,rsv,qsurvey,sisman}) for our study. Although TF-IDF has been popular, it fails to capture a term's contexts (\eg\ surrounding terms) during term weighting, which is a major issue \cite{rada,blanco}. \citet{saner2017masud} first capture semantic and syntactic dependencies among the words, employ PageRank algorithm \cite{pagerank} on the constructed graph, and then deliver appropriate search keywords from a bug report. They later adopt similar methodologies, and suggest search keywords from source code documents \cite{ase2017masud} and noisy bug reports containing stack traces \cite{fse2018masud}. To the best of our knowledge, these are the state-of-the-art approaches for graph-based keyword selection from bug reports and source code. We thus employ two of the above approaches \cite{saner2017masud,ase2017masud} for our empirical study. The detailed comparison between frequency-based and graph-based approaches can be found in Section \ref{sec:rq1}.

 In terms of methodology and research goals, our work is closely related to \citet{cmills-icsme2018}. Like ours, they also adopt a Genetic Algorithm to construct near-optimal search queries, and suggest that bug reports are often sufficient for constructing the right queries that can localize the bugs using Information Retrieval. Besides confirming and strengthening their finding, we conduct further investigation and report several major findings. First, bug reports could produce good or poor baseline queries
 %be of high-quality or low-quality 
 regardless of the presence of explicit bug localization hints (e.g., stack traces, program elements) in their texts (RQ$_1$). Second, state-of-the-art algorithms are not sufficient enough for constructing appropriate queries from the bug reports that lead to poor baseline queries (i.e., QE$>$10) (RQ$_2$). Third, optimal or near-optimal search queries might not be badly affected by the length of their candidate queries (Table \ref{table:ga-config}). Both Query Effectiveness and Mean Average Precision are found suitable as a fitness function for GA-based query construction (RQ$_2$). Fourth, optimal keywords are less frequent and less ambiguous than non-optimal ones. They are more likely to be found within the description section of a bug report and also more likely to be nouns(RQ$_3$). Fifth, our derived insights are actionable and they were able to improve both baseline and state-of-the-art queries significantly (Tables \ref{table:actionable-baseline}, \ref{table:actionable-strict}) (RQ$_3$).
 %(1) why the criticisms against IR-based bug localization concerning the natural language-only bug reports \cite{parninireval,bias-bug-loc-lo} could be biased (RQ$_1$, RQ$_3$), (2) that the state-of-the-art proxies of term importance (\eg\ TF-IDF \cite{refoqus}, TextRank \cite{saner2017masud}) are not sufficient enough (RQ$_1$), (3) that the optimal search queries are significantly different than state-of-the-art queries \cite{saner2017masud} (RQ$_3$).
 Such comprehensive investigations were not reported by the earlier studies, which makes our work \emph{novel}.

 There also exist a few other studies that employ query construction/reformulation in the bug localization \cite{observed,verbose}, concern/concept location \cite{shepherd,hillicse09,ccmapping,infer} and duplicate bug report detection \cite{oscar-duplicate}. However, they construct their queries through  natural language discourse analysis or software repository mining, which could be hard to replicate in an empirical study setting. We thus do not include them in our experiments although they are relevant. Besides, many of them were outperformed by the later studies. However, our study performs a detailed empirical study that examines the query construction practices in IR-based bug localization and also delivers several actionable insights, which could inspire future investigations.

 %that deal with query reformulation targeting concept/bug localization. However, they are not closely related to ours.

 %To the best of our knowledge, these are the state-of-the-art studies in frequency-based query construction for concept location and bug localization.

%  concept/feature/bug localization, and they employ term weighting \cite{}, heuristics \cite{kevic,observed,hillicse09,verbose}, natural language processing \cite{observed,saner2017masud}, pseudo-relevance feedback \cite{rocchio,sisman}, query quality analysis \cite{refoqus,ase2017masud} and machine learning \cite{ase2017masud,refoqus}. However, to the best of our knowledge, no studies employ genetic algorithms for the reformulation of a given poor or verbose query. Several studies on SE text retrieval use genetic algorithms mostly for parameter or configuration optimization \cite{paniICSE2013,paniSANER2016,trace-ga}, which makes our work \emph{novel}.

\section{Conclusion \& Future Work}\label{sec:conclusion}
IR-based bug localization approaches have been widely used because of being light-weight and cost-effective. However, quality of their search queries might impact their localization accuracies.
%However, the quality of the bug reports they are based on might impact the accuracy of the bug localization. 
In fact, recent studies report mixed findings on the performance of IR-based localization. While most studies suggest that IR-based approaches perform poorly with natural language-only bug reports as queries, 
%(containing no localization hints), 
there is one study suggesting that even these bug reports may have sufficient keywords for successfully detecting the bugs. These findings could potentially make one believe that natural language-only bug reports are a sufficient source of good queries. 
%the bug reports that are being considered poor for 
%IR-based bug localization, may not be really poor. 
In this study, we attempted to shed light on this pressing controversial issue by conducting an empirical study using 2,320 bug reports (939 natural language-only + 1,381 with localization hints), and ten existing approaches including the state-of-the-art on query construction for IR-based bug localization. We also employ a Genetic Algorithm-based approach for constructing optimal and near-optimal search queries from the bug reports. We conduct three comparative analyses, and answer three research questions. Our study reports several major findings. First, bug reports might lead to poor search queries despite containing explicit hints for localizing bugs (e.g., stack traces). 
On the other hand, bug reports containing only natural language texts might provide high-quality search queries for IR-based bug localization.
Second, even the state-of-the-art approaches are not sufficient enough to deliver appropriate queries from the bug reports that lead to poor baseline queries.
%natural language-only bug reports. 
Third, graph-based approaches are better than frequency-based ones in selecting query keywords.
%for search keyword selection. 
%Fourth, even though the bug reports might not contain explicit localization hints (a.k.a., natural), they still contain high-quality search keywords in their texts. 
Fourth, optimal search queries from bug reports are significantly different from the non-optimal queries in several aspects (\eg\ frequency, entropy, keyword position). 
Fifth, the comparison between optimal and non-optimal queries led us to several actionable insights that were found useful to significantly improve existing search queries.
Our study inspires several future research opportunities as follows.
\begin{itemize}
	\item Machine intelligence might be leveraged to recognize the right keywords from the natural language-only bug reports since they do not contain any explicit hints for localizing the bugs. Ours is a first attempt towards this direction that compares between optimal and non-optimal queries through feature importance analysis and machine learning.
	\item
	Genetic Algorithms (GA) might also be used to identify the optimal or near-optimal search keywords from a bug report given that an appropriate fitness function is available. Our fitness function discussed in the paper might not be feasible for practical use since it relies on ground truth information. Thus, future work can also focus on developing a practical fitness function for GA-based query construction that does not rely on the ground truth.
\end{itemize}

%Future work should (1) incorporate machine intelligence in identifying the keywords hidden within the texts of a bug report. The stat

\section*{Acknowledgement}
This research was supported by Tenure-track startup grant, Dalhousie University, Saskatchewan Innovation \& Opportunity Scholarship (2017--2018), and the Natural Sciences and Engineering Research Council of Canada (NSERC).

%\balance

\newpage
\appendix

\section{Query Attributes for the Comparative Analysis}

\small
\begin{longtable}[!t]{l|p{1.8in}|p{2in}}
	\caption{Query Attributes Used for the Comparative Analysis} \label{table:query-features} \\
				\hline
				\textbf{Attribute} & \textbf{Description} & \textbf{Formula} \\
				\hline
				$avgTF$ & Average frequency of all keywords from the search query $Q$ within the texts of a bug report & $\frac{1}{|Q|}\underset{q\in Q}{\sum} TF(q)$ \\
				\hline
				$medianTF$ & Median frequency of all keywords from the query $Q$ & $\underset{q\epsilon Q}{median}~(TF(q))$ \\
				\hline
				$maxTF$ & Maximum frequency of all keywords from the query $Q$ & $\underset{q\epsilon Q}{max}~(TF(q))$ \\
				\hline
				$stdTF$ & Standard deviation of all keyword frequencies from the query $Q$ & $\sqrt{\frac{1}{|Q|} \underset{q\in Q}{\sum}(TF(q)-\overline{TF})^2}$ \\
				\hline
				$avgIDF$ & Average Inverse Document Frequency of all keywords from the query $Q$ within the corpus & $\frac{1}{|Q|}\underset{q\in Q}{\sum} IDF(q)$ \\
				\hline
				$medianIDF$ & Median Inverse Document Frequency of all keywords from the query $Q$ & $\underset{q\epsilon Q}{median}~(IDF(q))$ \\
				\hline
				$maxIDF$ & Maximum Inverse Document Frequency of all keywords from the query $Q$ & $\underset{q\epsilon Q}{max}~(IDF(q))$ \\
				\hline
				$stdIDF$ & Standard deviation of Inverse Document Frequencies of all the keywords from the query $Q$ & $\sqrt{\frac{1}{|Q|} \underset{q\in Q}{\sum}(IDF(q)-\overline{IDF} )^2}$ \\
				\hline
				$avgTFIDF$ & Average TF-IDF score of all keywords from the query $Q$  & $\frac{1}{|Q|}\underset{q\in Q}{\sum} TF(q)\times IDF(q)$ \\
				\hline
				$medianTFIDF$ & Median TF-IDF score of all keywords from the query $Q$ & $\underset{q\epsilon Q}{median}~(TFIDF(q))$ \\
				\hline
				$maxTFIDF$ & Maximum TF-IDF score of all keywords from the query $Q$ & $\underset{q\epsilon Q}{max}~(TFIDF(q))$ \\
				\hline
				$stdTFIDF$ & Standard deviation of TF-IDF scores of all the keywords from the query $Q$ & $\sqrt{\frac{1}{|Q|} \underset{q\in Q}{\sum}(TFIDF(q)-\overline{TFIDF})^2}$ \\
				\hline
				$avgEntropy$ & Average entropy of all keywords from the query $Q$ & $\frac{1}{|Q|}\underset{q\in Q}{\sum} entropy(q)$ \\
				\hline
				$medianEntropy$ & Median entropy of all keywords from the query $Q$ & $\underset{q\epsilon Q}{median}~(entropy(q))$ \\
				\hline
				$maxEntropy$ & Maximum entropy of all keywords from the query $Q$ & $\underset{q\epsilon Q}{max}~(entropy(q))$ \\
				\hline
				$stdEntropy$ & Standard deviation of entropy measures of all the keywords from the query $Q$ & $\sqrt{\frac{1}{|Q|} \underset{q\in Q}{\sum}(entropy(q)-\overline{entropy})^2}$ \\
				\hline
				$JSD$ & Jensen Shannon Divergence is a symmetric version of Kullback-Leibler Divergence between two different probability distributions $P$ and $Q$ &
				$\frac{1}{2}(KLD(P||M) + KLD(Q||M))$
				 \\
				\hline
				$QSI$ & Query Specificity Index \cite{specificity} &   1 - $\underset{q\epsilon Q}{median}~(entropy(q))$ \\
				\hline
				$nounRatio$ & Fraction of all noun keywords $N$ (from the bug report) that are found in the query $Q$ &  $ \frac{1}{|N|}\underset{q\in Q}{\sum} isNoun(q)$  \\
				\hline
				$verbRatio$ & Fraction of all verb keywords $V$ (from the bug report) that are found in the query $Q$ &  $ \frac{1}{|V|}\underset{q\in Q}{\sum} isVerb(q)$  \\
				\hline
				$nounVerbRatio$ & Fraction of all the noun and verb keywords $NV$ (from the bug report) that are found in the query $Q$ & $\frac{1}{|NV|}\underset{q\in Q}{\sum} isNounOrVerb(q)$\\
				\hline
				$otherPOSRatio$ & Fraction of all the non-noun and non-verb keywords & $1-nounVerbRatio$\\
				\hline
				$titleKeywordRatio$ & Fraction of query keywords that are only found within the title of a bug report & $\frac{1}{|Q|}\underset{q\in Q}{\sum} inTitle(q)$\\
				\hline
				$bodyKeywordRatio$ & Fraction of query keywords that are only found within the body of a bug report & $\frac{1}{|Q|}\underset{q\in Q}{\sum} inBody(q)$
				\\
				\hline
				$titleBodyKWRatio$ & Fraction of query keywords that are found both in title and in body & $\frac{1}{|Q|}\underset{q\in Q}{\sum} inTitleBody(q)$ \\
				\hline
				$uniqueKeywords$ & Number of unique keywords in the query & $|removeDuplicates(Q)|$ \\
				\hline
				$gtTermRatio$ & Fraction of query keywords that are found in ground truth class names & $\frac{1}{|Q|}\underset{q\in Q}{\sum}inGroundTruth(q)$ \\
				\hline
				$avgPMI$ & Average Point-wise Mutual Information over all pairs of query keywords & $\frac{2(|Q|-1)!}{(|Q|)!}\underset{q1,q2\in Q}{\sum} PMI(q1,q2)$  \\
				\hline
				$medianPMI$ & Median PMI measure over all pairs of query keywords & $\underset{q1,q2\epsilon Q}{median}~(PMI(q1,q2))$ \\
				\hline
				$maxPMI$ & Maximum PMI measure over all pairs of query keywords & $\underset{q1,q2\epsilon Q}{max}~(PMI(q1,q2))$ \\
				\hline
				$stdPMI$ & Standard deviation of PMI measures over all pairs of query keywords & $\sqrt{\frac{1}{|Q|} \underset{q1,q2\in Q}{\sum}(PMI(q1,q2)-\overline{PMI})^2}$ \\
				\hline
%				$avgTR$ & Average TextRank of all keywords from the search query $Q$ within the texts of a bug report & $\frac{1}{|Q|}\underset{q\in Q}{\sum} TR(q)$ (check Equation \ref{eq:textrank})\\
%				\hline
%				$medianTR$ & Median TextRank of all keywords from the query $Q$ & $\underset{q\epsilon Q}{median}~(TR(q))$ \\
%				\hline
%				$maxTR$ & Maximum TextRank of all keywords from the query $Q$ & $\underset{q\epsilon Q}{max}~(TR(q))$ \\
%				\hline
%				$stdTR$ & Standard deviation of all TextRank scores from the query $Q$ & $\sqrt{\frac{1}{|Q|} \underset{q\in Q}{\sum}(TR(q)-\overline{TR})^2}$ \\
%				\hline
%				$avgPOSR$ & Average POSRank of all keywords from the search query $Q$ within the texts of a bug report & $\frac{1}{|Q|}\underset{q\in Q}{\sum} POSR(q)$ (check Equation \ref{eq:textrank}) \\
%				\hline
%				$medianPOSR$ & Median POSRank of all keywords from the query $Q$ & $\underset{q\epsilon Q}{median}~(POSR(q))$ \\
%				\hline
%				$maxPOSR$ & Maximum POSRank of all keywords from the query $Q$ & $\underset{q\epsilon Q}{max}~(POSR(q))$ \\
%				\hline
%				$stdPOSR$ & Standard deviation of all POSRank scores from the query $Q$ & $\sqrt{\frac{1}{|Q|} \underset{q\in Q}{\sum}(POSR(q)-\overline{POSR})^2}$ \\
%				\hline
				\multicolumn{3}{c}{$\mathbf{M(w)} = \frac{1}{2}(P(w)+Q(w))$, ~ $\mathbf{entropy(q)} = \underset{d\in D}{\sum} \frac{TF(q,d)}{TF(q,D)} \times \log_{D} \frac{TF(q,d)}{TF(q,D)}$}
				\\
				\multicolumn{3}{c}{ $\mathbf{PMI(q1,q2)}=log\frac{P_{q1,q2}(D)}{P_{q1}(D), P_{q2}(D)}$, $P_{q1,q2}(D)=\frac{|D_{q1}\cap D_{q2}|}{|D|}$, $P_q(D)=\frac{|D_q|}{|D|}$} \\
				\hline

			%\end{tabular}	
		%\end{threeparttable}
	%}
\end{longtable}
\normalsize

\newpage
\section{Comparison of Query Feature Distribution}

\begin{figure}[H]
	\centering
	\includegraphics[width=4.8in]{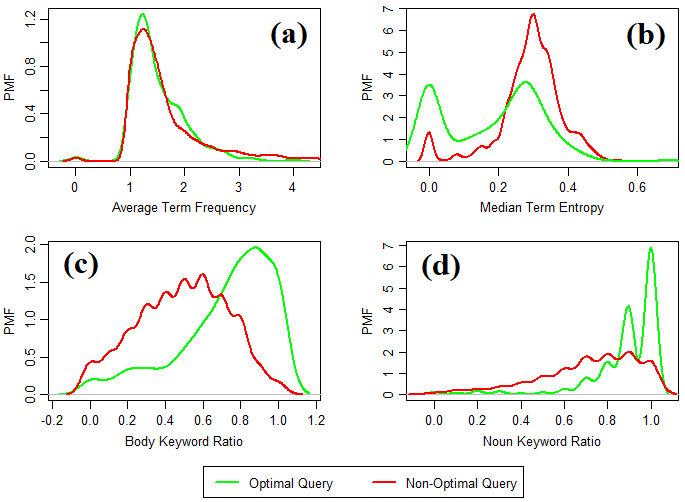}
	\vspace{-.3cm}
	\caption{Comparison between optimal and non-optimal keywords (from bug reports leading to poor baseline queries) using their distributions (e.g., PMF=probability mass function) of (a) frequency, (b) entropy, (c) percentage from the body/description section of a bug report, and (d) percentage of nouns}
	\label{fig:feature-distribution-lq}
	%\vspace{-.2cm}
\end{figure}

\newpage
\section{Comparison of Query Features using Box plots}

\begin{figure}[H]
	\centering
	\includegraphics[width=4.8in]{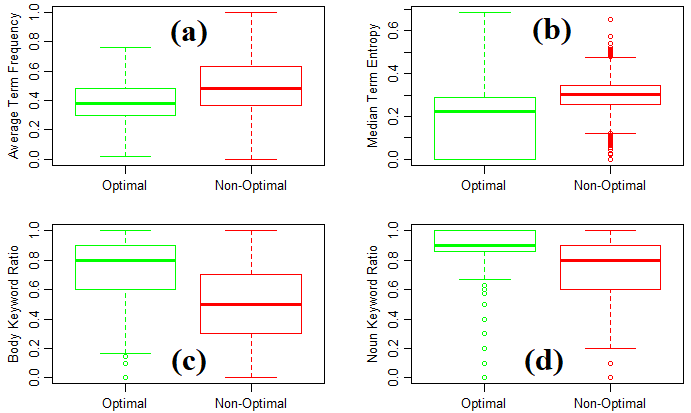}
	\vspace{-.3cm}
	\caption{Comparison between optimal and non-optimal keywords (from bug reports leading to poor baseline queries) using their box plot of (a) frequency, (b) entropy, (c) percentage from the body/description section of a bug report, and (d) percentage of nouns}
	\label{fig:boxplot-feature-distribution-lq}
	%\vspace{-.2cm}
\end{figure}

\bibliographystyle{plainnat}
\setlength{\bibsep}{0pt plus 0.3ex}
%\scriptsize
\bibliography{MyBibTex}  % sigproc.bib is the name of the Bibliography in this case
% ACM needs 'a single self-contained file'!
%\balancecolumns
\end{document}